\documentclass[final]{siamltex}
\usepackage[utf8x]{inputenc}
\usepackage{amsmath,amssymb,graphicx}
\usepackage{geometry}
 \newtheorem{remark}[theorem]{Remark}
 \newtheorem{conjecture}[theorem]{Conjecture}

\title{Numerical study of the long wavelength limit of the Toda lattice}
\author{C.~Klein\thanks{Institut de Math\'ematiques de Bourgogne,
		Universit\'e de Bourgogne, 9 avenue Alain Savary, 21078 Dijon
		Cedex, France
    ({\tt christian.klein@u-bourgogne.fr})}
\and
K.~Roidot\thanks{Fakult\"{a}t f\"{u}r Mathematik, Universit\"{a}t Wien - Wien Rossau,
Oskar-Morgenstern-Platz 1, 1090 Wien, \"{O}sterreich 
    ({\tt kristelle.roidot@univie.ac.at})}
}

\date{\today}

\begin{document}
\maketitle

\begin{abstract}
    We present the first detailed numerical study of the Toda equations in 
    $2+1$ dimensions in the limit of long wavelengths, both for the 
    hyperbolic and elliptic case. We first study 
    the formal dispersionless limit of the Toda equations and solve 
    initial value problems for the resulting system up to the point of
    gradient catastrophe. 
    It is shown that the break-up of the solution in the hyperbolic 
    case is similar to the shock formation in the Hopf equation, a 
     $1+1$ dimensional singularity. In 
    the elliptic case, it is found that the break-up is given by a 
    cusp as for the 
    semiclassical system of the focusing nonlinear Schr\"odinger 
    equation in $1+1$ dimensions. 
    
    The full Toda system is then studied for finite small values of 
    the dispersion parameter $\epsilon$ in the vicinity of the shocks 
    of the dispersionless Toda equations. We determine the 
    scaling in $\epsilon$ of the difference between the Toda solution 
    for small $\epsilon$ and the singular solution of the dispersionless 
    Toda system. In the hyperbolic case, the same scaling 
    proportional to
    $\epsilon^{2/7}$ is found as in the small dispersion limit of the 
    Korteweg-de Vries and the defocusing nonlinear Schr\"odinger 
    equations. In the elliptic case, we obtain the same scaling 
    proportional to
    $\epsilon^{2/5}$ as in the semiclassical limit for the focusing 
    nonlinear Schr\"odinger equation. We also study the formation of 
    dispersive shocks for times much larger than  the break-up 
    time in the hyperbolic case. In the elliptic case, an $L_{\infty}$ blow-up is 
    observed instead of a dispersive shock for finite times greater 
    than the break-up time. The $\epsilon$-dependence 
    of  the blow-up time is determined.
    
\end{abstract}

\section{Introduction}

The Toda lattice  \cite{Tod1, Fla741, Fla742, Mana2}, 
\begin{equation}
    \dot{q}_{n}=p_{n},\quad \dot{p}_{n} = 
    e^{q_{n+1}-q_{n}}-e^{q_{n}-q_{n-1}},\quad 
     n\in\mathbb{Z}
    \label{toda1}.
\end{equation}
is a completely integrable variant of the Fermi-Pasta-Ulam (FPU) system. The 
latter is given by the Hamiltonian,
\begin{equation}
    H(p,q) = 
    \sum_{n\in\mathbb{Z}}\left(\frac{1}{2}p_{n}^{2}+  V( q_{n}-q_{n-1})\right),
    \label{HFPU}
\end{equation}
which describes the interaction of a chain of particles with equal 
mass $m$ (here we have put $m=1$) via nearest-neighbor interactions. The displacement of the $n$-th particle from its equilibrium position is denoted by 
$q_n$, and $p_n$ represents its momentum.
This more general model provides an important example for nonlinear
 Hamiltonian many-particles dynamics
which exhibits a vast collection of complex phenomena, see for instance \cite{FPU55, ICh66} for first observations, and 
\cite{DrHe08} for energy transfer to short-wave modes via dispersive 
shocks. For comprehensive reviews on the subject
we refer 
to \cite{Gallavotti, Ford92, CaRoZa05}. In this paper, we are 
interested in the long wavelength limit of the Toda lattice.

The FPU dynamics have been widely investigated both numerically and analytically.
A substantial amount of understanding was obtained via
 the approximation of the FPU model by completely integrable systems, 
 like the KdV equation, see \cite{ICh66, ZK, BambPo1, FlPo08, 
 BeLiPo}, and from the integrable Toda model (\ref{toda1}), for which 
the potential $V$ is
of exponential form, see e.g.~\cite{FerFlMcL82, IsoLiRuVu, CaSoPeCo, GiPaPe, ZaSuPe06}.
It was shown in  \cite{Saitoh80} that the KdV equation 
can be viewed as a continuum approximation to the Toda lattice in the 
limit of long wavelengths and small amplitude \cite{Tod1}, and that 
 both systems 
are derived as limiting cases of a generalized equation, see also the recent work 
\cite{BamKapPau}.
%
The integrability of (\ref{toda1}) implies that many explicit 
solutions can be constructed, see for example 
\cite{Tod1, Tod2} for exact cnoidal wave solutions, and \cite{Tesch1, Tesch2}
for the construction of quasi-periodic finite-gap solutions.
 The limit of 
small dispersion for PDEs corresponds to 
the limit of long wavelength or an infinite number of particles for a lattice system. Following 
\cite{Dub08}, we treat the latter in the following way: we put
\begin{equation}
    u(n\epsilon) = q_{n+1}-q_{n},\quad v(n\epsilon)=p_{n},\quad 
     n\in\mathbb{Z},
    \label{uv1}
\end{equation}
and $x=n\epsilon$, so that $x$ becomes a  continuous variable in the 
limit $\epsilon\to0$. 
Rescaling $t\to t/\epsilon$,  we get for (\ref{toda1})
\begin{equation}
     \rho \, u_{tt}(x)=\frac{1}{\epsilon^2} \left( e^{u(x+\epsilon)} -2e^{u(x)}+e^{u(x-\epsilon)} \right).
    \label{todacont3}
\end{equation}
Here we have introduced a parameter $\rho=\pm1$ mainly for 
mathematical reasons.
The original equation (\ref{todacont3}) with $\rho=1$ is hyperbolic, 
but  we will also consider in this paper its 
elliptic variant $\rho=-1$ as for instance in \cite{MS09Tod}.
The energy for this equation reads 
\begin{equation}
    E = \sum_{n\in \mathbb{Z}}^{}\left(\frac{\rho}{2}v^{2}+e^{u}\right).
    \label{toda1en}
\end{equation}
The formal limit $\epsilon \to 0$ called the \emph{dispersionless Toda 
equation} 
in the following yields 
\begin{equation}
\rho \,    u_{t}=v_{x},\quad v_{t}= \left(e^{u}\right)_{x} ,\quad \rho=\pm 1,
    \label{toda1e0}
\end{equation}
for $u$ and $v$ defined in (\ref{uv1}). The equation has a conserved  
energy of the form (\ref{toda1en}), where the sum is replaced by an integral over 
$x$. In the present paper we concentrate on the Toda system since it 
appears more conceivable 
 that proofs for the presented conjectures can 
be obtained in the context of a completely integrable PDE. Note, 
however, that the used numerical techniques can be directly applied 
to general FPU systems.

One motivation to study the long wavelength limit of the Toda 
equation is the fact that it corresponds to the dispersionless limit of
dispersive partial differential equations (PDEs). It is well known 
that solutions to such equations  stay in 
general regular in $(x,t)$ near the point of gradient catastrophe of 
the solution to the corresponding dispersionless equation for the 
same initial data, but develop a zone of rapid modulated 
oscillations there, called \emph{dispersive shocks}. Dubrovin 
\cite{Dub06} identified a  class of 
Hamiltonian regularizations of the  Hopf equation which are 
integrable up to finite order in the small dispersion parameter 
$\epsilon$. This class of equations contains many classical 
integrable PDEs as the Korteweg-de 
Vries (KdV) equation. It was conjectured in \cite{Dub06} that the solutions 
to these equations near the point of gradient catastrophe of the 
corresponding Hopf solution can be described asymptotically via a
special solution to the second equation in the Painlev\'e I 
hierarchy denoted by PI2, see \cite{CV07,KKG}. The difference between the 
dispersionless and the dispersive solution near the critical point 
scales as $\epsilon^{2/7}$ at the critical time, the corrections via the PI2 solution 
appear in order $\epsilon^{4/7}$. This conjecture was numerically studied in 
\cite{GK08,DGK11} for various equations and proven for the KdV case by Claeys and Grava in 
\cite{CG09}.

In \cite{DGK,DGK13} this approach was generalized to dispersive 
regularizations of two-component systems, an important example of 
which are $1+1$ dimensional nonlinear Schr\"odinger (NLS) equations. 
It was conjectured that in hyperbolic such systems, the behavior of 
the solutions near a point of gradient catastrophe of the 
corresponding dispersionless solution is similar to the KdV case. But 
for elliptic systems, the solutions near the break-up of the 
corresponding dispersionless solution are asymptotically given by the 
\emph{tritronqu\'ee} solution \cite{Bou13} of the Painlev\'e I (PI)
equation. The difference between dispersionless and dispersive 
solution scales as $\epsilon^{2/5}$ in this case, the PI solution appears in order 
$\epsilon^{4/5}$ of the asymptotic description. Numerical evidence 
for these conjectures was provided in  \cite{DGK,DGK13}.

There are essentially no analytic results on dispersive 
shocks for nonlinear dispersive PDEs in $2+1$ dimensions. For a 
completely integrable generalization of the KdV equation 
to $2+1$ dimensions, the Kadomtsev-Petviashvili (KP) equation, the 
appearance of dispersive shocks was first numerically shown in 
\cite{KSM}. In \cite{dkpsulart} a detailed numerical study of the 
shock formation in the dispersionless KP (dKP) was presented which was 
based on the tracking of a singularity in the complex plane via the 
Fourier coefficients of the dKP solution. It was shown that this 
method first applied numerically by Sulem, Sulem and Frisch 
\cite{SSF} allows to quantitatively identify the critical time and 
the break-up solution. This made it possible to obtain the scaling of the 
difference between KP and dKP solutions close to the critical point. 
It was shown that the break-up 
behavior is as in the $1+1$ dimensional case of KdV and Hopf 
equation, the difference between both solutions scales as 
$\epsilon^{2/7}$. A similar study has been presented in \cite{DSdDS} 
for the Davey-Stewartson (DS) II equation, an integrable $2+1$ dimensional 
generalization of the NLS equation. The latter has a focusing 
(corresponding to an elliptic system) and a defocusing (corresponding 
to a hyperbolic system) variant. It was found in \cite{DSdDS} that the 
defocusing DS II solutions behave near the break-up of the corresponding
dispersionless system (also called \emph{semiclassical} in this case) 
as the KdV and the  KP solution. The focusing solutions on the other 
hand show the same behavior as solutions to the focusing $1+1$ 
dimensional NLS equation. This could indicate that these two  
cases represent some universal behavior of break-up in nonlinear 
dispersive PDEs not only in $1+1$, but higher dimensions. 
Note that the class of equations studied in \cite{Dub06} also contains 
discrete systems as the Volterra and Toda lattice (\ref{toda1}) in 
the limit of long wavelengths.

To show that the numerical results on critical behavior 
in $2+1$ dimensional dispersive systems are not limited to continuous PDEs only, we 
present in this paper a numerical study of 
the $2+1$ dimensional Toda system,
\begin{equation}
    \dot{q}_{n}=p_{n},\quad \dot{p}_{n} = 
    e^{q_{n+1}-q_{n}}-e^{q_{n}-q_{n-1}}+q_{n,yy},\quad 
     n\in\mathbb{Z},
    \label{toda2}
\end{equation}
which is known to be also completely 
integrable, see \cite{MatWat88}. The two-dimensional Toda system  follows from the Hamiltonian
\begin{equation}
    H = 
    \sum_{n\in\mathbb{Z}}^{}\int_{\mathbb{R}}^{}\left(\frac{1}{2}p_{n}^{2}+\frac{1}{2}q_{n,y}^{2}
    +e^{q_{n}-q_{n-1}}-1\right)dy
    \label{H2}.
\end{equation}

Again we are interested in the limit of long wavelengths. With 
(\ref{uv1}) we get
after the change of scale  $y\to y/\epsilon$,
for (\ref{toda2})
\begin{equation}
\begin{array}{ccc}
  \rho \,  u_{t}(x) & = & \frac{1}{\epsilon}(v(x+\epsilon)-v(x)),\\
    v_{t}(x+\epsilon)-v_{t}(x) & = & \frac{1}{\epsilon}\left(e^{u(x+\epsilon)}+e^{u(x-\epsilon)}-2e^{u(x)}+\epsilon^{2}u_{yy}\right), 
    \end{array}
    \label{toda2cont}
\end{equation}
where we have once more introduced the parameter $\rho=\pm1$. 
The plus sign corresponds to the 
hyperbolic case, and the minus sign to 
the elliptic case.
In the limit $\epsilon\to0$ we get the dispersionless two-dimensional Toda equation 
\begin{equation}
 \rho \,   u_{tt}=(e^{u})_{xx}+u_{yy}, \, \rho = \pm 1
    \label{toda2e0}.
\end{equation}
The latter equation is not  completely integrable in the classical 
sense, i.e., it cannot be 
solved via a linear 
Riemann-Hilbert problem (RHP). But it can be treated with the nonlinear 
RHP
approach by Manakov and Santini \cite{MS09Tod} (note that in 
\cite{MS09Tod} the 
elliptic variant of (\ref{toda2e0}) is studied with $x$ and $t$ 
interchanged), which allowed in \cite{MS09Tod} to study the long 
time behavior of the solutions. It is also possible to solve equation (\ref{toda2e0}) 
with methods from the theory of infinite dimensional Frobenius 
manifolds. The corresponding manifold for (\ref{toda2e0}) was 
constructed in 
\cite{CarDubMer11}. The dispersionless Toda equation (\ref{toda2e0}) 
is equivalent to the Boyer-Finley equation \cite{BF}
$$u_{\xi\eta}=(e^{u})_{tt},$$
which follows from (\ref{toda2e0}) by interchanging the coordinates 
$x$ and $t$ as in \cite{MS09Tod} and using characteristic 
coordinates. The Boyer-Finley equation appears in the theory of general 
relativity as the self-dual Einstein equations with a Killing vector. 
In addition to the above mentioned techniques, it can be treated with 
Twistor methods \cite{DMT} and hydrodynamic reductions \cite{FKS}.

These approaches as for instance the nonlinear RHP and infinite 
dimensional Frobenius manifolds
are rather implicit if a Cauchy 
problem has to be solved, and so far such a program has not been 
successfully implemented. Therefore we numerically integrate equation 
(\ref{toda2e0}) up to the critical time $t_{c}$ at which 
the first point of gradient catastrophe appears. Then 
we solve the two-dimensional Toda equation (\ref{toda2cont}) for  small, nonzero
$\epsilon$ for the same initial data up to the time $t_{c}$ and study 
the scaling of the difference between the solution to the 
dispersionless and the two-dimensional Toda equation with small $\epsilon$. The 
latter is also solved for larger times. In the hyperbolic case we 
find a dispersive shock, in the elliptic case generically an 
$L_{\infty}$ blow-up. The results of the numerical study can be 
summarized in the following 
\begin{conjecture}
Consider initial data which are the $x$ derivative of a rapidly 
decreasing smooth function in $L_{2}(\mathbb{R}^{2})$ with a single 
maximum. Then
\begin{itemize}
    \item  Solutions to the 2d hyperbolic dispersionless Toda equation 
    ($\rho=1$ in
    (\ref{toda2e0})) will have one or more points of gradient 
    catastrophe at finite times $t_{c}$. Generically these will be  
    cubic singularities at which the solution has a finite $L_{\infty}$ 
    norm. As for dKP (see \cite{dkpsulart} and references therein), 
    the solution becomes singular at $t_{c}$
    only in one direction in the $x,y$-plane and stays regular in the 
    second (these directions only coincide with the coordinate axes 
    for special initial data). 
    
    \item  Solutions to the 2d elliptic  dispersionless Toda equation 
     ($\rho=-1$ in (\ref{toda2e0}) ) will have a point of gradient 
    catastrophe at a finite time $t_{c}$. Generically this will be a 
    square root singularity at which the solution has a finite $L_{\infty}$ 
    norm. The solution becomes singular at $t_{c}$
    only in one direction in the $x,y$-plane and stays regular in the 
    second (these directions only coincide with the coordinate axes 
    for special initial data). 

    \item    The difference between  solutions to the 2d hyperbolic Toda equation 
    (\ref{toda2cont}) and the dispersionless Toda equation 
    (\ref{toda2e0}) (both with $\rho=1$) will scale as 
    $\epsilon^{2/7}$ at the critical time $t_{c}$.

    \item    The difference between  solutions to the 2d elliptic Toda equation 
    (\ref{toda2cont}) and the dispersionless Toda equation 
    (\ref{toda2e0}) (both with $\rho=-1$) will scale as 
    $\epsilon^{2/5}$ at the critical time $t_{c}$.

    \item    Solutions to the 2d elliptic Toda equation 
    (\ref{toda2cont})  ($\rho=-1$) will blow up in finite time 
    $t^{*}>t_{c}$ for $\epsilon\ll 1$. In the limit $\epsilon\to0$ 
    the difference $t^{*}-t_{c}$ tends to zero as $\epsilon^{0.9}$.

\end{itemize}
\end{conjecture}

The paper is organized as follows: in section 2 we present a 
convenient formulation of the Toda equations with and without 
dispersion in two dimensions and collect the used numerical 
approaches to integrate them. In section 3 we treat the hyperbolic 
case both with and without dispersion in $1+1$ and $2+1$ dimensions. 
The same treatment for the 
elliptic case is presented in section 4. We add some concluding 
remarks in section 5.

\section{Numerical Methods}

In this section we present a formulation of the Toda 
equations convenient for the numerical treatment and the numerical tools 
to efficiently integrate these equations up to possibly appearing 
singularities. The task is to resolve strong gradients in dispersive 
shocks and in break-up or $L_{\infty}$ blow-up of the solutions. 

For the spatial dependence of the solution we use a Fourier spectral 
method. We denote the two-dimensional Fourier transform of a function 
$f(x,y)\in L_{2}(\mathbb{R}^{2})$ by
$$\hat{f}(k_{x},k_{y})=\int_{\mathbb{R}^{2}}^{}f(x,y)\exp(-ik_{x}x-ik_{y}y)dxdy.$$
The choice of a Fourier method is  
convenient here because 
we want to identify 
singularities on the real axis by tracing them in the complex plane 
via the asymptotic behavior of the Fourier coefficients as in 
\cite{SSF}. In addition we are interested in the excellent 
approximation properties of spectral methods for smooth functions, 
and what is especially interesting in the context of dispersive 
equations, the minimal introduction of numerical dissipation by 
spectral methods. Thus we approximate the spatial dependence via a 
discrete Fourier transform computed with a \emph{fast Fourier 
transform}. For the resulting finite dimensional system of ordinary 
differential equations, we use the standard explicit fourth order 
Runge-Kutta method. The reason why we do not use the stiff integrators 
applied in a similar context in \cite{KassT,ckkdvnls,KR} is that the 
dispersion and thus the stiffness does not appear only in the linear 
part of the equation here. 

To treat the two-dimensional Toda equation (\ref{toda2cont}), we introduce the 
operator $T$ acting on a function of $x$ via 
\begin{equation}
    Tu(x)=\frac{1}{\epsilon}(u(x+\epsilon)-u(x))
    \label{T}.
\end{equation}
 This means it has
the Fourier symbol
$$
    \hat{T}=\frac{1}{\epsilon}\left(e^{ik_{x}\epsilon}-1\right).
$$
In the limit $\epsilon\to0$, this operator obviously becomes the 
derivative with respect to $x$. With (\ref{T}), equation (\ref{toda2cont}) 
can thus be put into the form
\begin{equation}
   \rho \, u_{t}=Tv,\quad 
    v_{t}=T e^{u(x-\epsilon)}+T^{-1}u_{yy}, \, \, \rho = \pm 1.
    \label{toda2T}
\end{equation}
The conserved energy for this equation can be 
written as
\begin{equation}
    E = \sum_{n\in\mathbb{Z}}^{}\int_{\mathbb{R}}^{}dy \left(\frac{\rho}{2}v^{2}
    +\frac{1}{2}(T^{-1}u_{y})^{2}+e^{u}\right).
    \label{toda2E}
\end{equation}

The appearance of the operator $T^{-1}$ indicates a nonlocality in 
this form of the equation which becomes the anti-derivative in the 
limit $\epsilon\to0$. It corresponds to a division by $k_{x}$ in 
Fourier space. To address potential problems near $k_{x}=0$, we consider only initial data 
which are  the $x$-derivative of some rapidly decreasing function and 
thus are proportional to $k_{x}$ which allows for  
a division by $k_{x}$ in Fourier space. We 
introduce the function $U$ via $u=U_{x}$. Then we get for the Toda 
equation in Fourier space a form of the equation without division by $k_{x}$,
\begin{equation}
 \rho \,   \hat{U}_{t}=\phi\hat{v},\quad 
    \hat{v}_{t}=\phi 
    ik_{x}\widehat{e^{u(x-\epsilon)}}-k_{y}^{2}\phi^{-1}\hat{U},
    \label{toda2f}
\end{equation}
where $\phi=\phi_{1}(ik_{x}\epsilon)$, with $\phi_1(z)=(e^z-1)/z$. 
This is the first $\phi$ function appearing also in \emph{ exponential 
time differencing} schemes, see \cite{KassT} and references therein.  
The efficient and accurate numerical evaluation of this function is a well known numerical problem, 
because of \emph{cancellation errors} for $|z|\sim0$. A possible way to 
avoid such errors are complex contour integrals as in \cite{KassT,Schme} or 
Taylor series expansions for small $z$. We apply both methods here to 
ensure that the function is computed with machine precision. 

In the limit $\epsilon\to0$, we get 
\begin{equation}
  \rho \,    U_{tt}=(e^{U_{x}})_{x}+U_{yy}, \, \rho = \pm 1,
    \label{toda2e0int}
\end{equation}
with conserved energy
\begin{equation}
    E = 
    \int_{\mathbb{R}^{2}}^{}\left(\frac{\rho}{2}U_{t}^{2}
    +e^{U_{x}}+\frac{1}{2}U_{y}^{2}\right)dxdy
    \label{E2d0}.
\end{equation}
To control the accuracy of the numerical solution to (\ref{toda2T}) 
and (\ref{toda2e0int}), we use the relative numerically
computed energy,
\begin{equation}
    \Delta_{E}:=\frac{E(t)}{E(0)}-1,
    \label{Delta}
\end{equation}
which will depend on time due to unavoidable numerical errors. It was 
shown in \cite{ ckkdvnls,KR} that this quantity typically overestimates the 
difference between numerical and exact solution by two orders of 
magnitude. It is crucial in this context that sufficient resolution 
in Fourier space is provided since  $\Delta_{E}$ cannot indicate 
reliably a higher accuracy than imposed by the spatial resolution. 
Therefore we always present the Fourier coefficients at the last 
computed time to ensure that the coefficients decrease to the wanted 
precision. 

\begin{remark}
\emph{    Equation (\ref{toda1en}) can be treated analytically with the 
    hodograph method as explained for instance in the case of the 
    semiclassical NLS system in \cite{DGK}. 
Interchanging dependent and independent variables and writing 
\begin{equation}
    x = f_{u}(u,v),\quad t = f_{v}(u,v)
    \label{hodo1}
\end{equation}
we find that $f(u,v)$ must satisfy in the 
hyperbolic case $\rho=1$ the linear equation
\begin{equation}
    f_{uu}=f_{vv}e^{u}
    \label{hodof}.
\end{equation}
With the coordinate change $u=2\ln r$ and $\nu = v/2$, we get for 
(\ref{hodof})
\begin{equation}
    f_{rr}+\frac{1}{r}f_{r}=f_{\nu\nu}
    \label{ED},
\end{equation}
i.e., the hyperbolic Euler-Darboux equation. It would be attractive to find 
localized smooth initial data $u_{0},v_{0}$ which solve 
(\ref{hodof}), since these would provide an analytic test for our 
numerical approach in $1+1$ dimensions. A possible way is to 
construct data such that $f_{v}=0$ 
for $t=0$ by choosing $v_{0}=0$ and $f(u,v)$ to be an even function of $v$. 
A solution to (\ref{ED}) can be given in the form
\begin{equation}
    f(u,v) = \int_{-1}^{1}\frac{g(r\mu+\nu)}{\sqrt{1-\mu^{2}}}d\mu
    \label{int},
\end{equation}
where $g$ is some H\"older continuous function.
But we are interested here for numerical reasons in solutions which 
are rapidly decreasing in $x$. Since $u=2\ln r$, we did not find an 
analytic expression with the wanted properties. Therefore we also use 
a purely numerical approach for the $1+1$ dimensional case. 
}\end{remark}
\\
\\

In this paper we essentially observe two types of blow-up, a gradient 
catastrophe at which the solution itself stays bounded, and an 
$L_{\infty}$ blow-up. Both singularities are identified with a method 
from asymptotic 
Fourier analysis first applied to 
numerically identify singularities in solutions to PDEs in 
\cite{SSF}. 
The idea is to 
use that a singularity at $z_{0}$ of a real 
function in the complex plane of the form $U\sim (z-z_{0})^{\mu}$ 
($\mu$ not an integer)
leads for $|k|$ large to a Fourier transform of the form (if this 
is the only singularity of this type in the complex plane)
\begin{equation}
    |\hat{U}|\sim 
    \frac{1}{k^{\mu+1}} e^{-k\delta} e^{ i\alpha k}, \,\, |k| \to \infty
    \label{fourasymp}
\end{equation}
where $\delta=\Im z_{0}$ and $\alpha=\Re z_{0}$.
Through the analysis of its Fourier spectrum, one can thus 
determine the 
the first time $t_c$ where a function $U(t)$ develops a 
singularity on the real axis, i.e., where the real solution becomes 
singular, as given by the first $t_c$ at which $\delta(t_c )$ vanishes. 
%
The real part $\alpha$ of $z_0$, the location of the singularity, can be determined by 
estimating the period of the oscillations of the spectrum. 
In addition this method provides the quantity $\mu$ which 
characterizes the type of the singularity. 
Concretely the Fourier 
coefficients of a function $U$ are fitted to 
\begin{equation}
    \ln |\hat{U}|\sim A-B\ln |k|-\delta |k|
    \label{fourasymp2}.
\end{equation}
This method has been used for both ordinary and 
partial differential equations \cite{SSF, PS98, RLSS, CR, MBF, FMB, SSP, BMS85, GaSS09}, and 
more recently, in \cite{dkpsulart, DSdDS, KNLSsol, GaSSC}.
This approach requires high numerical precision in the simulations to 
avoid interference of the round-off errors.
It was shown in 
\cite{dkpsulart} that the quantity $\delta$ can be identified reliably 
from a fitting of the Fourier coefficients, whereas there is a larger 
uncertainty in the quantity $\mu=B-1$. In \cite{DSdDS} these techniques 
were applied to NLS equations. It was shown by comparison with exact 
solutions that in the focusing 
case, the best results are obtained when the code is stopped once the 
singularity is closer to the real axis than the minimal resolved 
distance via Fourier methods,  
\begin{equation}
    m:=2\pi L/N
    \label{mres}
\end{equation}
with $N$ being the number of 
 Fourier modes and $2\pi L$ the length of the computational domain in 
 physical space.  All values of $\delta$ below this 
 threshold cannot be distinguished numerically from 0.
 We perform the study here only in the 
$x$-direction, as done in \cite{dkpsulart}, firstly because 
we are expecting only one-dimensional gradient catastrophe
and secondly because it turns out (see \cite{DSdDS}) that 
in the case of $L_{\infty}$ blow-up phenomena
the study of the 
Fourier coefficients in only one direction is sufficient to determine 
the appearance of the blow-up.
Moreover, following the $2/3$ dealiasing rule (the coefficients of 
the highest one third 
of the wave numbers $|k|$ is put equal to zero to address the 
\emph{aliasing error} of the Fourier method whilst calculating 
nonlinear terms), we usually consider the interval $10 < k < 2 \max(k)/3$ for the fitting of the Fourier coefficients.

\section{Numerical study of the hyperbolic Toda equations}
In this section we study the hyperbolic $2+1$ dimensional Toda equations in the 
limit of long wavelengths for initial data being the $x$-derivative 
of a rapidly decreasing function with a single hump. The initial data 
are supposed to being invariant with respect to $y\to-y$. The latter 
condition will lead to a break-up on the $x$-axis and facilitates the 
identification of the shock formation. The reason for this choice is 
that one can identify the singularity by using one-dimensional 
techniques which are more reliable than a fully two-dimensional 
approach as for instance in \cite{DSdDS}. 
We first consider the one-dimensional case to test the numerical 
approaches via analytic expectations. 
For the dispersionless equation we show that the solutions for the 
considered initial data lead to a point of gradient catastrophe with 
a cubic singularity as known from solutions to the Hopf equation. The 
difference between the singular solution to the dispersionless 
equation and the corresponding Toda solution is shown to scale as 
$\epsilon^{2/7}$ as expected from \cite{Dub08}. For times much larger 
than the critical time $t_{c}$, a dispersive shock is observed. As in 
the case of the KP equation \cite{dkpsulart} and the defocusing DS II 
equation \cite{DSdDS}, where also the found singularities were 
one-dimensional, the same behavior as in one dimension 
is  found for solutions 
to the hyperbolic Toda equation in two dimensions: 
\begin{itemize}
    \item  the solution to 
the dispersionless equation has a point of gradient catastrophe on 
the $x$-axis of cubic type, whereas it stays regular in 
$y$-direction;

    \item  the difference between the solutions to the 
dispersionless and the full Toda equation scales as $\epsilon^{2/7}$ 
as in the one-dimensional case;

\item for times much larger than $t_{c}$, a dispersive 
shock is observed in the vicinity of the shock of the dispersionless 
solution. 
\end{itemize}
The behavior of the Toda solution for small $\epsilon$ at the 
critical time $t_{c}$ indicates that the PI2 equation 
might play a role in the asymptotic description of break-up also in 
two dimensions. 
 
%
%
%
%
 
\subsection{One-dimensional hyperbolic dispersionless Toda equation}
In this subsection we study the formation of a point of gradient 
catastrophe in solutions to the hyperbolic dispersionless Toda 
equation (\ref{toda1e0}) with $\rho=1$ in one 
spatial dimension. 
We consider initial data of the form
\begin{equation}
u(x,0)=u_0(x)=-2x \exp(-x^2), \,\,\, v(x,0)=v_0(x)=0.
\label{uini1}
\end{equation}
Note that we use here also initial data being a derivative of a 
rapidly decreasing function though the nonlocality, which is the 
reason for applying  
such data to avoid numerical problems, only appears in the 
two-dimensional case (\ref{toda2T}). But since we see the 
one-dimensional case as a test ground for two dimensions, we treat 
similar data in both settings.
The computations are carried out for $x  \in [-5\pi, 5\pi]$ with a 
number $N$ of Fourier modes and time step $\delta_{t}$. 
We perform the fitting of the Fourier coefficients to the asymptotic 
formula (\ref{fourasymp2}) on the interval $10<k<2*\mbox{max}(k)/3$. This is 
done for the Fourier coefficients of both $u$ and $v$, denoted in the following by $uf$ and $vf$.

\begin{figure}[htb!]
\begin{center}
\includegraphics[width=0.4\textwidth]{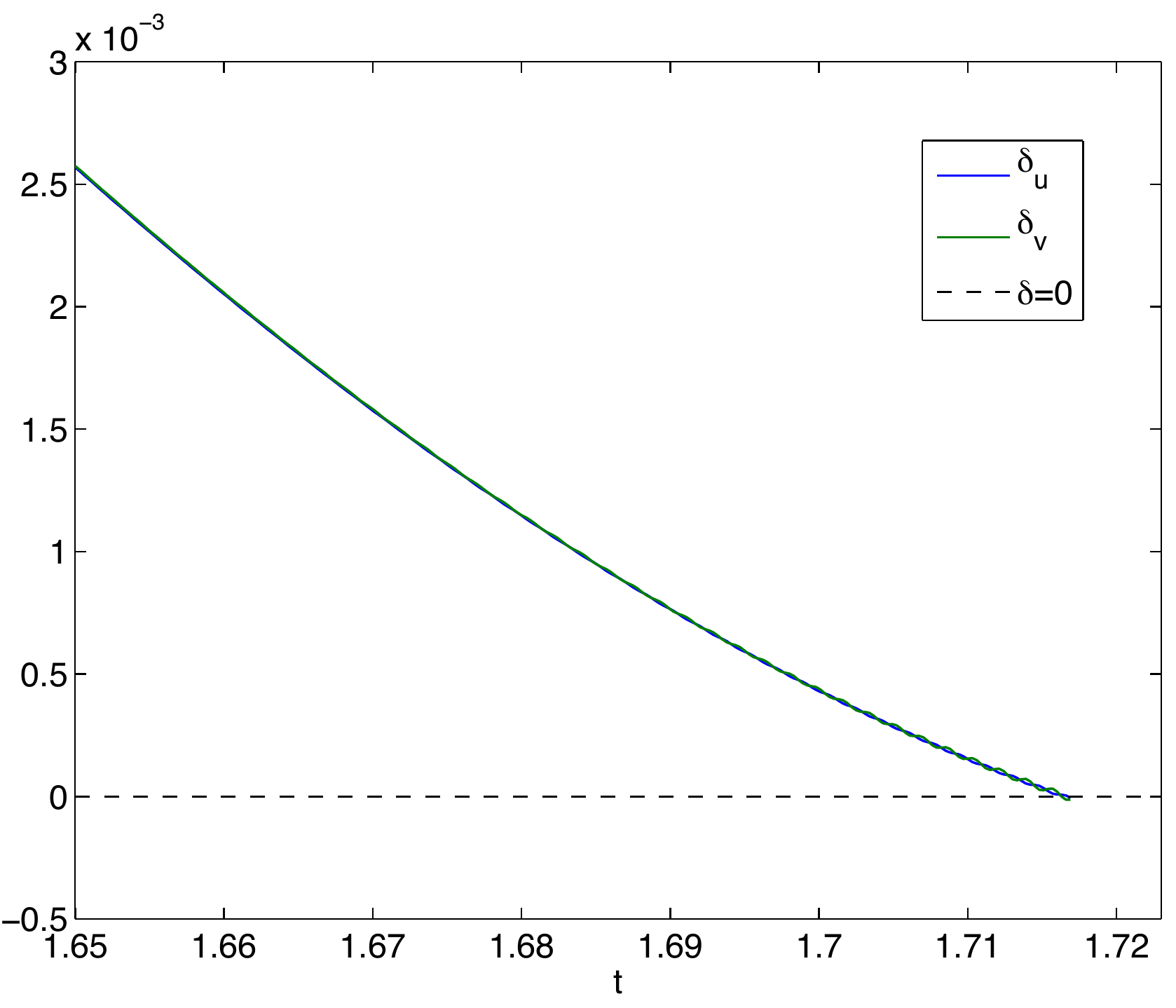}
\caption{Fitting parameters $\delta_u$ and $\delta_{v}$ corresponding 
to the fitting to (\ref{fourasymp2}) of the Fourier coefficients of $u$ and 
 $v$ respectively. Here $u $ and $v$ are the solution to the 1d 
 hyperbolic dispersionless Toda equation (\ref{toda1e0}) with $\rho=1$ for
 initial data of the form (\ref{uini1}).}
\label{dels}
\end{center}
\end{figure}

In the hyperbolic case, i.e. $\rho=1$ in (\ref{toda1e0}), we use  
$N=2^{14}$ and $\delta_t = 3*10^{-4}$ and determine $\delta_u$ and 
$\delta_{v}$,
 corresponding to the $\delta$ parameter in (\ref{fourasymp2}) for the 
 fitting of the Fourier coefficients of $u$ and  of $v$ respectively. 
 We find that they vanish at the same time, $t=t_c=1.717$, as can be 
 seen in Fig.~\ref{dels}. Note that there will be a second point of 
 gradient catastrophe for negative $x$ for $t>t_{c}$. Since  equation 
 (\ref{toda1e0}) does not respect the symmetry of the initial 
data (\ref{uini1}), $u_{0}(-x)=-u_{0}(x)$, the break-up does not occur 
at the same time for positive and negative $x$. 

\begin{figure}[htb!]
\begin{center}
\includegraphics[width=0.4\textwidth]{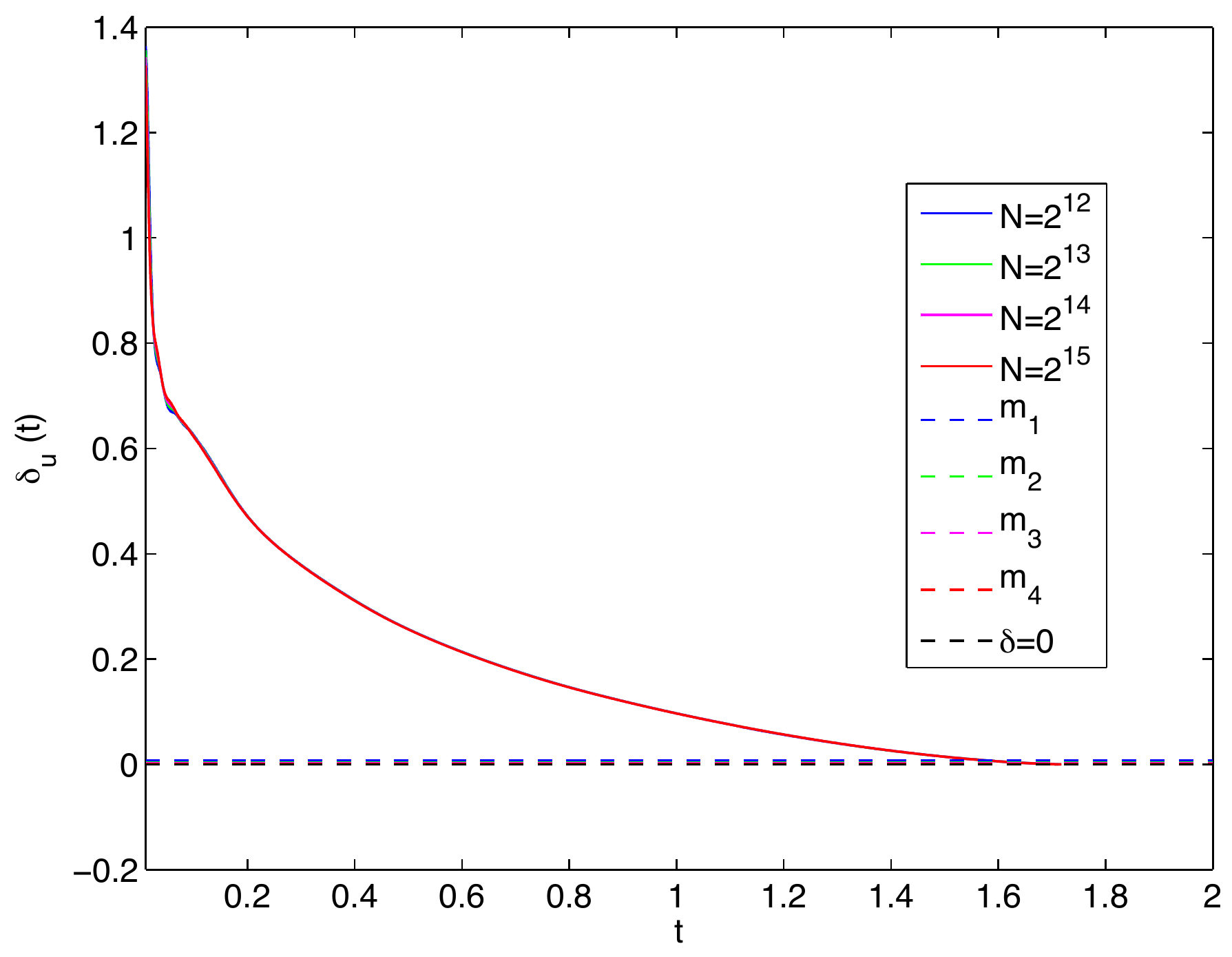}
\includegraphics[width=0.4\textwidth]{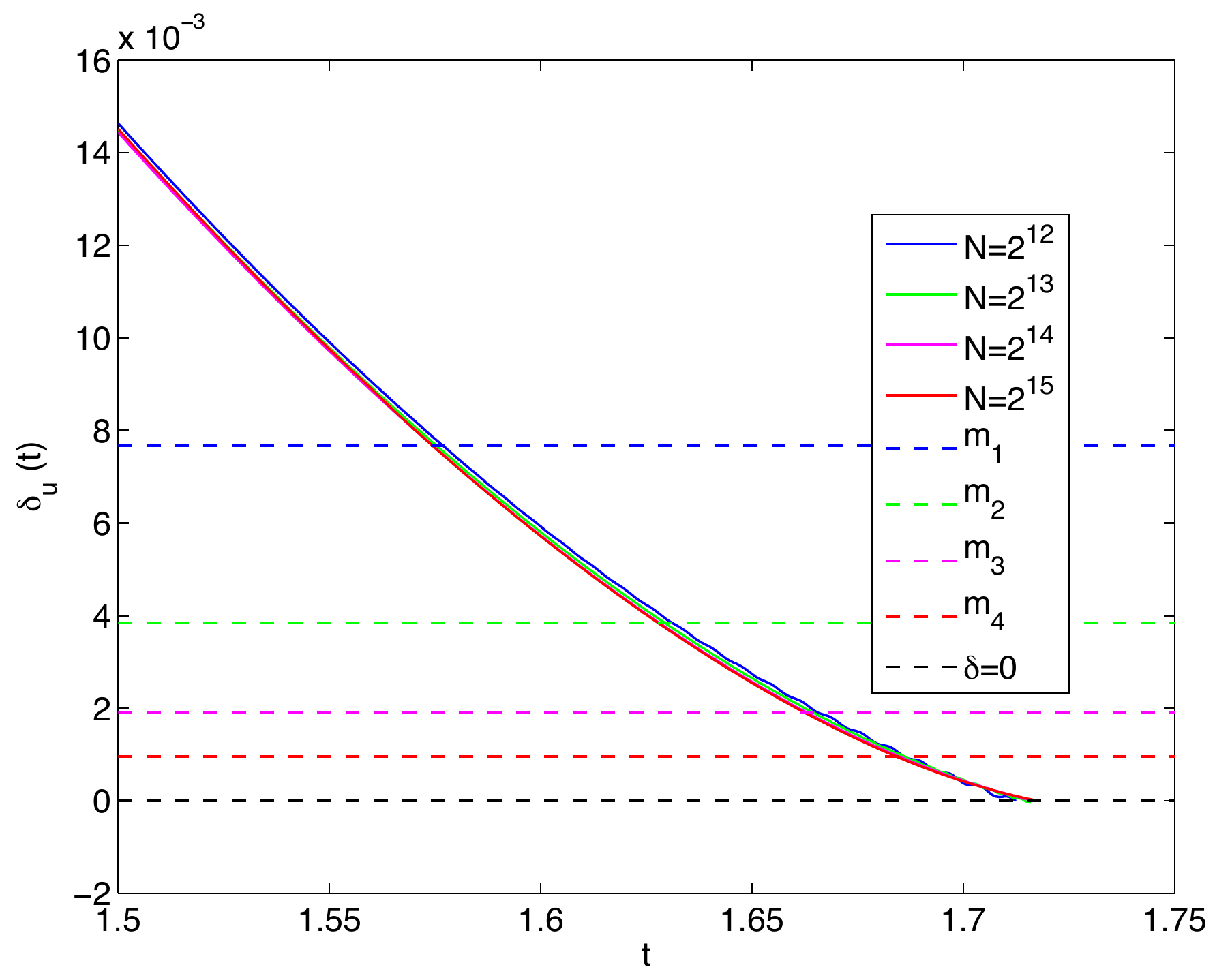}
\caption{Fitting parameter $\delta_u$ for the solution of the 1d hyperbolic dispersionless Toda equation (\ref{toda1e0}) with $\rho=1$ for initial data of the form \ref{uini1}. 
The fitting is done for $10<k<2*max(k)/3$ for different resolutions 
$N=2^{12},2^{13},2^{14},2^{15}$. The figure on the right shows a 
close-up of the left figure for $t\sim t_{c}$. }
\label{deluNs}
\end{center}
\end{figure}

To test the influence of resolution in Fourier space on the found 
results, we present in Fig.~\ref{deluNs} the time evolution of $\delta_u$ for 
different resolutions, namely $N=2^{12},2^{13},2^{14},2^{15}$.

\begin{figure}[htb!]
\begin{center}
\includegraphics[width=0.6\textwidth]{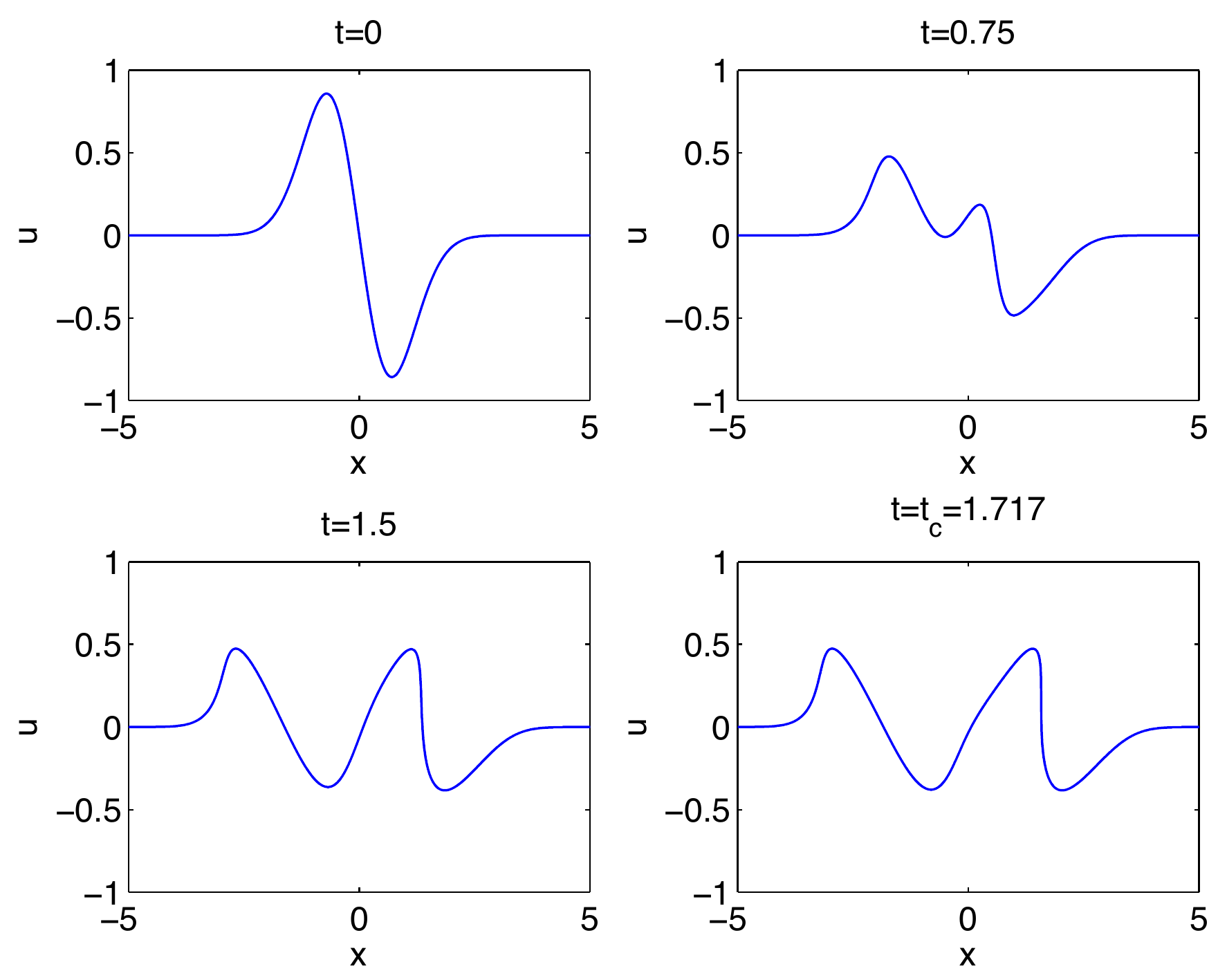}
\caption{Solution $u$ of the 1d hyperbolic dispersionless Toda equation  (\ref{toda1e0}) with $\rho=1$ at different times for initial data of the form (\ref{uini1}).}
\label{Soluts}
\end{center}
\end{figure}

In the figure, $m_i$ denotes the minimal distance (\ref{mres}) in physical space 
for a given resolution, $m_i=10\pi /N_i, i=1,2,3,4$ corresponding to 
$N=2^{12},2^{13},2^{14},2^{15}$.
As discussed in \cite{DSdDS}, the computation  can be stopped once 
$\delta_u$ vanishes in the case of a cubic singularity as expected 
here. Visibly the  dispersionless Toda equation  
behaves as the Hopf equation, i.e., its solutions develop a cubic shock. 
As observed before, the vanishing of $\delta_u$ occurs at $t_c \sim 1.717$. 
At this time, the solution $u$ has indeed a cubic shock, as can 
seen in Fig. \ref{Soluts}, where we present $u$ at several times.

\begin{figure}[htb!]
\begin{center}
\includegraphics[width=0.6\textwidth]{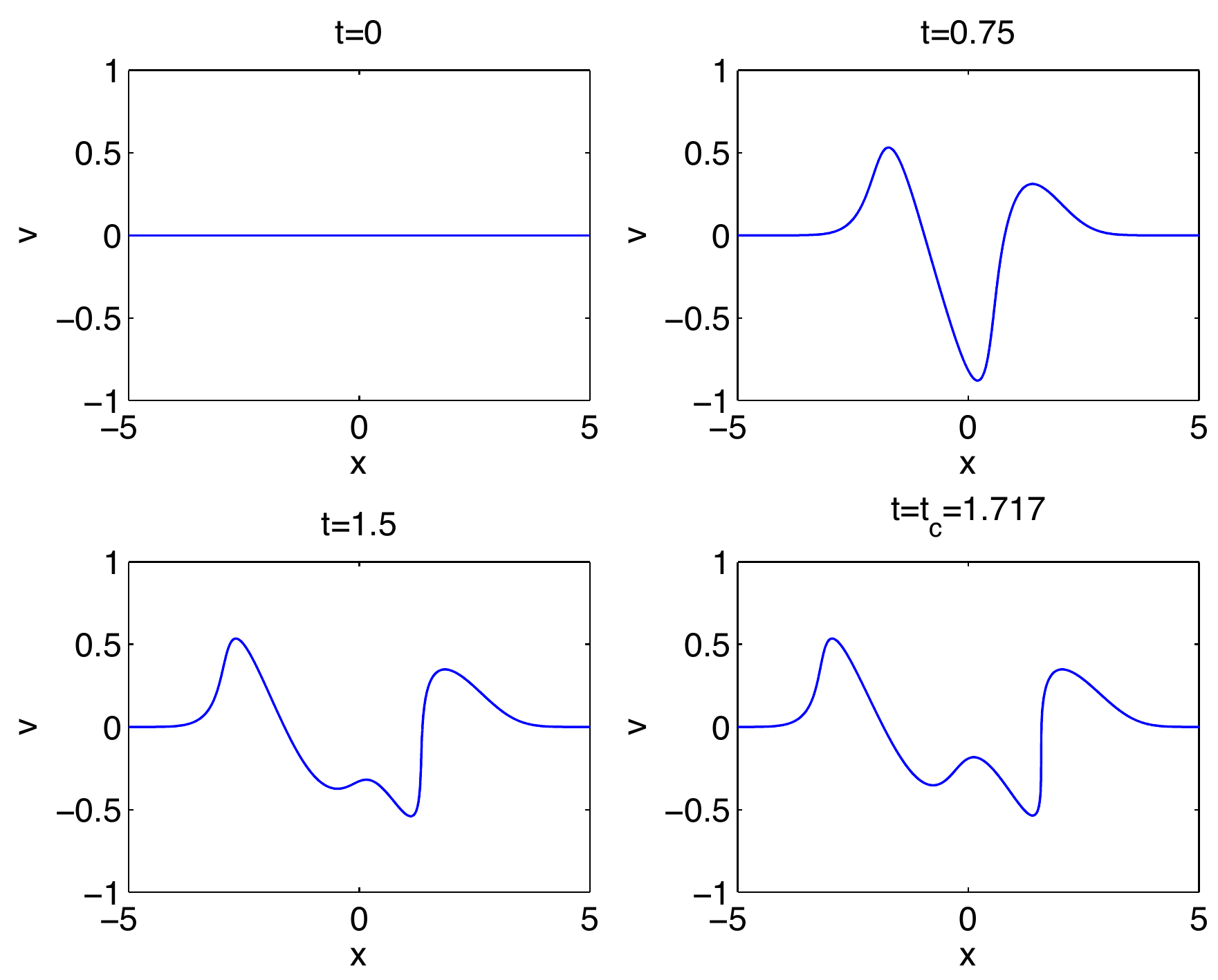}
\caption{Solution $v$ of the 1d hyperbolic dispersionless Toda equation  (\ref{toda1e0}) with $\rho=1$ at different times for initial data of the form (\ref{uini1}). 
}
\label{Solvts}
\end{center}
\end{figure}

As was shown in \cite{dkpsulart, DSdDS}, the numerical approach based 
on asymptotic Fourier analysis   is very efficient in the case of a 
cubic singularity, and the fitting parameter $B_u$ 
converges to the theoretical value ($4/3$) as $N$ increases. This 
indicates that the  fitting is reliable. Moreover, we are able to 
determine also the location of the singularity as $x_c=1.5808$ in the 
present example. 
It can be seen from Fig. \ref{Solvts} that $v$ as well has a cubic 
singularity at $t=t_c$.  
This can be also obtained from the 
fitting of the Fourier coefficients $vf$. It had been noted that 
$\delta_v$ vanishes at the same time as $\delta_u$, moreover 
the values of $B_u$ and $B_v$ are almost identical ($\sim 1.34$), 
indicating shock formation at the critical time, $t_c=1.717$.

We present in Table \ref{NtcBuv} the values of $t_c$, $B_u(t_c)$ and $B_v(t_c)$ for each resolution used. 
 As $N$ increases, $B_u(t_c)$ respectively $B_v(t_c)$ approach $4/3$.
 \begin{table}
\centering
\begin{tabular}{|c|ccc|}
\hline

  $N$ & $t_c$ & $B_u(t_c)$ & $B_v(t_c)$ \\
  \hline
 
  $2^{12}$ & 1.7121 & 1.3518 & 1.3570 \\
  $2^{13}$ & 1.7157 & 1.3515 & 1.3447\\
   $2^{14}$ & 1.7166 & 1.3473 & 1.3488\\
  $2^{15}$ & 1.7175 & 1.3461 & 1.3440\\
  \hline
\end{tabular}
\label{NtcBuv}
\caption{Critical times of the solution to the 1d hyperbolic dispersionless Toda equation for initial data of the form (\ref{uini1})
for several values of $N$. The values of the fitting parameters 
$B_u$, $B_v$ at $t_c$ are also given.}
\end{table}

\begin{figure}[htb!]
\begin{center}
\includegraphics[width=0.4\textwidth]{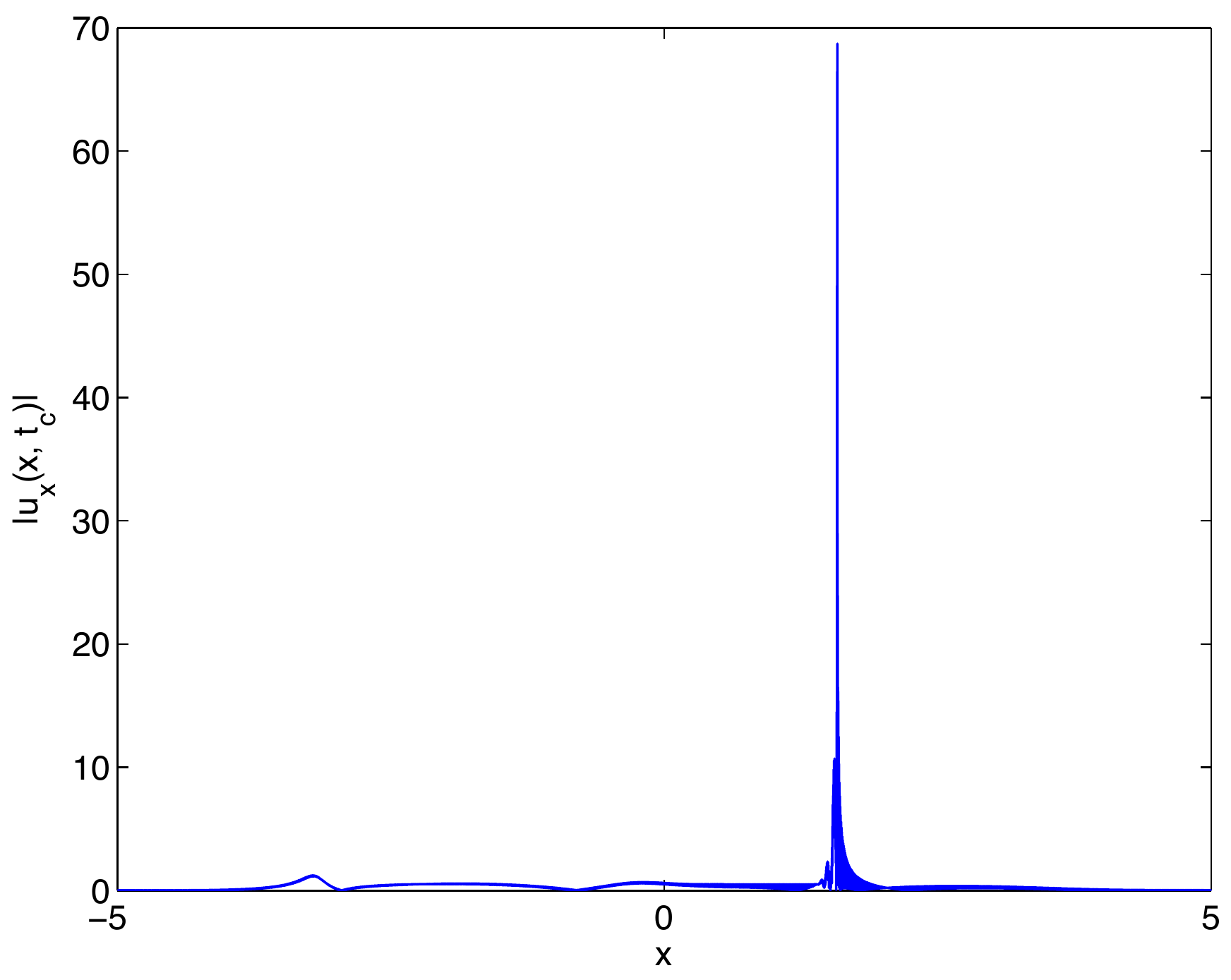}
\includegraphics[width=0.4\textwidth]{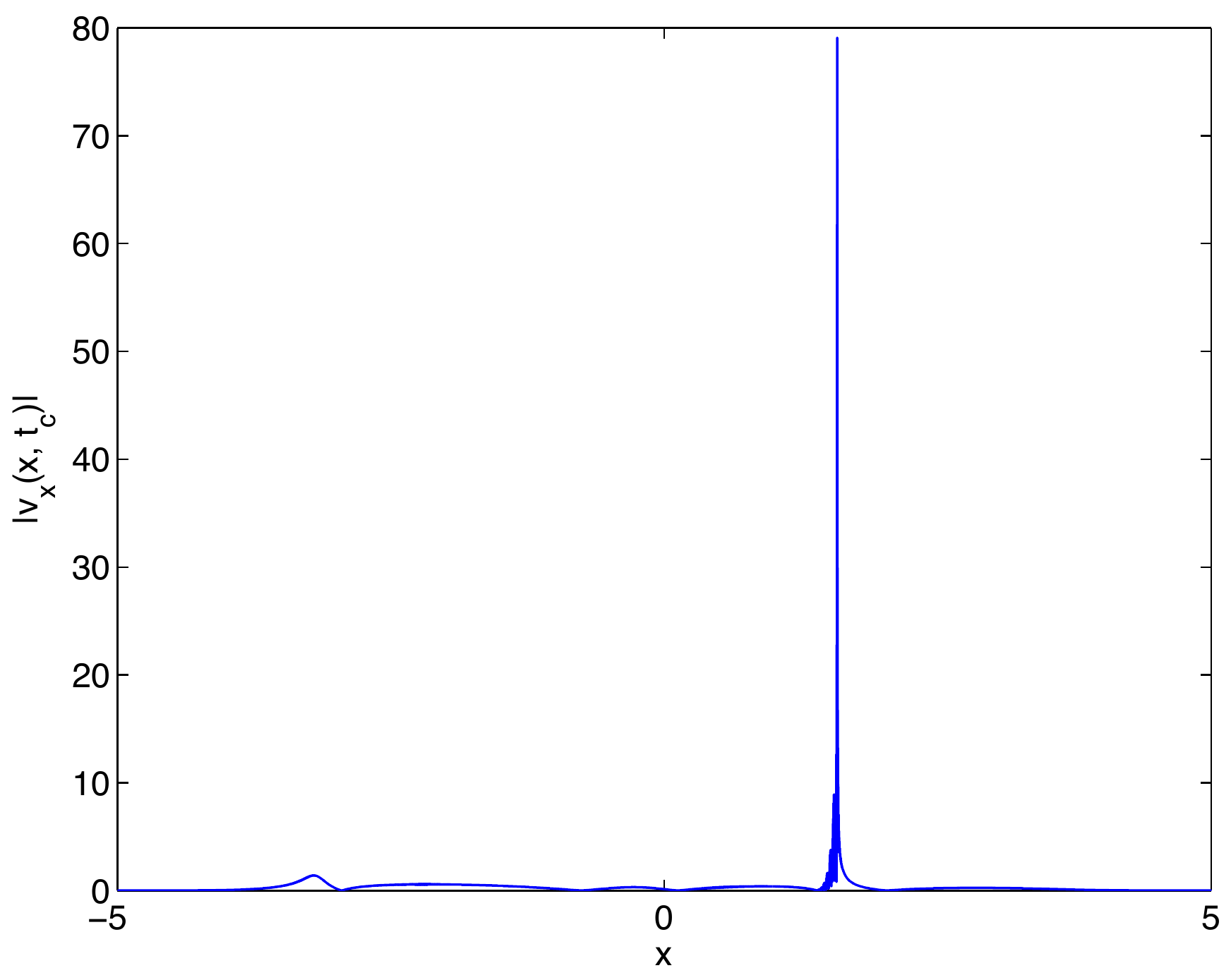}
\caption{Profile of $|u_x|$ on the left, and of $|v_x|$ on the right, at $t=t_c=1.717$ for $(u,v)$ being the solution of 
the 1d hyperbolic dispersionless Toda equation (\ref{toda1e0}) with 
$\rho=1$ for initial data of the form (\ref{uini1}).}
\label{gradtcuv}
\end{center}
\end{figure}

As expected, at $t_c=1.717$ the gradients of both $u$ and $v$ blow 
up, with $\| u_x \|_{\infty} \sim 70$ and $\| v_x \|_{\infty} \sim 95$, see Fig. \ref{gradtcuv} for the profiles of $|u_x|$ and $|v_x|$ at the critical time.

\begin{figure}[htb!]
\begin{center}
\includegraphics[width=0.4\textwidth]{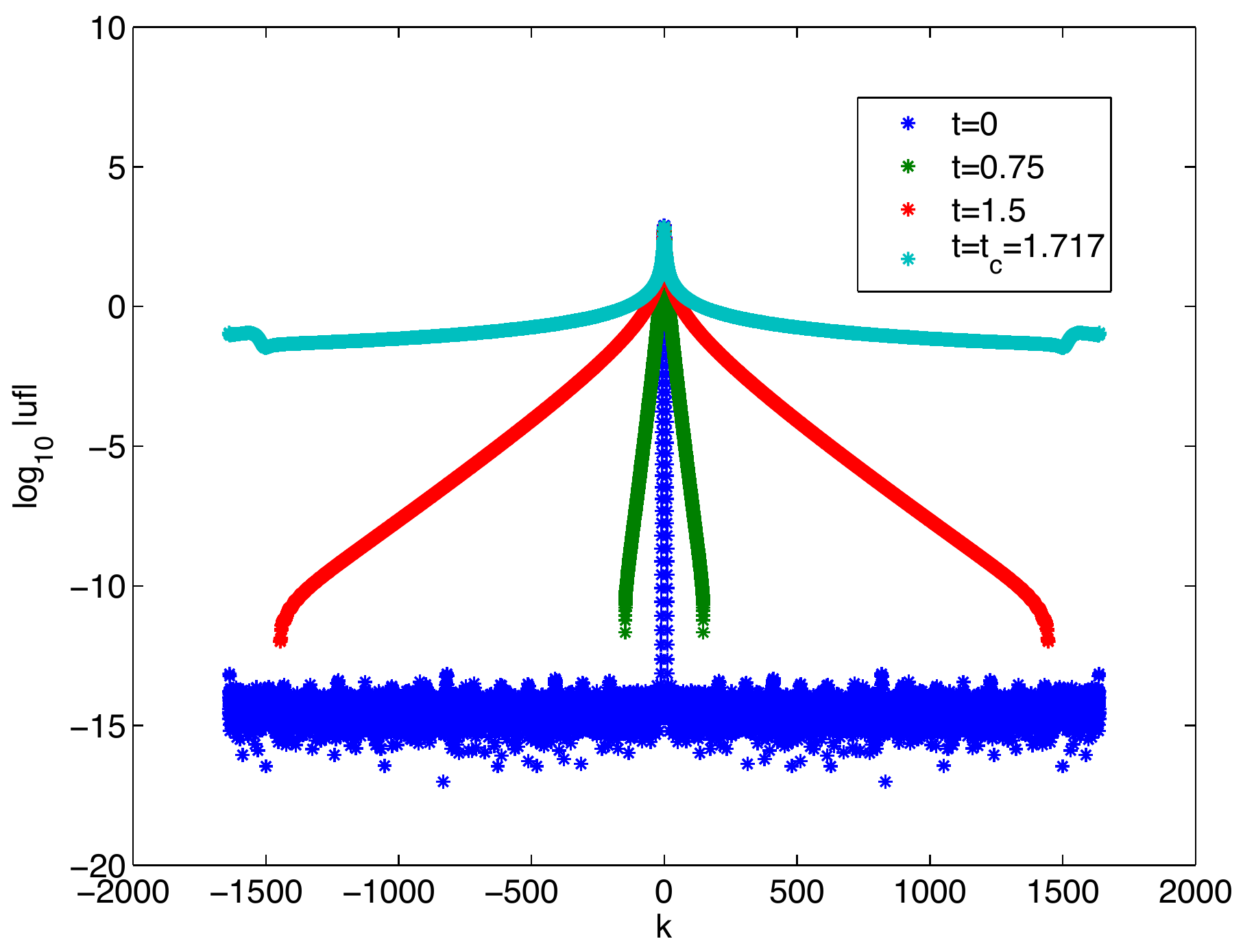}
\includegraphics[width=0.4\textwidth]{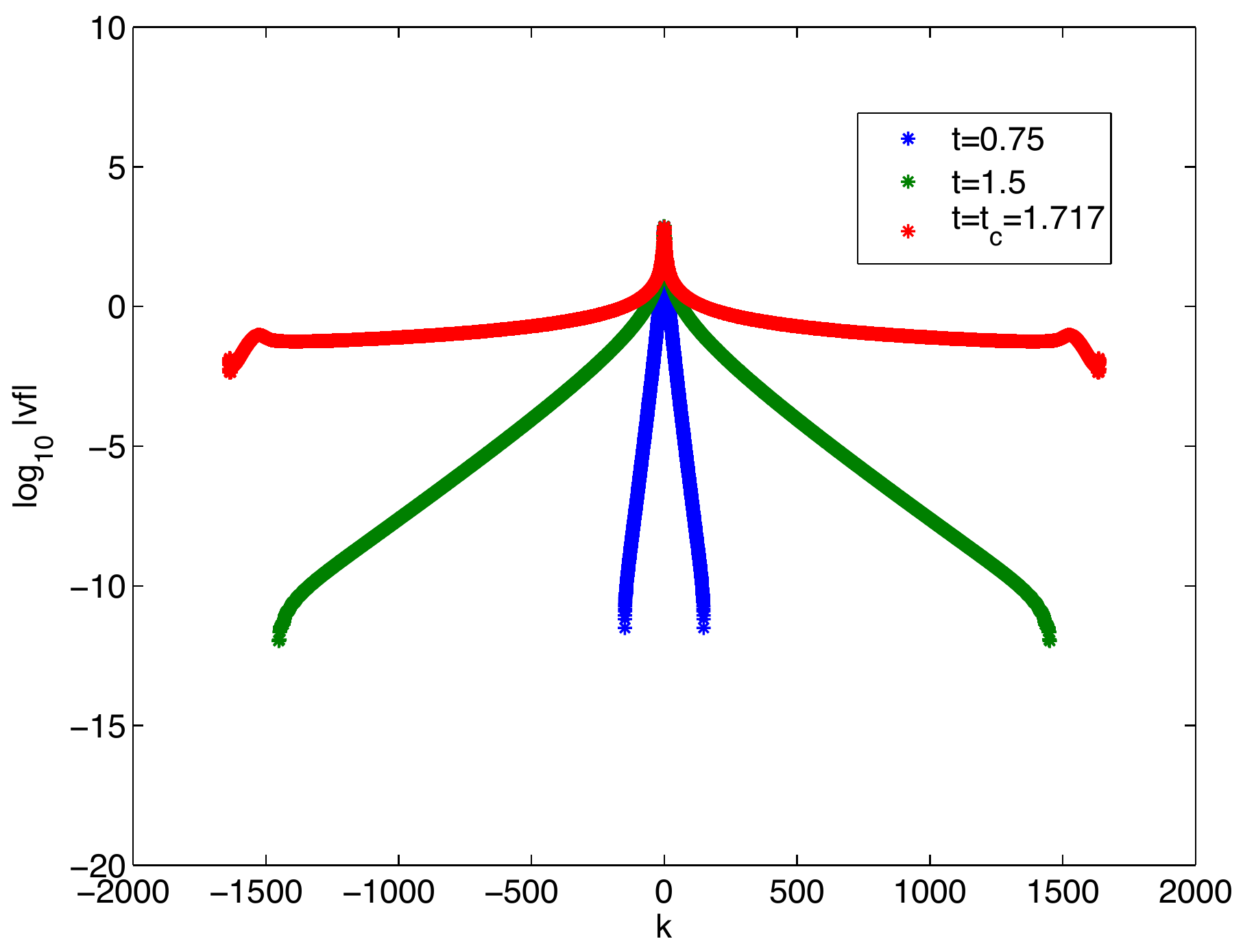}
\caption{Fourier coefficients $uf$ of $u$ (left) and $vf$  of $v$ 
(right)
at different times corresponding to the situations in Fig. 
\ref{Soluts} and \ref{Solvts}.}
\label{Coefstsuv}
\end{center}
\end{figure}

We ensure the system is numerically well resolved by 
checking the decay of the Fourier coefficients during the whole 
computation, see Fig. \ref{Coefstsuv}; the situation for $uf$ is shown on the left, and for $vf$ on the right. Moreover the time evolution of the numerically computed energy does not show any sudden changes, and reaches a value of $\Delta_E \sim 10^{-10} $ at the end of the computation at $t=t_c$.

The solutions of the one-dimensional hyperbolic dispersionless Toda 
equation thus behave like solutions to the Hopf equation for localized 
initial data, i.e., they develop a cubic shock.

\subsection{One-dimensional hyperbolic Toda equation in the limit of small 
dispersion}

In this subsection we study solutions to the one-dimensional 
hyperbolic Toda 
equation (\ref{todacont3})  ($\rho=1$) for small nonzero $\epsilon$. 
We choose as before the initial data (\ref{uini1}) and a time step $\delta_t = 3*10^{-4}$.

\begin{figure}[htb!]
\begin{center}
\includegraphics[width=0.45\textwidth]{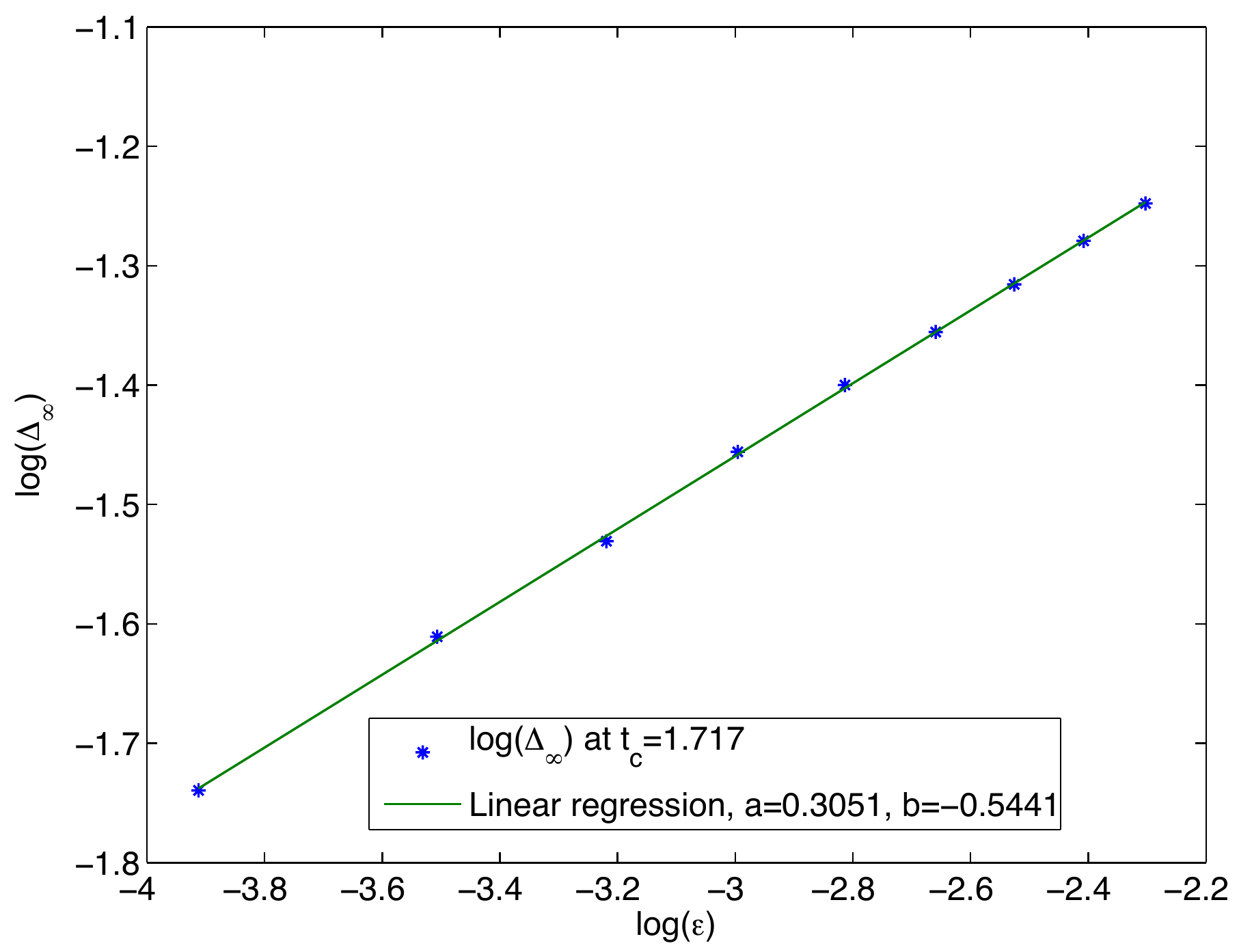}
\caption{$L_{\infty}$ norm  $\Delta_{\infty}$ of the difference 
 between 1d hyperbolic dispersionless Toda and 1d Toda solutions for the initial data (\ref{uini1}) in dependence of $\epsilon$ at $t_c=1.717$ for several values of $\epsilon$.}
\label{scalingtc}
\end{center}
\end{figure}

First we study the scaling with $\epsilon$ of the 
$L_{\infty}$ norm of the difference between the solution to the 1d
dispersionless Toda equation (\ref{toda1e0})
and the Toda equation (\ref{todacont3}) for small $\epsilon$ for the same initial data at the critical time of the former, here at $t_c=1.717$. 
The $L_{\infty}$ norm $\Delta_{\infty}$ of this difference is shown in 
Fig.~\ref{scalingtc} at $t_c=1.717$ 
 in dependence of $\epsilon$ for  $0.02\leq  \epsilon \leq 0.1$. 
A linear regression analysis ($\log_{10} \Delta_{\infty} = a \log_{10} \epsilon + b$ ) shows that $\Delta_{\infty}$ decreases as 
\begin{align}
\mathcal{O} \left( \epsilon^{0.30} \right)  \sim \mathcal{O} \left( \epsilon^{2/7} \right) \,\, \mbox{at} \,\, t=t_c=1.717, \,\, \mbox{with}\,\, a=0.3051  \,\,\mbox{and} \,\,  b= -0.5441.
\end{align}
The correlation coefficient is $r = 0.999$. 
It is thus again similar to the results for the KdV equation and for the defocusing cubic NLS 
equation in \cite{DGK13}.

\begin{figure}[htb!]
\begin{center}
\includegraphics[width=0.6\textwidth]{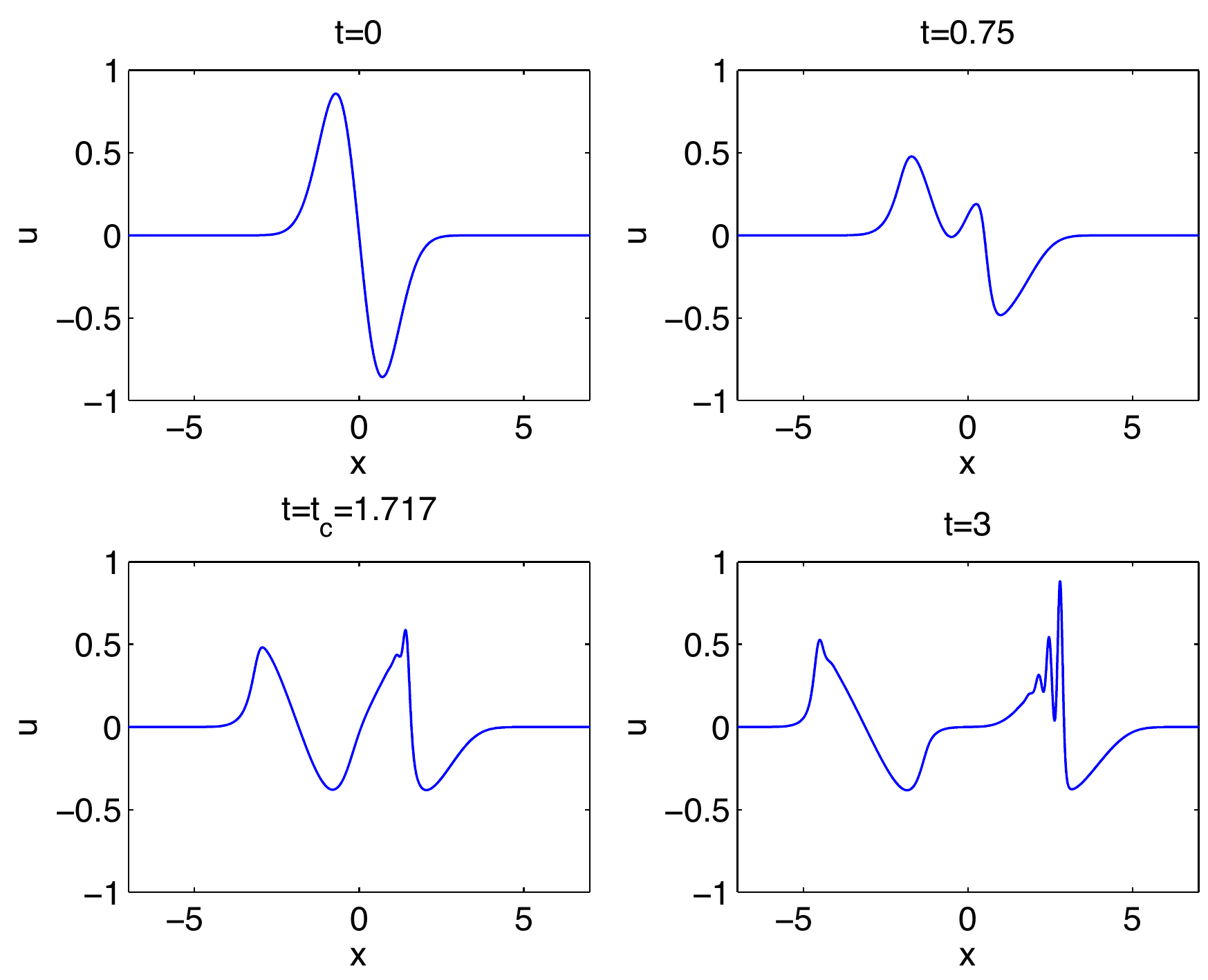}
\caption{Solution $u$ of the 1d hyperbolic Toda equation  (\ref{todacont3}) with $\rho=1$ at different times 
for initial data of the form (\ref{uini1}) and $\epsilon=0.1$. 
}
\label{eps01uts}
\end{center}
\end{figure}

\begin{figure}[htb!]
\begin{center}
\includegraphics[width=0.6\textwidth]{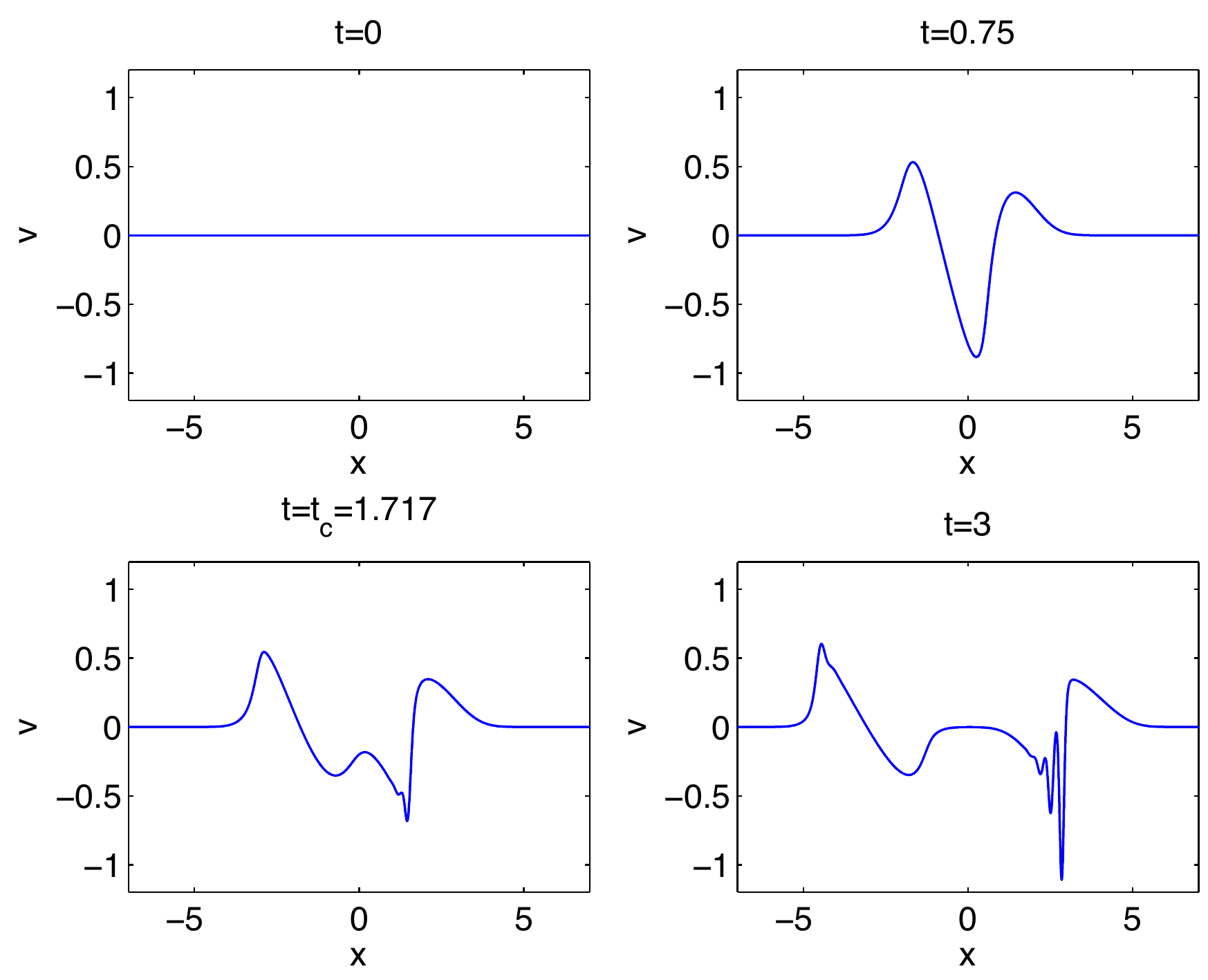}
\caption{Solution $v$ of the 1d hyperbolic Toda equation  (\ref{todacont3}) 
with $\rho=1$ at different times for initial data of the form (\ref{uini1}) and $\epsilon=0.1$. 
}
\label{eps01vts}
\end{center}
\end{figure}

For larger times and $\epsilon=0.1$ we observe as expected the development of rapid 
oscillations from $t\sim t_{c}=1.717$ 
where the solution to the corresponding dispersionless system becomes singular, see Fig. \ref{eps01uts} for $u$ and Fig. \ref{eps01vts} for $v$.

\begin{figure}[htb!]
\begin{center}
\includegraphics[width=0.4\textwidth]{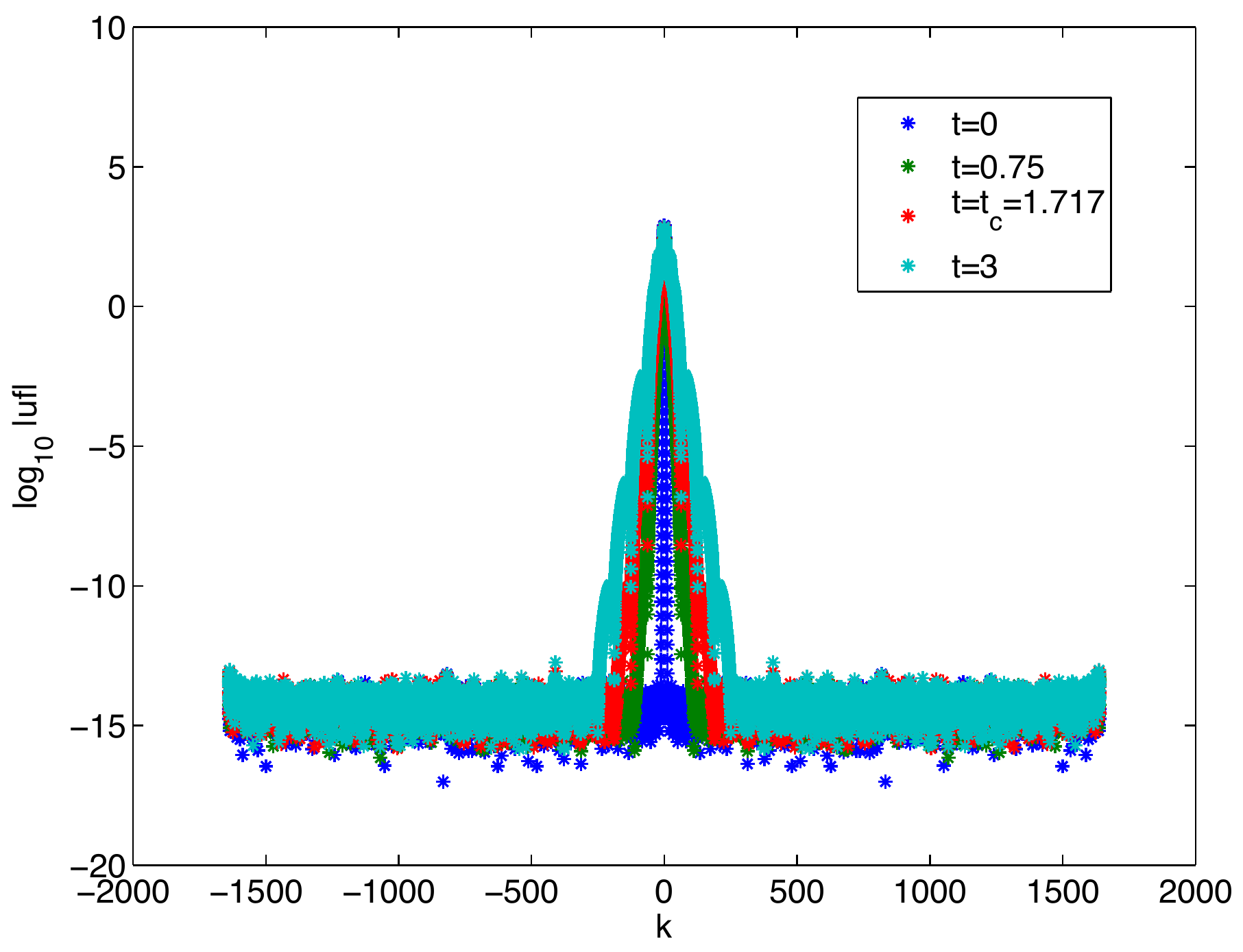}
\includegraphics[width=0.4\textwidth]{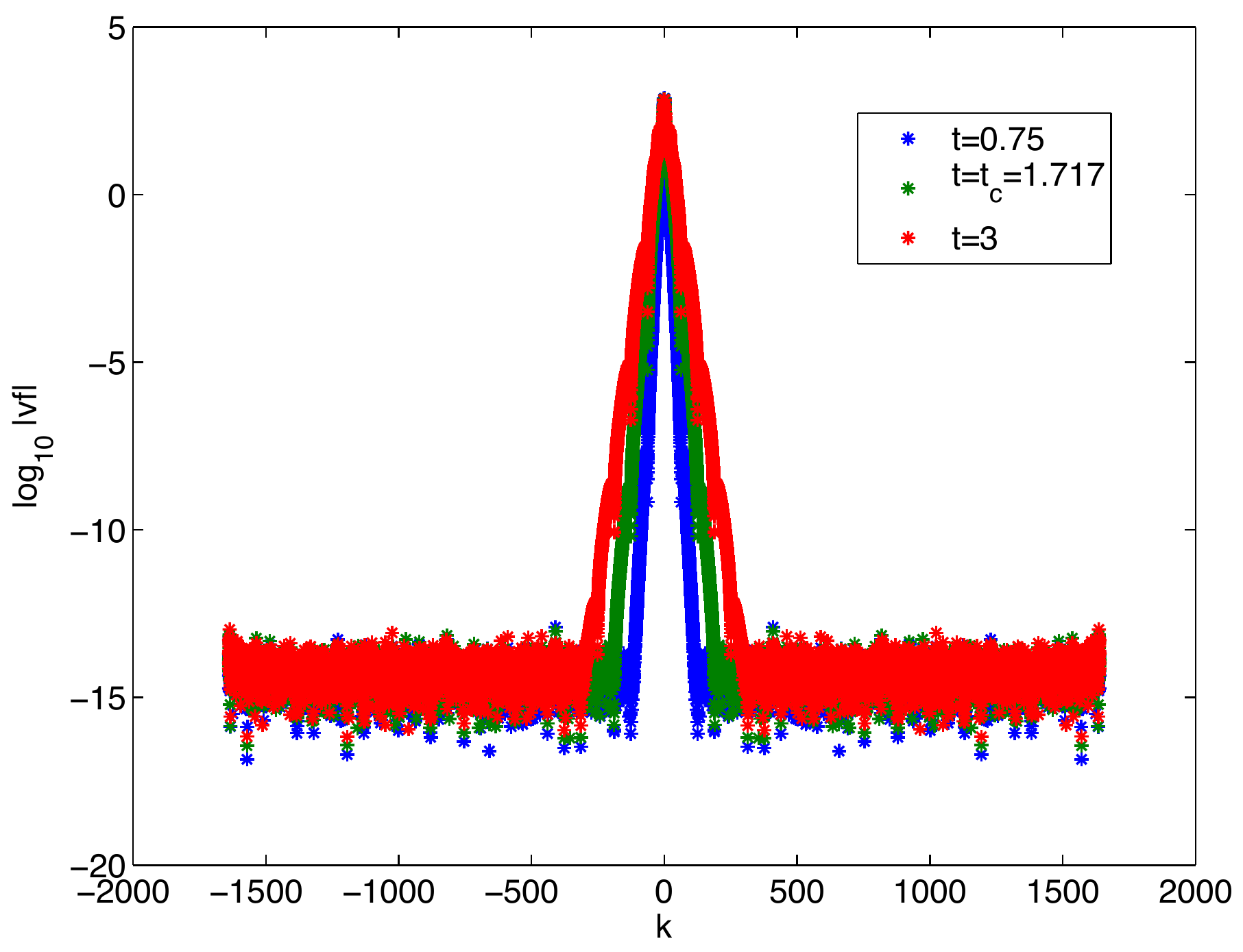}
\caption{Fourier coefficients of $u$ (left) and of $v$, (right)
 at different times, corresponding to the situations in Fig. 
 \ref{eps01uts} and \ref{eps01vts}.}
\label{eps01ufvfts}
\end{center}
\end{figure}

We show in Fig. \ref{eps01ufvfts} the corresponding Fourier 
coefficients at several times. They decrease to machine precision 
$\sim 10^{-15}$ during the whole computation.
In addition the numerically computed energy reaches also the same 
precision $\Delta_E \sim 10^{-15}$ until the maximal time of 
computation $t_{max}=3$, indicating that the system is well resolved, 
and that sufficient resolution is provided both in space and in time.

We have identified in the previous subsection the time where the first 
shock occurs ($t_{c}\sim1.717$), but we can infer from 
Fig.~\ref{eps01uts} and Fig.~\ref{eps01vts} that a second shock 
occurs at a later time $t\sim3$ for negative $x$. It can be observed 
there that small oscillations begin to form in the region $x<0$.

\begin{figure}[htb!]
\begin{center}
\includegraphics[width=0.6\textwidth]{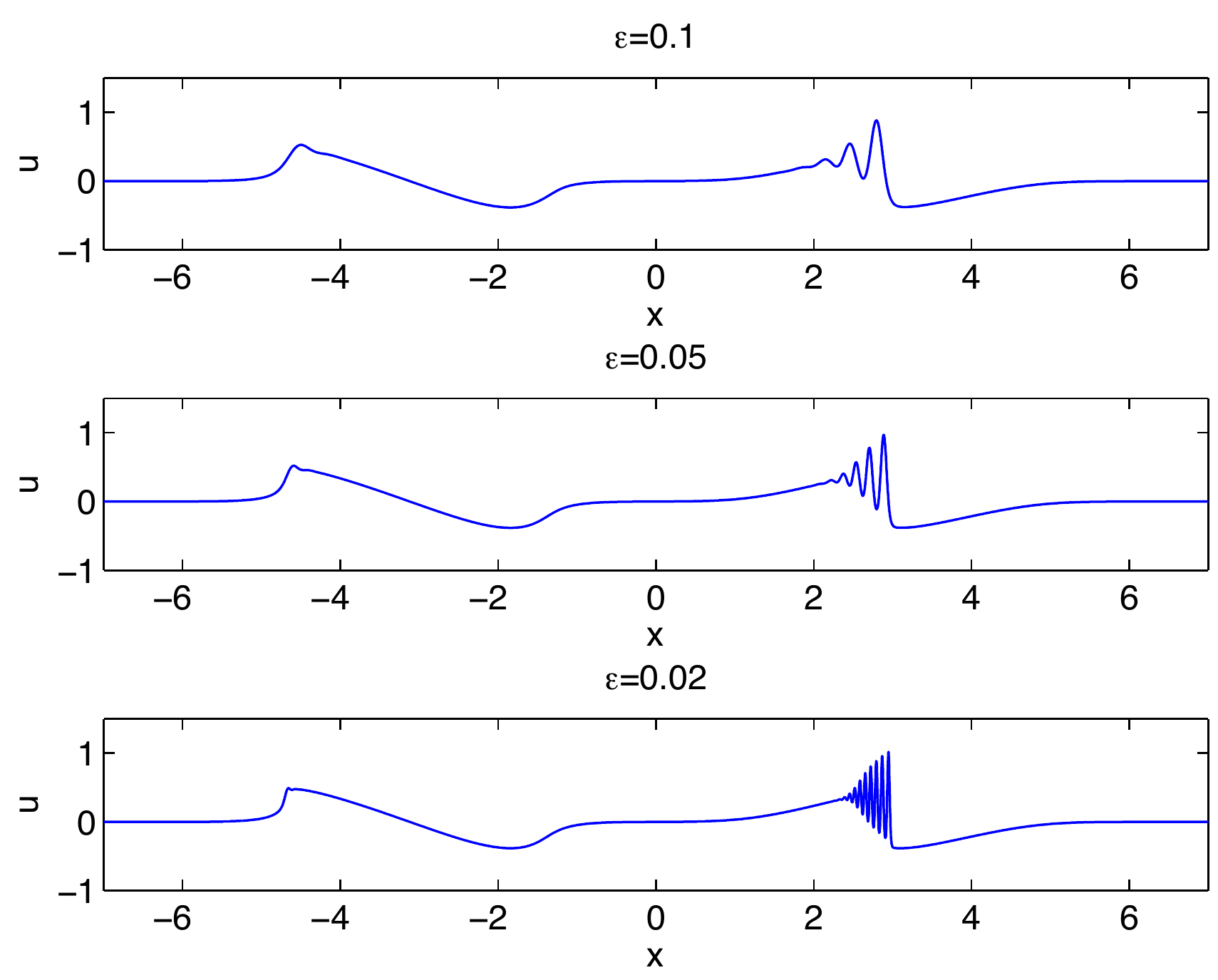}
\caption{Solution $u$ of the 1d hyperbolic Toda equation at $t=3$ for 
different values of $\epsilon$.}
\label{epssut3}
\end{center}
\end{figure}

As $\epsilon \to 0$, the number of oscillations increases as 
expected as  can be seen in Fig. \ref{epssut3} for $u$; the situation 
is similar for $v$ which is therefore not shown. 

\subsection{Hyperbolic dispersionless Toda equation in $2+1$ dimensions}
We now perform the same study as in the $1+1$ dimensional case in 
$2+1$ dimensions. 
We consider 
initial data of the form
\begin{equation}
u(x,y,0)=u_0(x,y)=\partial_x \exp(-R^2), \,\, R=\sqrt{x^2+y^2}, \,\,\, v(x,y,0)=v_0(x,y)=0.
\label{uini2}
\end{equation}
for (\ref{toda2e0}) corresponding to 
\begin{equation}
U(x,y,0)=U_0(x,y)=\exp(-R^2), \,\, R=\sqrt{x^2+y^2}, \,\,\, v(x,y,0)=v_0(x,y)=0
\label{uini2b}
\end{equation}
for (\ref{toda2e0int}).
The computations are carried out with $2^{14} \times 2^{9}$ points 
for $x \times y \in [-5\pi, 5\pi] \times [-5\pi, 5\pi]$ and time step $\delta_t=4*10^{-4}$.
We find that the solutions will develop in $x$-direction a 
singularity as in the $1+1$-dimensional case, but that they stay 
smooth in $y$-direction. The results described here are thus similar 
to the ones of \cite{DSdDS} for the defocusing semiclassical DS II system.

\begin{figure}[htb!]
\begin{center}
\includegraphics[width=0.45\textwidth]{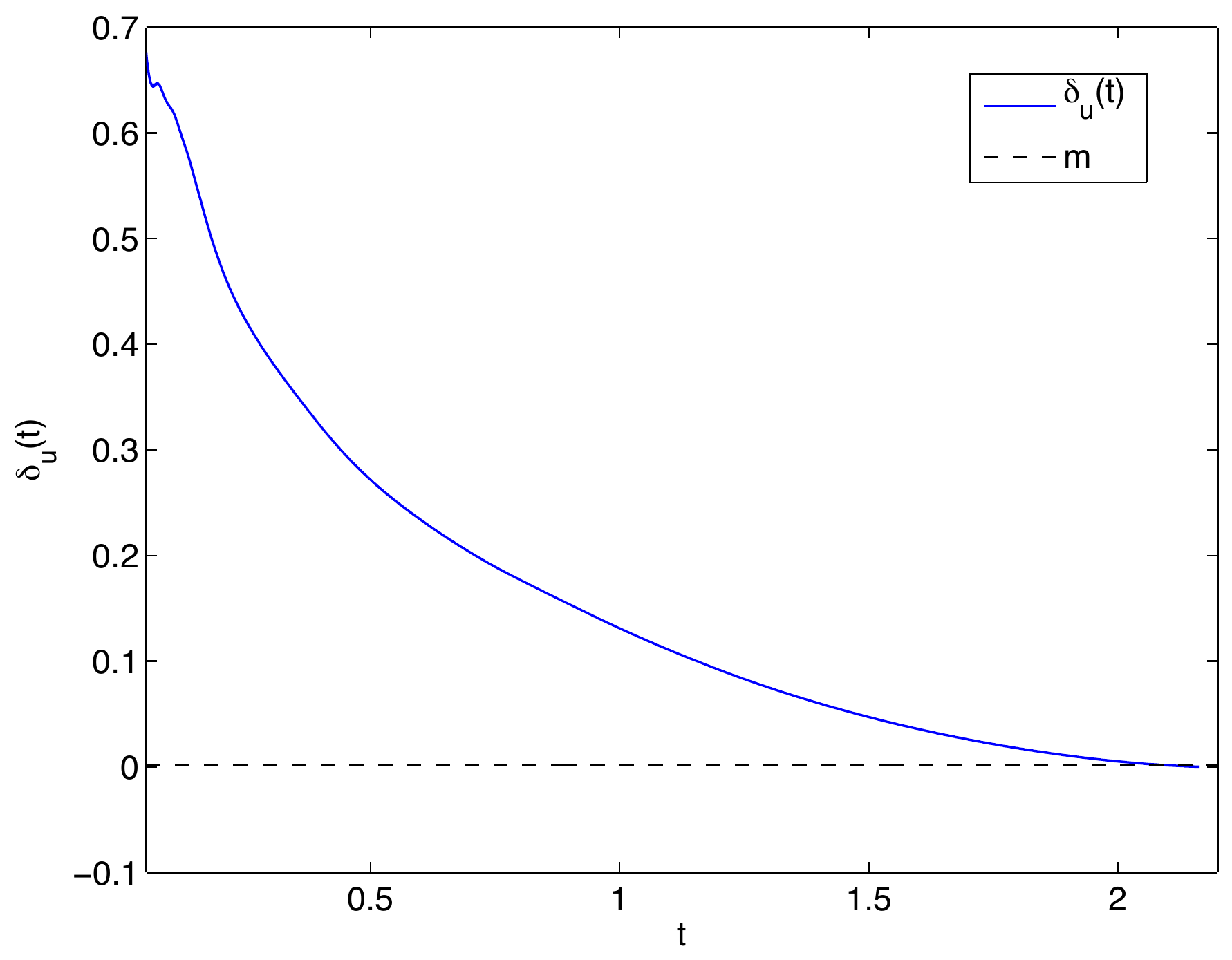}
\caption{Fitting parameter $\delta$ in (\ref{fourasymp2}) for the 
solution $u$ (denoted by $\delta_{u}$) of the two-dimensional 
hyperbolic dispersionless Toda equation (\ref{toda2e0})
with $\rho=1$ for initial data of the form (\ref{uini2}). 
The fitting is done for $10<k<2*max(k)/3$.}
\label{deluhyp2d}
\end{center}
\end{figure}

We will first identify numerically the appearance of break-up in 
solutions to the 2d dispersionless Toda equations (\ref{toda2e0}) for 
the initial data (\ref{uini2}) for $\rho=1$.
The numerical study of the asymptotics of the Fourier coefficients of the resulting solution $u$
 leads to a vanishing of the quantity $\delta$ in (\ref{fourasymp}) 
 for $u$, denoted by $\delta_u$, at $t=t_c=2.162$, see Fig. \ref{deluhyp2d}. 

\begin{figure}[htb!]
\begin{center}
\includegraphics[width=0.45\textwidth]{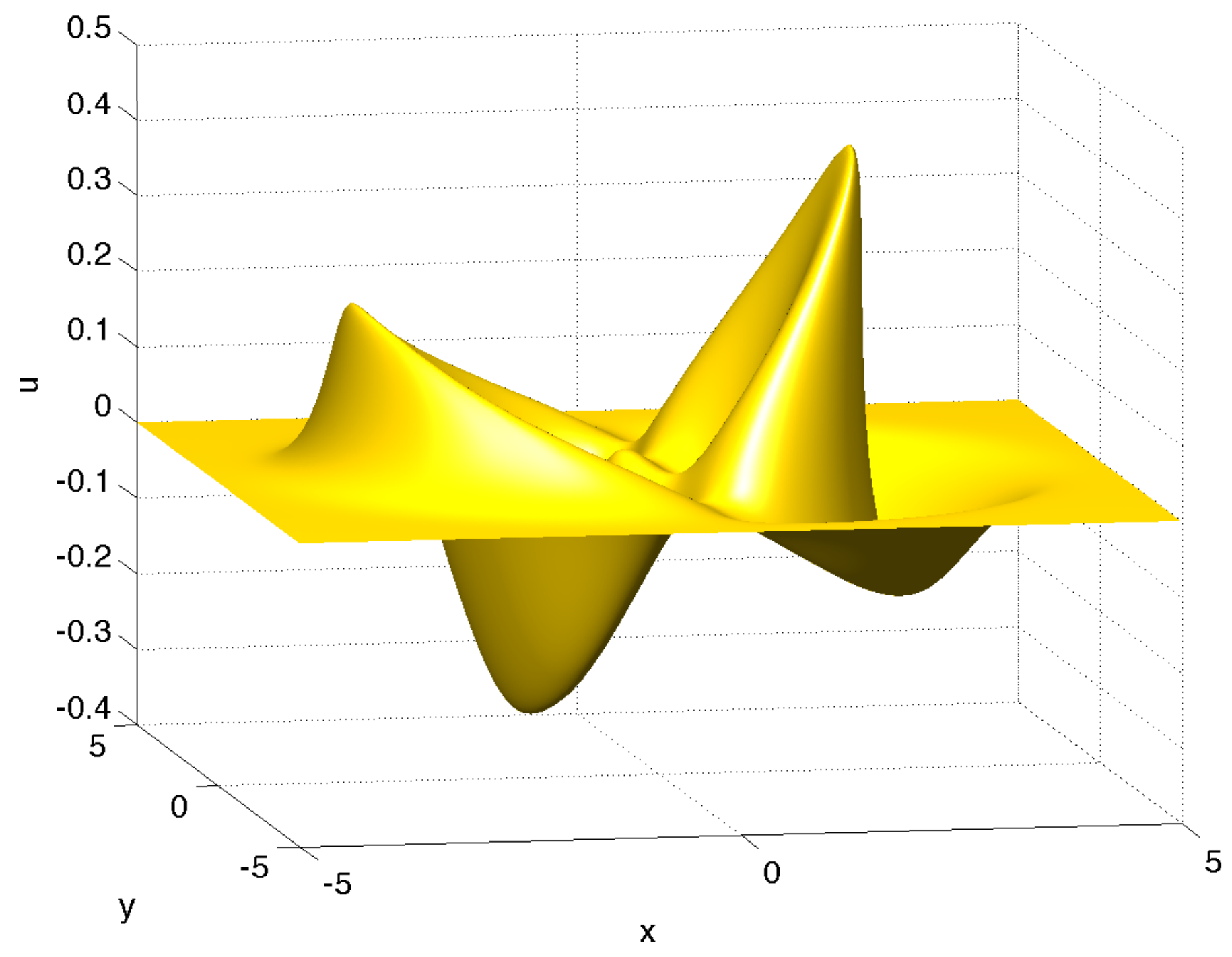}
\includegraphics[width=0.45\textwidth]{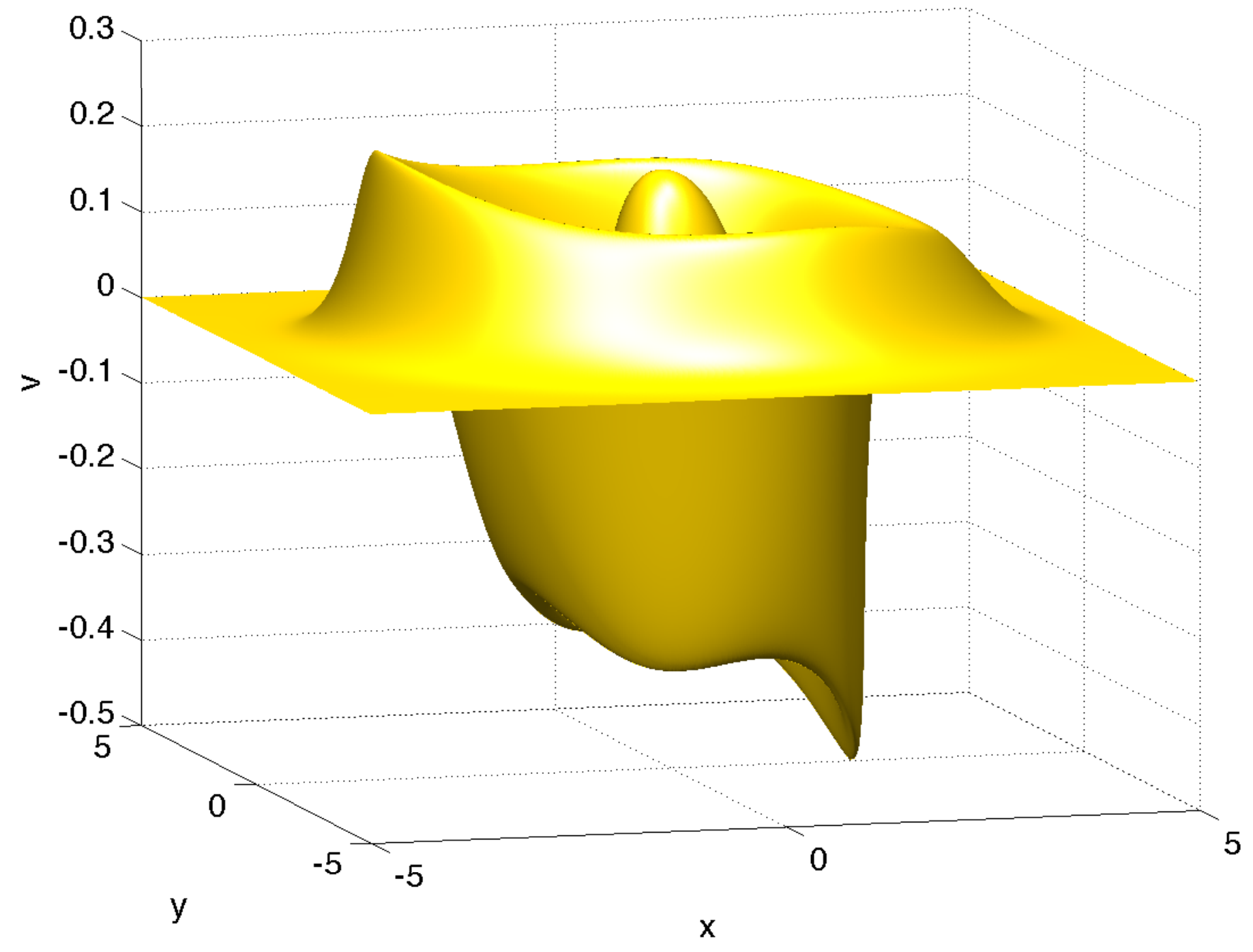}
\caption{Solution of 
the $2+1$ dimensional dispersionless hyperbolic Toda equation (\ref{toda2e0})
with $\rho=1$ for initial data of the form (\ref{uini2}) at $t=t_c=2.162$.}
\label{Solvutchyp2d}
\end{center}
\end{figure}

At this time the solution develops a shock in $x$-direction as in the 
$1+1$ dimensional case as can be seen in Fig. \ref{Solvutchyp2d} where we show the solution ($u,v$) at $t=t_c=2.162$.

\begin{figure}[htb!]
\begin{center}
\includegraphics[width=0.4\textwidth]{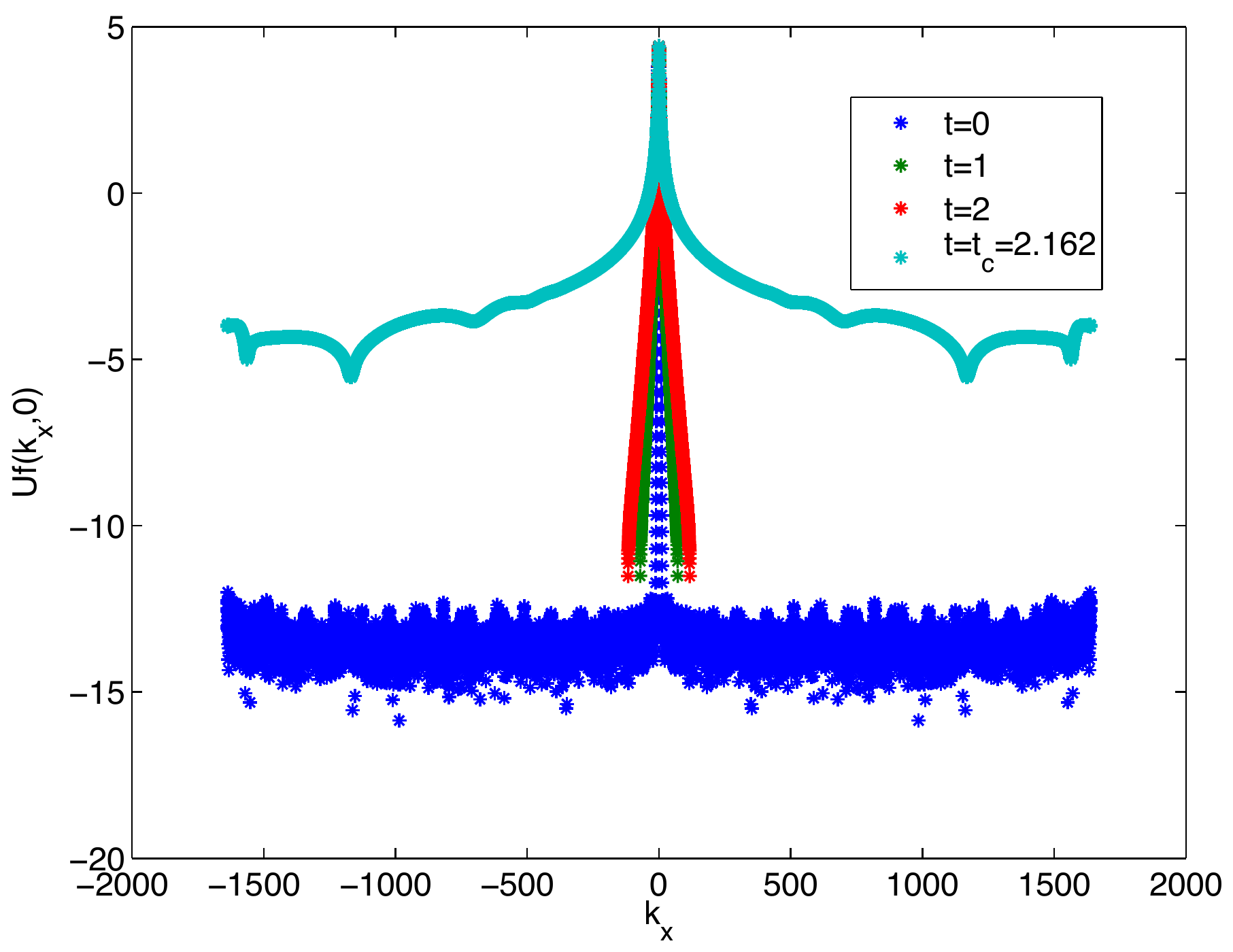}
\includegraphics[width=0.4\textwidth]{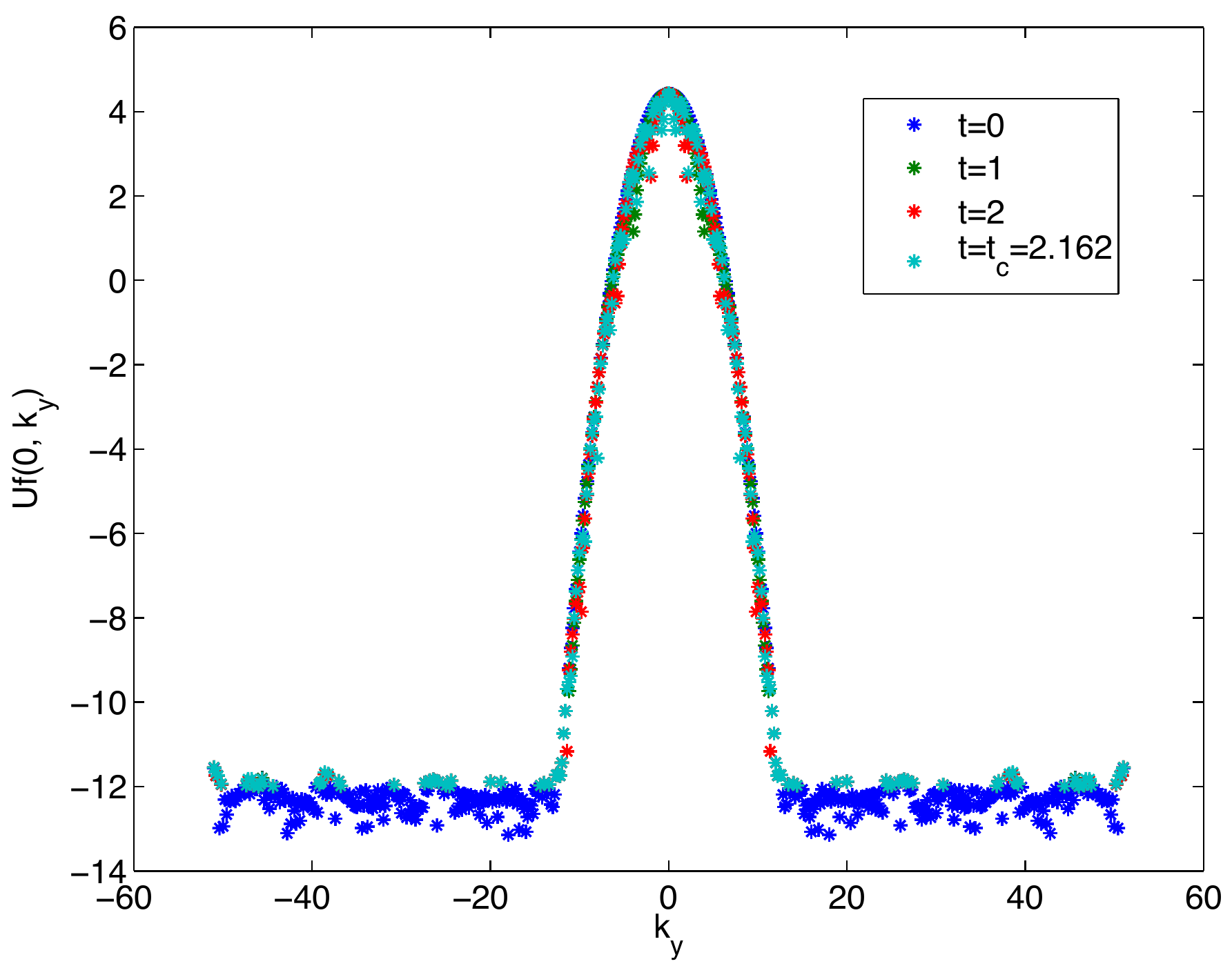}
\caption{Fourier coefficients of the solution $U$
 of 
the $2+1$ dimensional hyperbolic dispersionless Toda equation (\ref{toda2e0int})
with $\rho=1$ for initial data of the form (\ref{uini2b})
 at several times plotted on the $k_x$-axis on the left and on the $k_y$-axis on the right.}
\label{coefshyp2de0}
\end{center}
\end{figure}

The cubic singularity can be also inferred from the value of the fitting parameter $B_u$ at this time, which reaches a value of 
$B_u(t_c) = 1.353$. We get here less precision than in the one 
dimensional case (exactly as in the cases reported in \cite{dkpsulart}) 
because of the lower space resolution used. The latter is however 
high enough in terms of accuracy as can be seen in Fig. 
\ref{coefshyp2de0} from
the Fourier coefficients of $U$ at several times plotted on the $k_x$-axis on the left, and on the $k_y$-axis on the right. 
It can be seen that the chosen resolution for the $y$-direction 
allows to reach machine precision. In the $x$-direction, i.e., the 
direction, where the gradient catastrophe occurs, the Fourier 
coefficients show the expected algebraic decay at $t=t_c$. 

\begin{figure}[htb!]
\begin{center}
\includegraphics[width=0.4\textwidth]{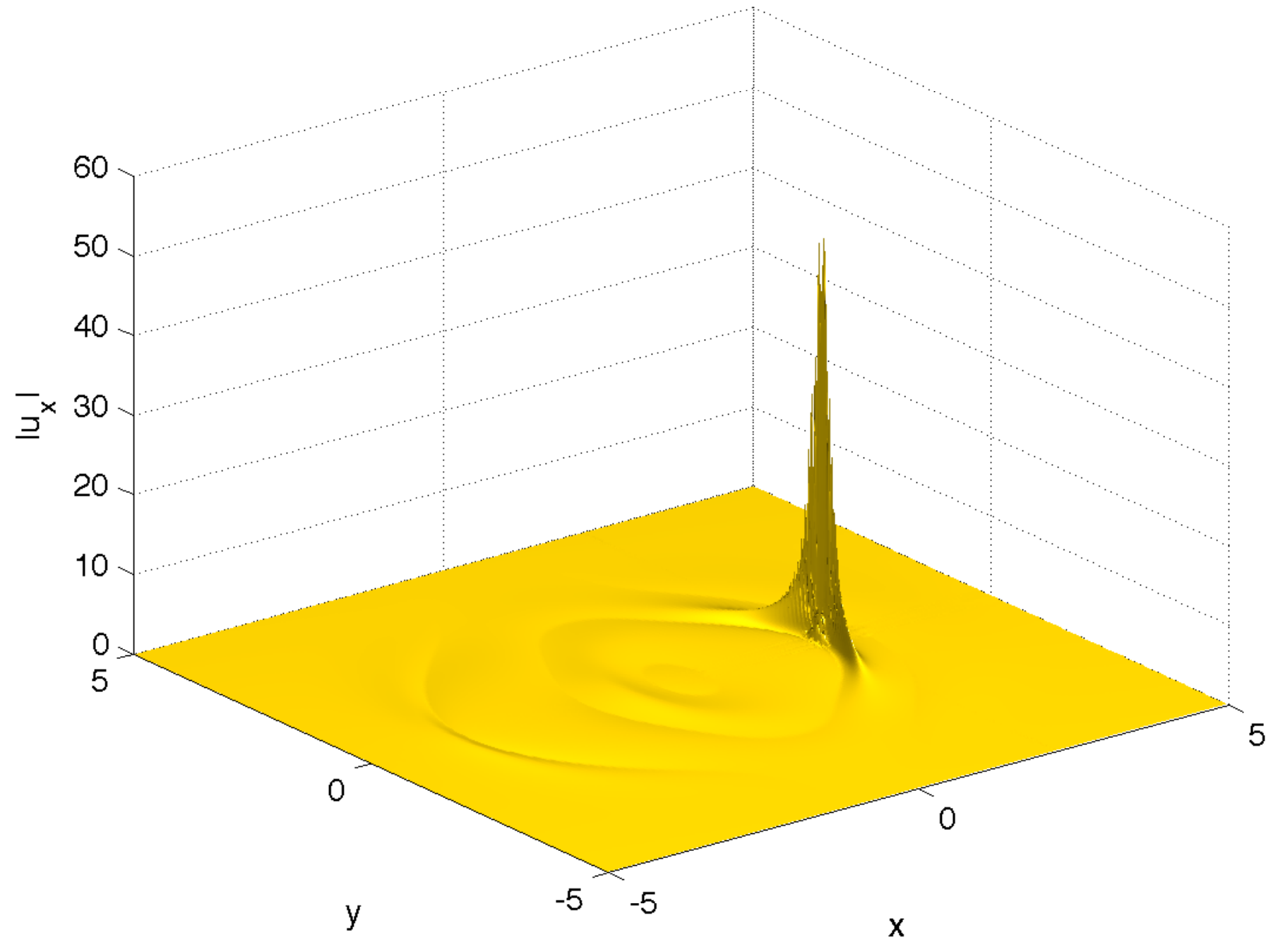}
\includegraphics[width=0.4\textwidth]{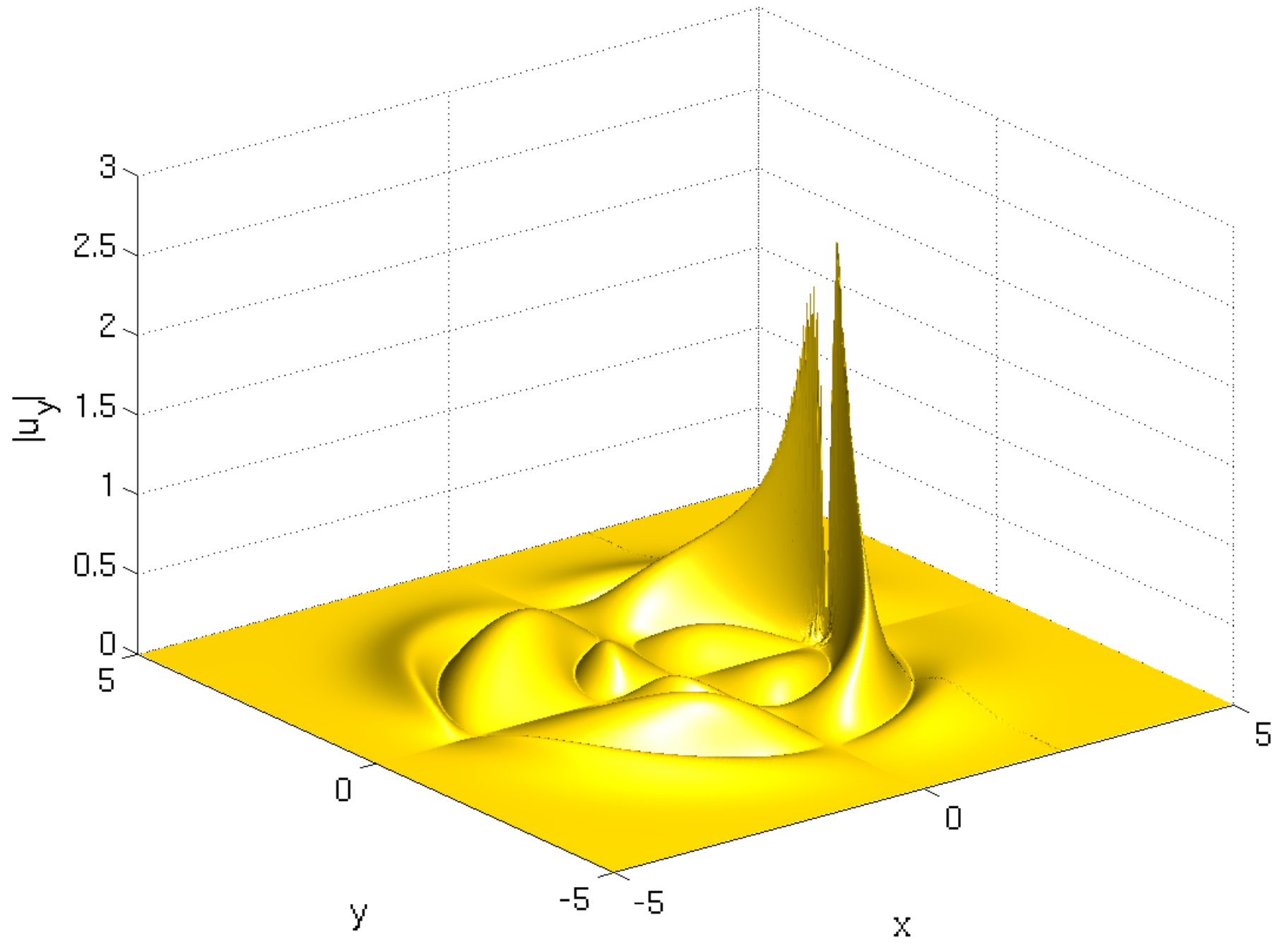}
\caption{Absolute value of the $x$-derivative of $u$ of 
Fig.~\ref{Solvutchyp2d} at $t=t_c=2.162$ on the left, and of the $y$-derivative on the right.}
\label{gradtcuetuhathyp2d}
\end{center}
\end{figure}

%
%
%
The numerically computed energy $\Delta_E$, where $E$ is defined in 
(\ref{E2d0}) reaches a precision of $\sim 10^{-11}$ at the end of the 
computation, where the $x$-derivative of $u$ begins to blow up,  see 
Fig. \ref{gradtcuetuhathyp2d}, with $\| u_x  \|_{\infty} \sim 50$. In 
the same way one gets for $|v_x|$ that $\| v_x  \|_{\infty} \sim 65$.
The shock is clearly a one-dimensional phenomenon. It can be seen in 
Fig. \ref{gradtcuetuhathyp2d} that $u_y$ stays small, with $\| u_y  
\|_{\infty} \sim 2$. The location of the singularity is found to be 
$\alpha(t_c)=2.2615$ on the $x$-axis.

\subsection{Small dispersion limit of the hyperbolic Toda equation in $2+1$ 
dimensions}
In this subsection we study the solutions for the hyperbolic Toda equation in 
$2+1$ dimensions (\ref{todacont3}) for the initial data 
(\ref{uini2b}) for small nonzero $\epsilon$.

\begin{figure}[htb!]
\begin{center}
\includegraphics[width=0.45\textwidth]{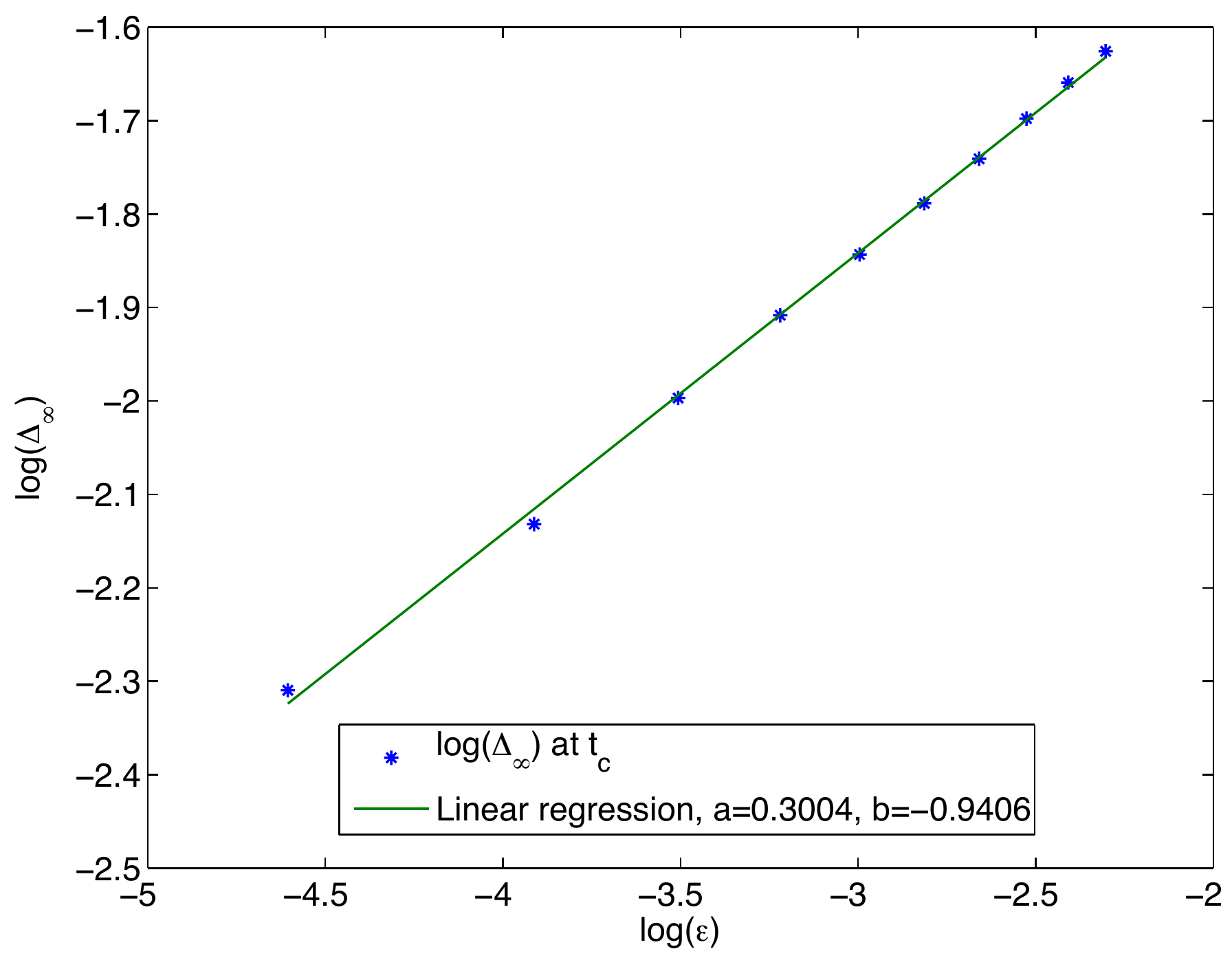}
\caption{$L_{\infty}$ norm  $\Delta_{\infty}$ of the difference 
 between solutions to the 2d hyperbolic dispersionless Toda and the Toda 
 equation for the initial data (\ref{uini2})
  in dependence of $\epsilon$ at $t_c=2.162$.}
\label{hyp2dscaltc}
\end{center}
\end{figure}

First we investigate the scaling with $\epsilon$ of the 
$L_{\infty}$ norm $\Delta_{\infty}$ of the difference between the 
solutions to the 2d dispersionless Toda (\ref{toda2e0})
and the Toda (\ref{toda2cont}) equation for the same initial data in 
dependence of $\epsilon$. 
The $L_{\infty}$ norm of this difference is shown in 
Fig.~\ref{hyp2dscaltc} at $t_c=2.162$ 
for  $0.01\leq  \epsilon \leq 0.1$. 

A linear regression analysis ($\log_{10} \Delta_{\infty} = a \log_{10} \epsilon + b$ ) shows that $\Delta_{\infty}$ decreases as 
\begin{align}
\mathcal{O} \left( \epsilon^{0.30} \right)  \sim \mathcal{O} \left( \epsilon^{2/7} \right) \,\, \mbox{at} \,\, t=t_c=2.162, \,\, \mbox{with}\,\, a=0.3004  \,\,\mbox{and} \,\,  b= -0.9406.
\end{align}
The correlation coefficient is $r = 0.99$. 
As expected, the situation is similar to the $1+1$ dimensional case 
which itself is as for KdV and the defocusing NLS equation.

\begin{figure}[htb!]
\begin{center}
\includegraphics[width=0.6\textwidth]{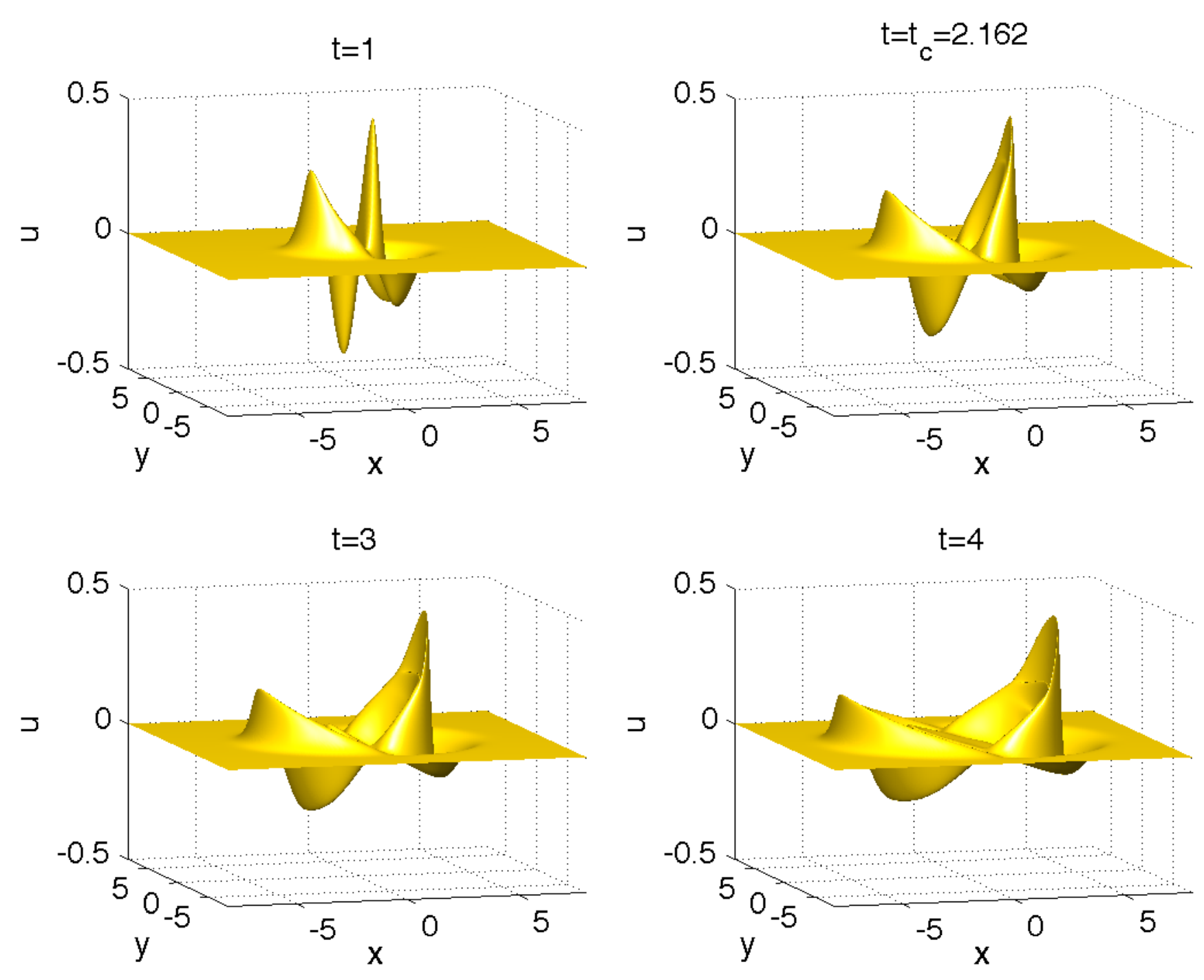}
\caption{Solution $u$ of the $2+1$ dimensional hyperbolic Toda equation (\ref{toda2cont}) with $\rho=1$ at different times 
for initial data of the form (\ref{uini2}) and $\epsilon=0.1$. }
\label{hyp2deps01uts}
\end{center}
\end{figure}

\begin{figure}[htb!]
\begin{center}
\includegraphics[width=0.6\textwidth]{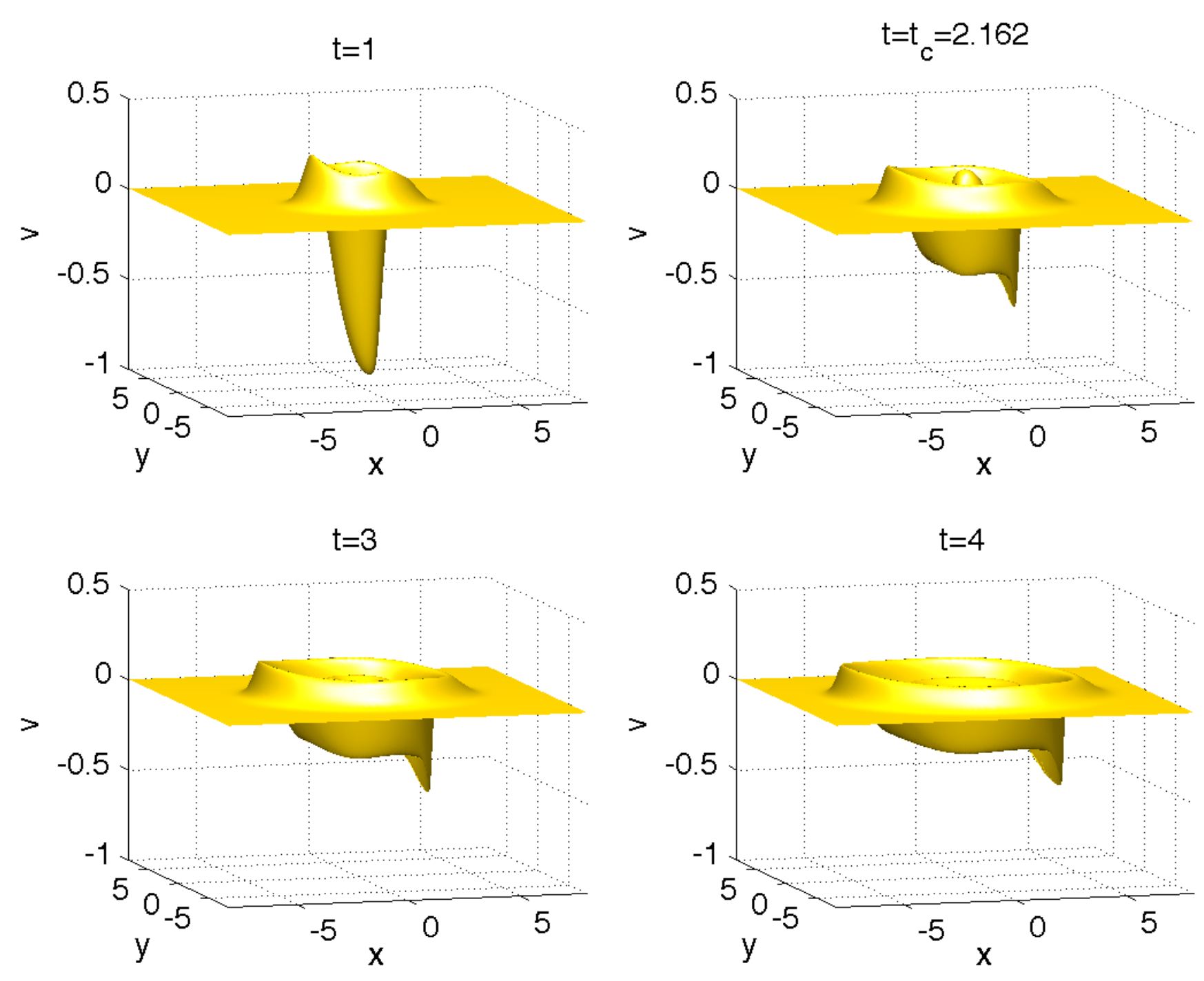}
\caption{Solution $v$ of the $2+1$ dimensional hyperbolic Toda equation (\ref{toda2cont}) with $\rho=1$ at different times 
for initial data of the form (\ref{uini2}) and $\epsilon=0.1$. }
\label{hyp2deps01vts}
\end{center}
\end{figure}

For $t\gg t_{c}$ the solution of the hyperbolic Toda equation in $2+1$ 
dimensions (\ref{toda2cont}) with $\rho=1$ for initial data of 
the form (\ref{uini2}) develops as expected a zone of rapid modulated oscillations, where the solution of the corresponding 
dispersionless system admits a shock (as identified in the previous subsection).
We show the numerical solution $u$ of (\ref{toda2cont}) with 
$\epsilon=0.1$  in Fig. \ref{hyp2deps01uts} and $v$ in Fig. \ref{hyp2deps01vts} at several times.

\begin{figure}[htb!]
\begin{center}
\includegraphics[width=0.4\textwidth]{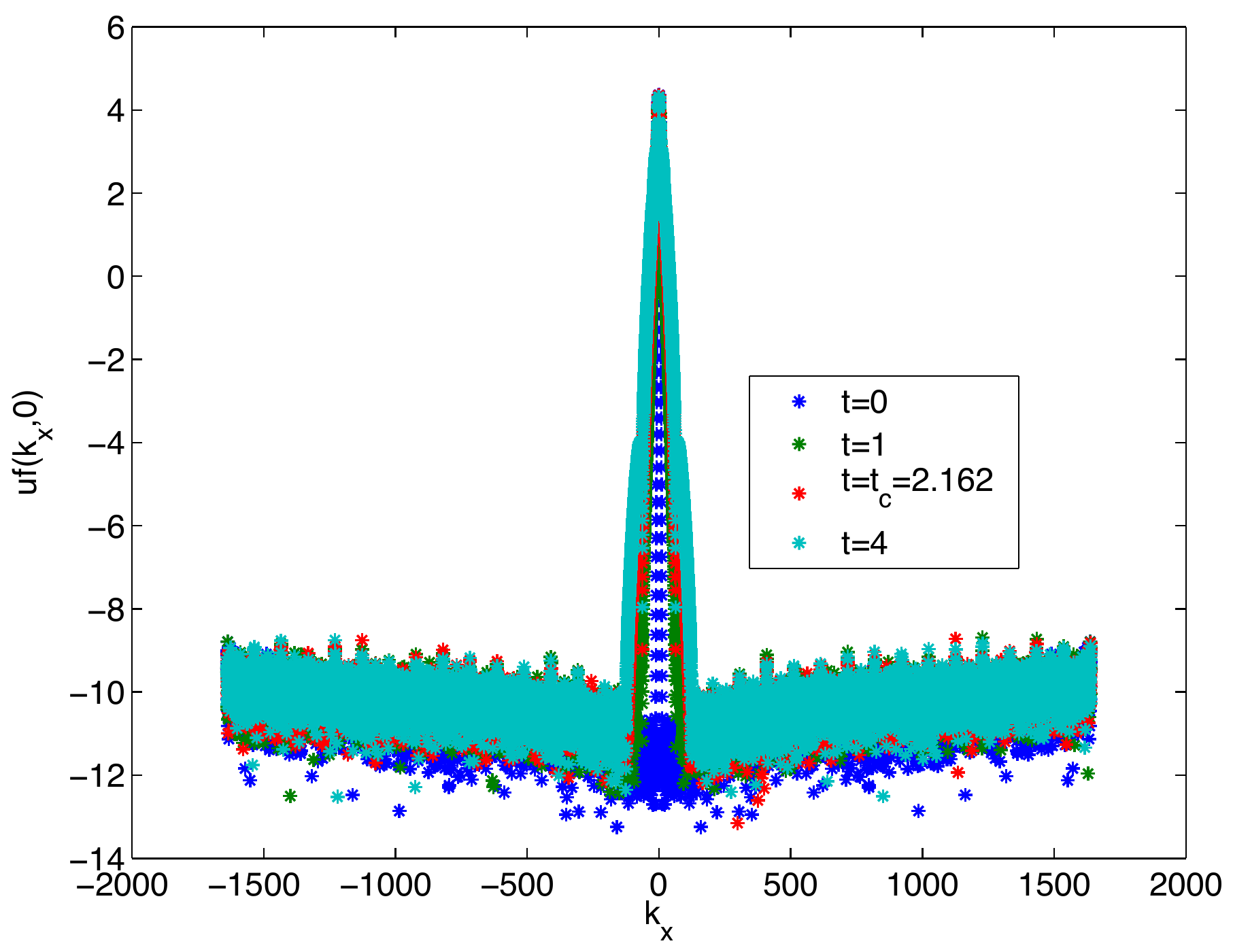}
\includegraphics[width=0.4\textwidth]{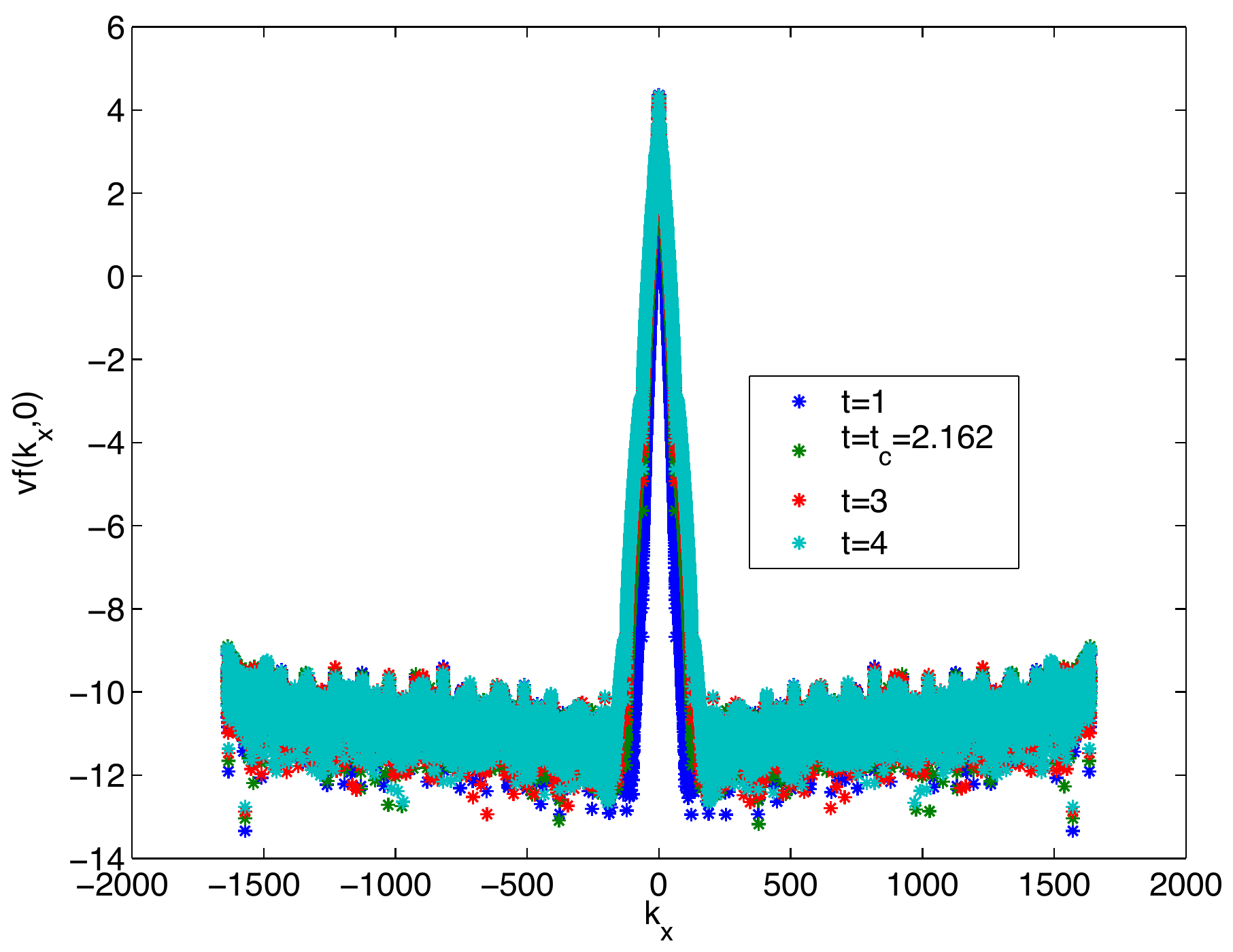}
\caption{Fourier coefficients of $u$ (left) and of $v$ (right) on 
the $k_{x}$-axis
 at different times, corresponding to the situations in Fig. 
 \ref{hyp2deps01uts} and \ref{hyp2deps01vts}.}
\label{hyp2deps01ufvfts}
\end{center}
\end{figure}

The Fourier coefficients decrease to machine precision during the 
whole computation in both spatial directions. 
As an example, we show in Fig. \ref{hyp2deps01ufvfts} the Fourier coefficients of $u$ on the left and of $v$
on the right plotted on the $k_x$-axis at different times.

\begin{figure}[htb!]
\begin{center}
\includegraphics[width=0.6\textwidth]{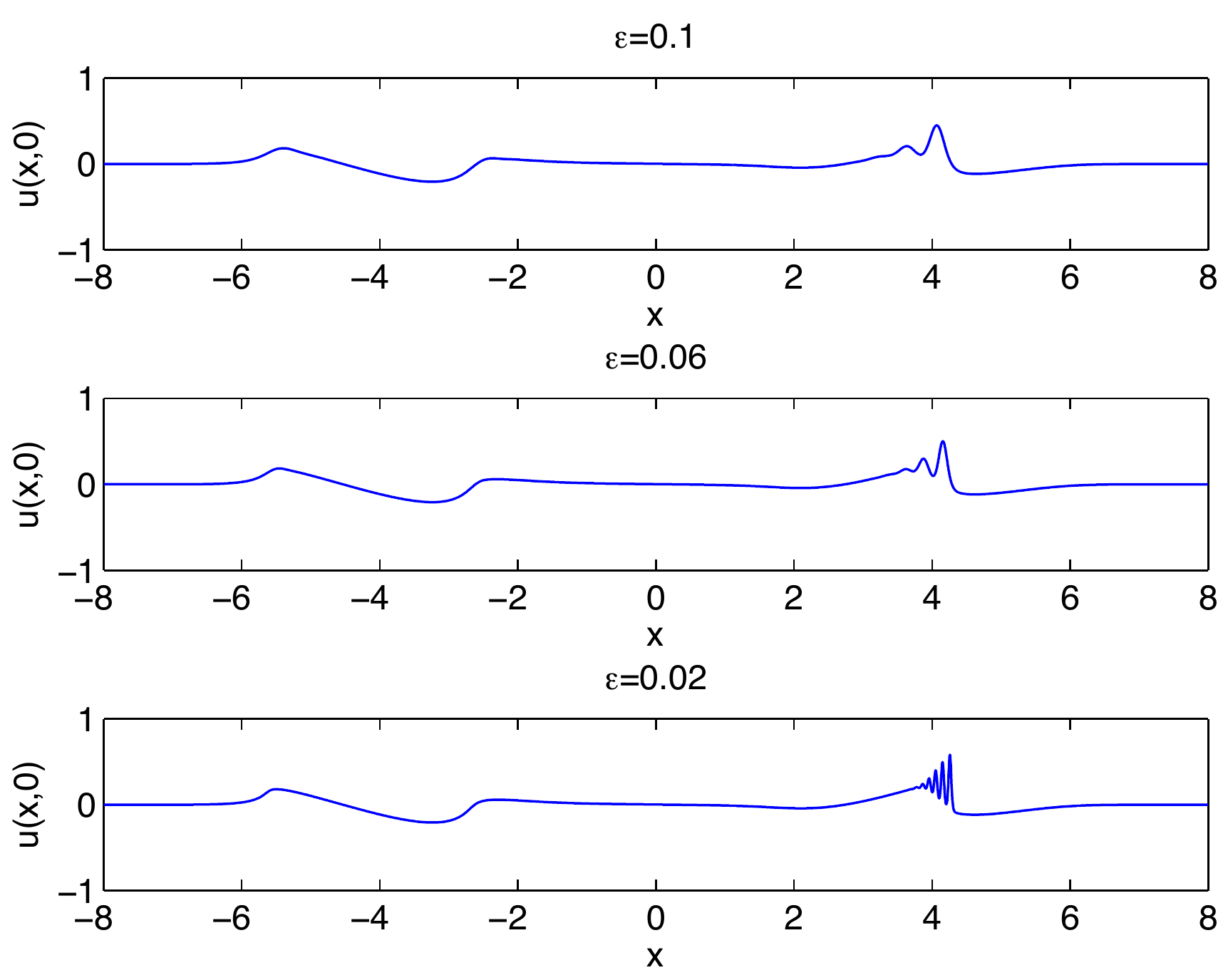}
\caption{Solutions $u$ of the $2+1$ dimensional hyperbolic Toda 
equation (\ref{toda2cont}) with $\rho=1$ at $t=4$ on the $x$-axis for different values of $\epsilon$.}
\label{hyp2depssut3}
\end{center}
\end{figure}

\begin{figure}[htb!]
\begin{center}
\includegraphics[width=0.6\textwidth]{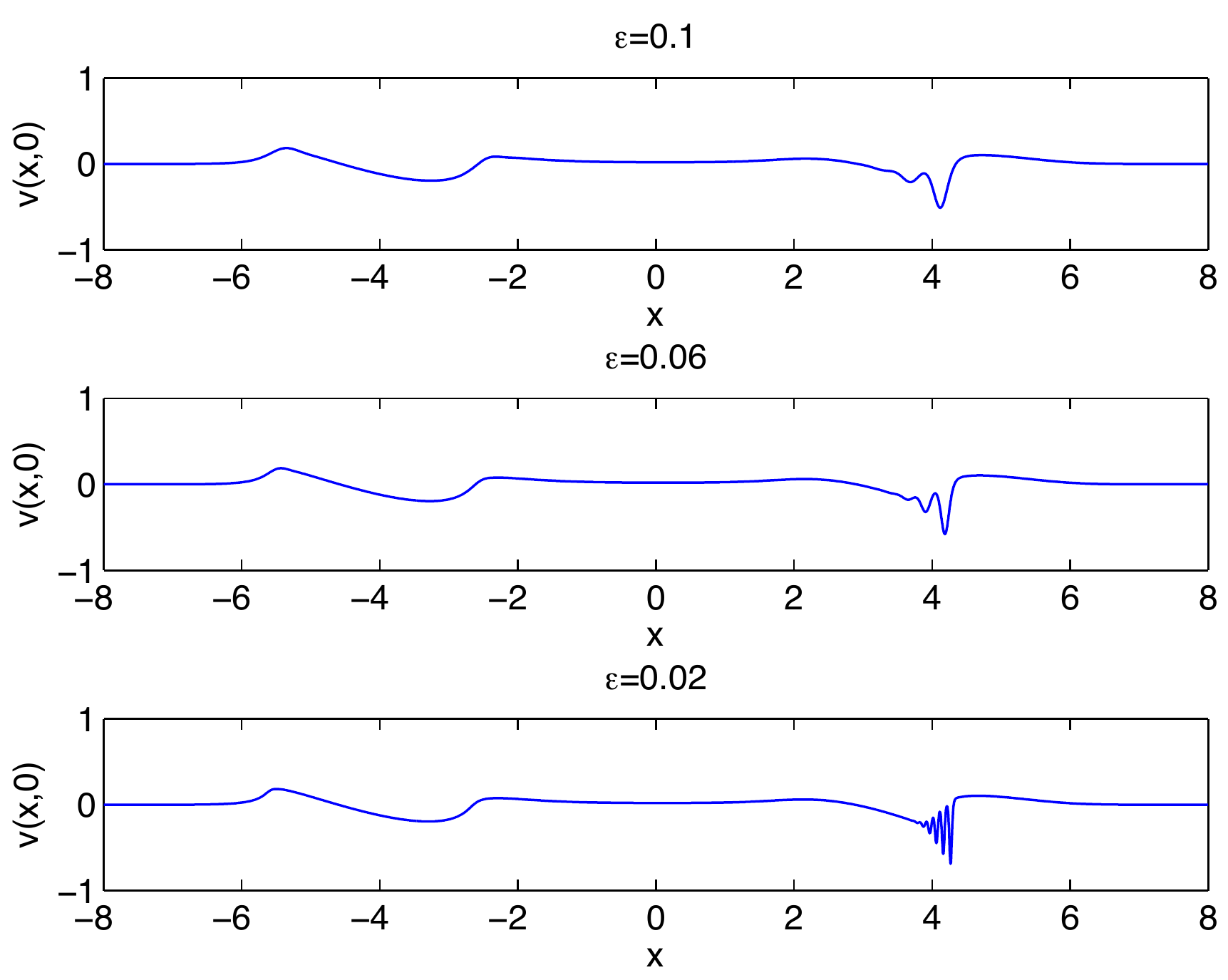}
\caption{Solutions $v$ of the $2+1$ dimensional hyperbolic Toda 
equation (\ref{toda2cont}) with $\rho=1$ at $t=4$ on the $x$-axis for different values of $\epsilon$}
\label{hyp2depssvt3}
\end{center}
\end{figure}

As $\epsilon$ becomes smaller, the number of oscillations increases 
as can be seen in Fig. \ref{hyp2depssut3} for $u$ and in Fig. \ref{hyp2depssvt3} for $v$.

\begin{figure}[htb!]
\begin{center}
\includegraphics[width=0.6\textwidth]{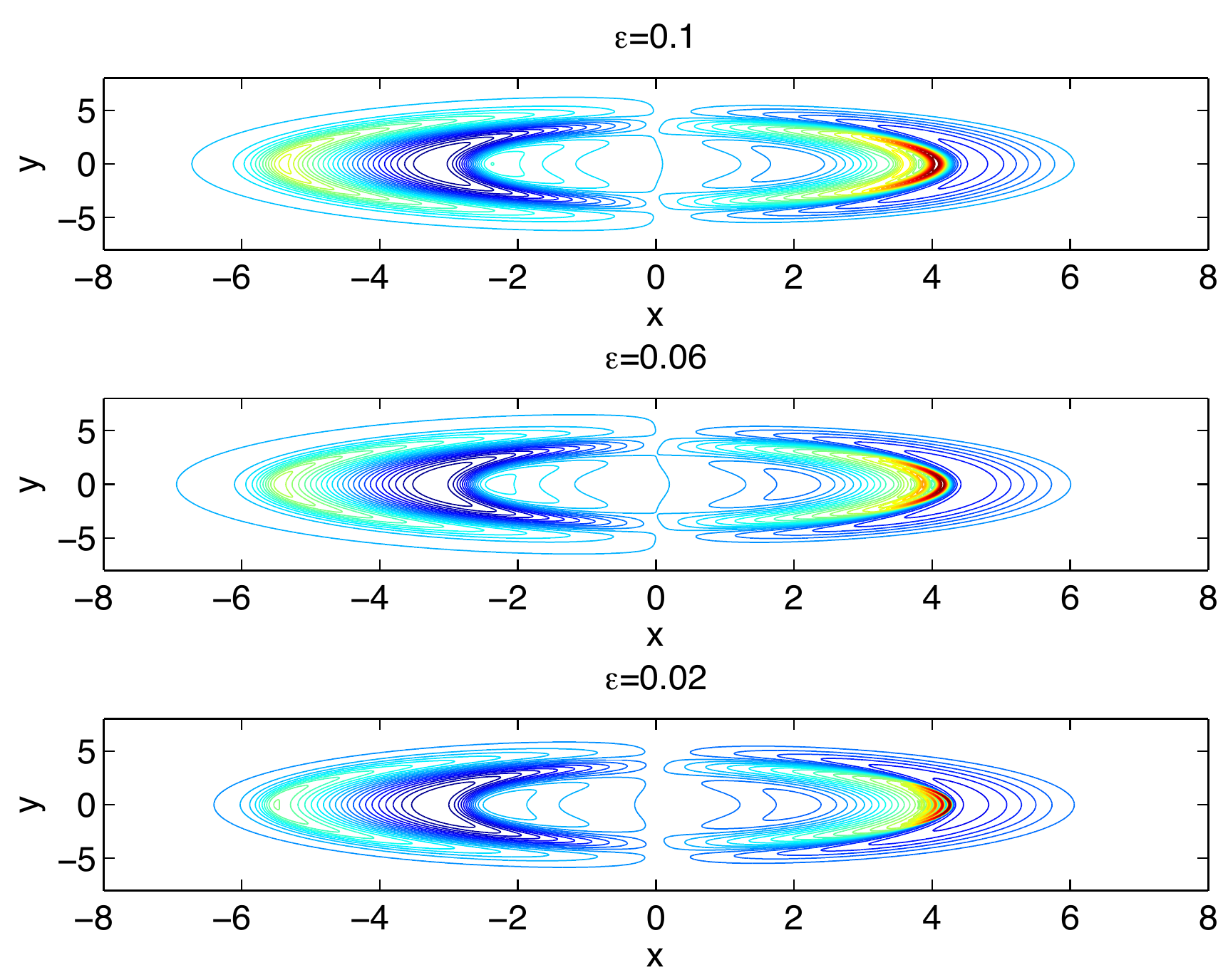}
\caption{Contour plot of the solutions $u$ of the $2+1$ dimensional 
hyperbolic Toda (\ref{toda2cont}) with $\rho=1$ equation at $t=4$ for different values of $\epsilon$.}
\label{hyp2depssut3cont}
\end{center}
\end{figure}

The contour plots of the solutions $u$ at $t=4$ of (\ref{toda2cont}) 
with $\rho=1$ for several values of $\epsilon$ can be seen in Fig. 
\ref{hyp2depssut3cont}. Note again the formation of an oscillatory zone
which becomes more clearly delimited for smaller $\epsilon$. 
%
%
%
%
%
%
%
%
%

\section{Numerical study of the elliptic Toda equations}
In this section we study the small dispersion limit of the elliptic Toda 
equations in $1+1$ and $2+1$ dimensions for the same initial data as 
in the hyperbolic case. We find for the one-dimensional dispersionless system that a cusp forms as in the case of 
the focusing semiclassical Schr\"odinger equations in $1+1$ 
dimensions. For small nonzero $\epsilon$ this cusp is not regularized via a 
dispersive shock as in the hyperbolic case, but leads to an 
$L_{\infty}$ blow-up in finite time $t^{*}>t_{c}$. The situation is 
thus as in the semiclassical limit of NLS equations with critical or 
supercritical nonlinearity. In $2+1$ dimensions the situation is similar to what was observed for the 
focusing DS II equations in \cite{DSdDS}. Concretely we find:
\begin{itemize}
    \item  the solution to 
the dispersionless equation has a point of gradient catastrophe on 
the $x$-axis of square root type, whereas it stays regular in 
$y$-direction;

    \item  the difference between the solutions to the 
dispersionless and the full Toda equation scales as $\epsilon^{2/5}$ 
as in the one-dimensional case;

\item for times larger than $t_{c}$, the solution develops for 
$\epsilon\ll1$ an $L_{\infty}$ blow-up in finite 
time $t^{*}>t_{c}$.
\end{itemize}
The scaling of the difference between the solutions to the 
dispersionless and the full Toda equation for small $\epsilon$ as 
$\epsilon^{2/5}$ could indicate that the \emph{tritronqu\'ee} 
solution of the PI equation might play a role also in the asymptotic 
description of  the Toda solution near the break-up of the 
corresponding solution 
to the dispersionless equation. 

\subsection{Dispersionless elliptic Toda equation in $1+1$ dimensions}

In this subsection we study again the initial data (\ref{uini1}), but 
this time for equation (\ref{toda1e0}) with  $\rho=-1$. The Fourier 
coefficients for the solution are fitted to the asymptotic 
formula (\ref{fourasymp2}). It was shown at the example of exact 
solutions  for the semiclassical 
NLS equations in $1+1$ dimensions in \cite{DSdDS} that 
for elliptic systems, the break-up is best identified from the 
asymptotic behavior of the Fourier coefficients if the code is 
stopped once the $\delta$ in (\ref{fourasymp}) becomes smaller than the 
smallest resolved distance (\ref{mres}) in physical space. Since we 
deal with an elliptic system here as well, we use the same criterion 
as in \cite{DSdDS} to identify the formation of a singularity.

\begin{figure}[htb!]
\begin{center}
\includegraphics[width=0.6\textwidth]{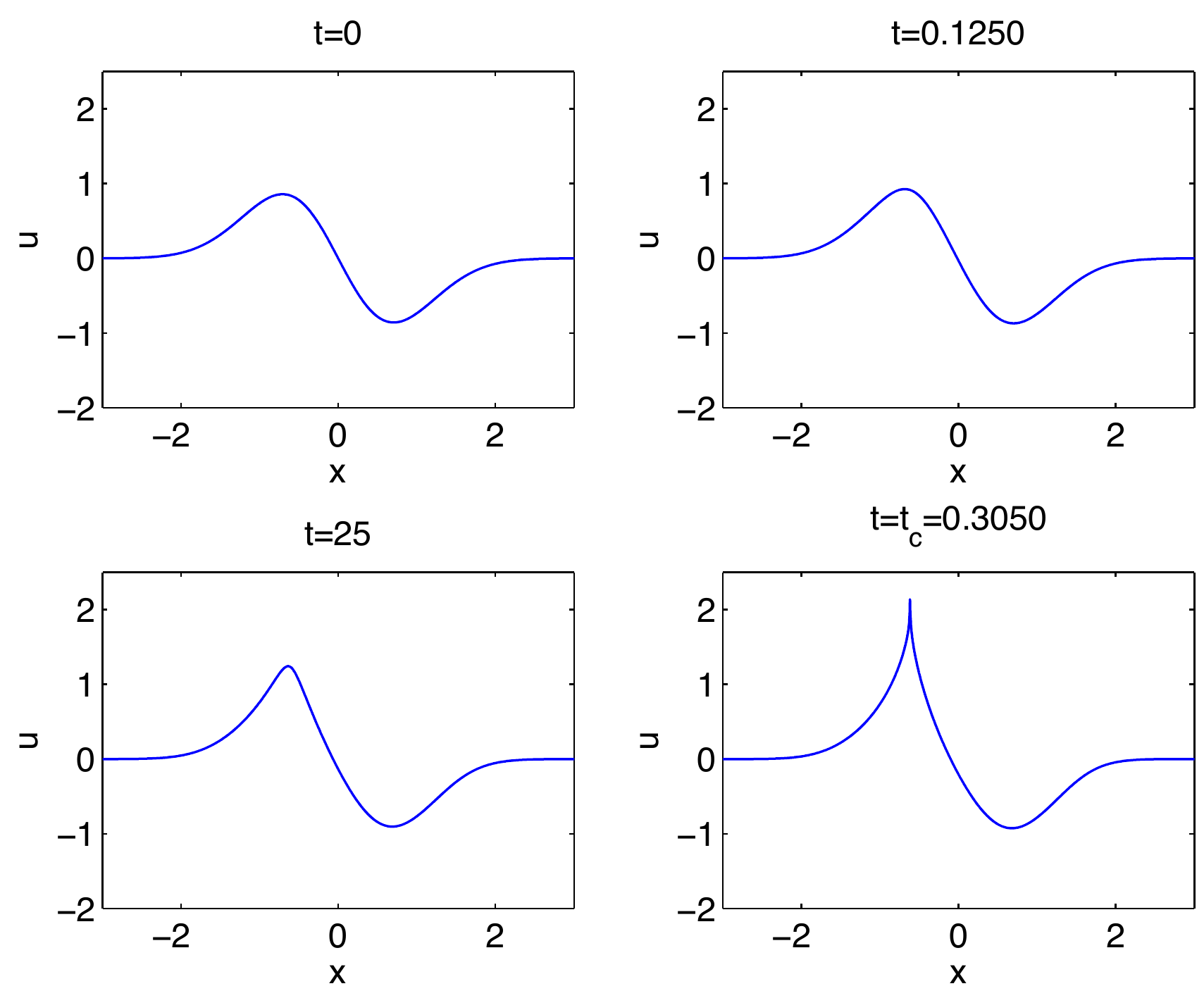}
\caption{Solution $u$ of the 1d elliptic dispersionless Toda equation  (\ref{toda1e0}) with $\rho=-1$ at different times for initial data of the form (\ref{uini1}).}
\label{Solutsellip}
\end{center}
\end{figure}

\begin{figure}[htb!]
\begin{center}
\includegraphics[width=0.6\textwidth]{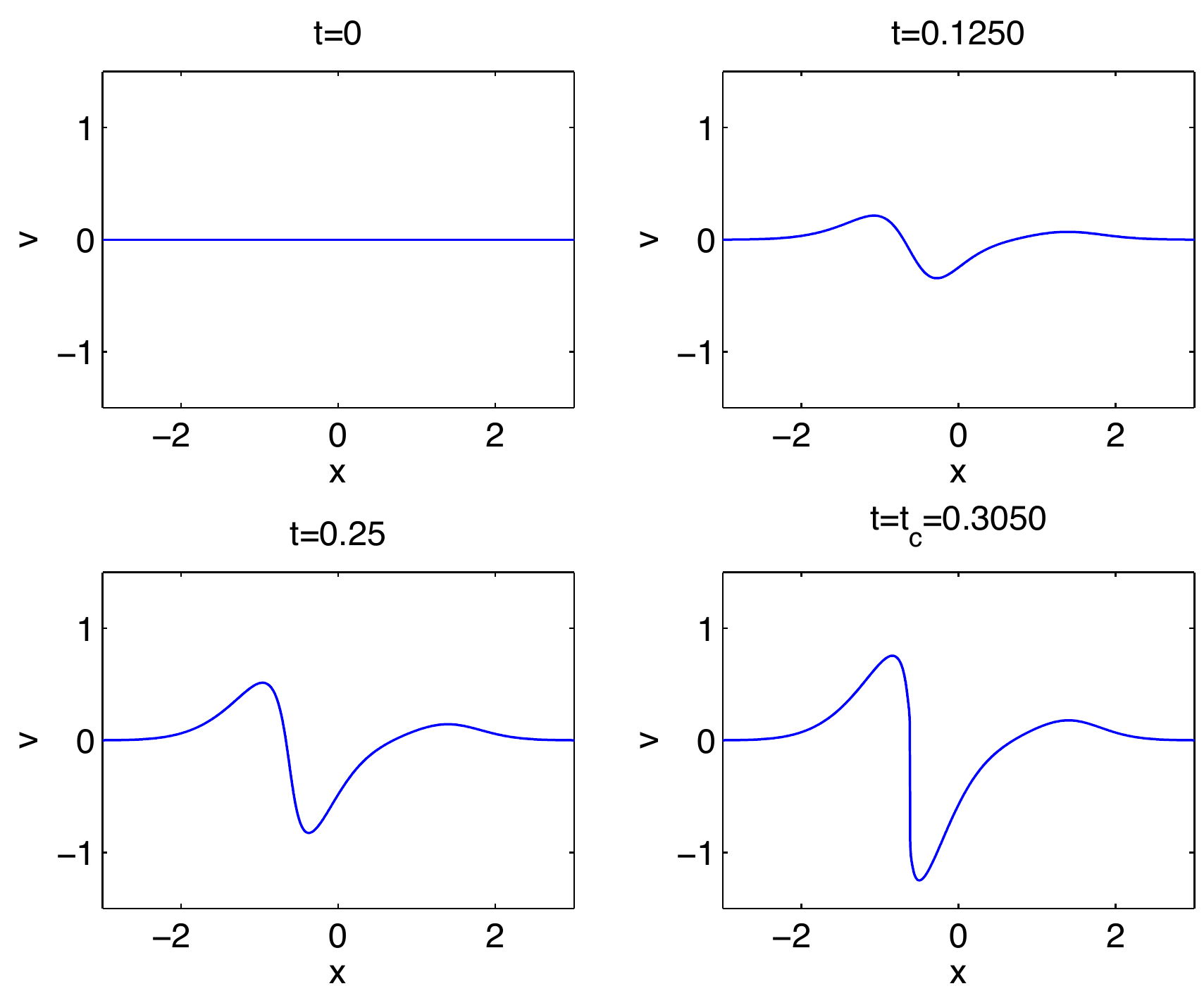}
\caption{Solution $v$ of the 1d elliptic dispersionless Toda equation  (\ref{toda1e0}) with $\rho=-1$ at different times for initial data of the form (\ref{uini1}).}
\label{Solvtsellip}
\end{center}
\end{figure}

To solve the dispersionless Toda equation (\ref{toda1e0}) with 
$\rho=-1$ for the initial data (\ref{uini1}), we use $N=2^{14}$ 
Fourier modes and the time step 
$\delta_t = 5*10^{-5}$. The resulting solutions $u$ and $v$ are shown for several 
times up to the critical time $t_{c}$  in Fig. \ref{Solutsellip}, respectively \ref{Solvtsellip}.
The former clearly develops a cusp, as in the case of the focusing 
semiclassical NLS equation. As in the latter case in \cite{DSdDS},  we  get a value of 
$B_u$ close to one instead of the expected value $1.5$. The reason 
for this is again that the quantity $B=\mu+1$ in (\ref{fourasymp2}), 
corresponding to an algebraic decrease of the Fourier coefficients,  is much 
more sensitive to the fitting procedure than the exponential part 
parametrized by $\delta$. 
Thus the errors in the Fourier coefficients for the high wave numbers 
as discussed in \cite{DSdDS} affect the determination of $\mu$ much more than 
the vanishing of $\delta$. 

\begin{figure}[htb!]
\begin{center}
\includegraphics[width=0.4\textwidth]{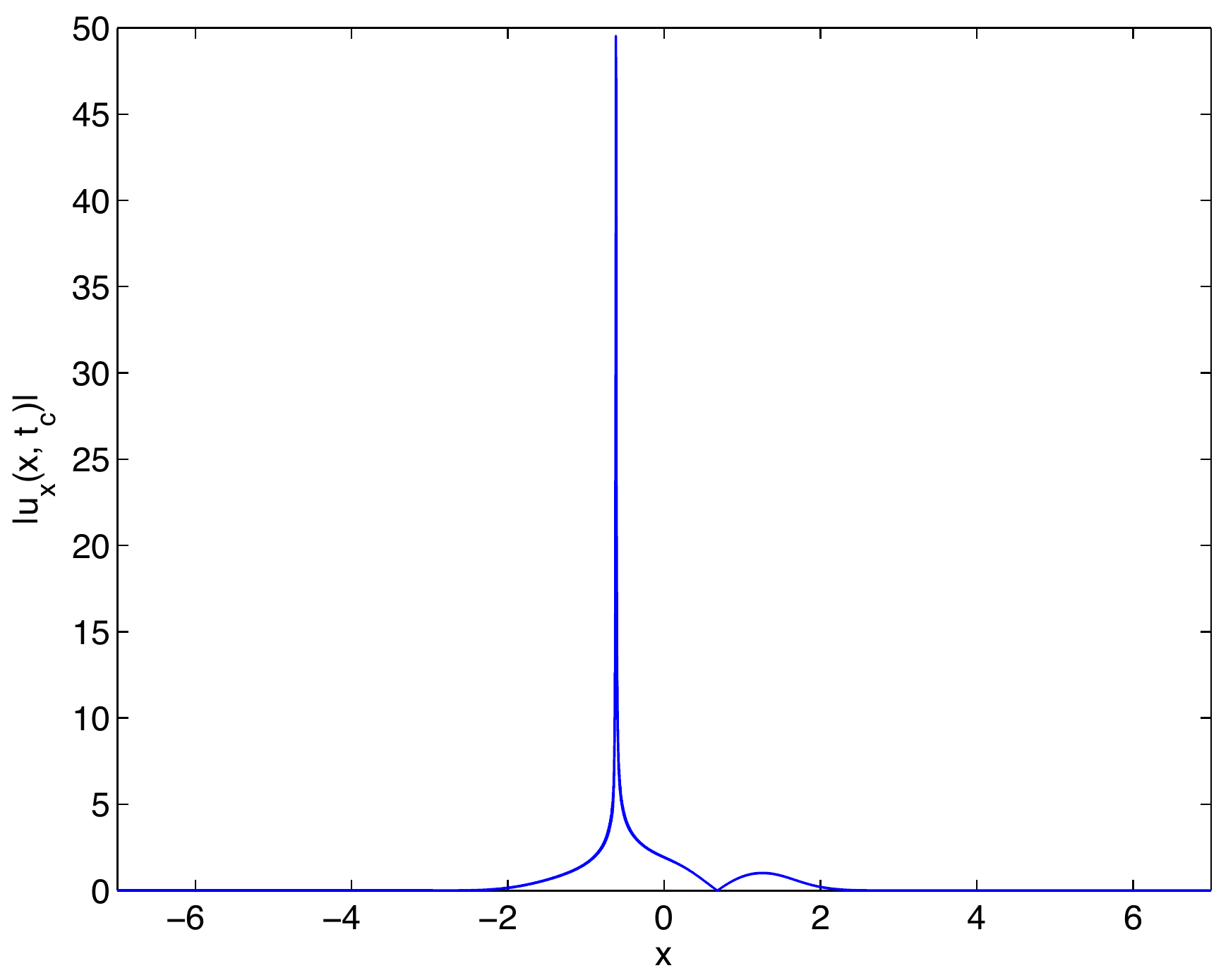}
\includegraphics[width=0.4\textwidth]{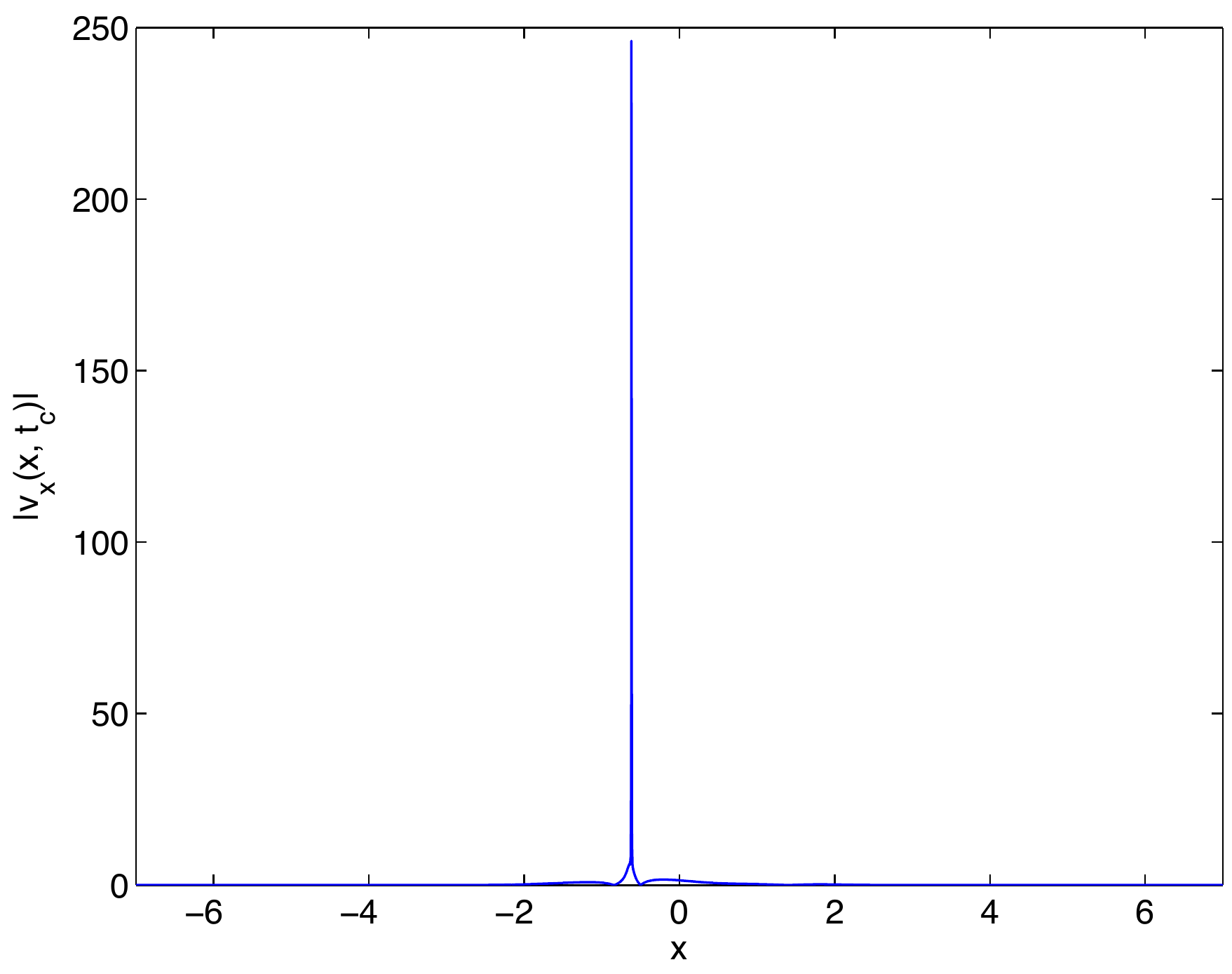}
\caption{Gradient of $u$ on the left and of $v$ on the right at the 
critical time $t_c=0.3050$; $(u,v)$ being the solution of 
the 1d elliptic dispersionless Toda equation (\ref{toda1e0}) with $\rho=-1$ for 
initial data of the form (\ref{uini1}).}
\label{gradtcuvellip}
\end{center}
\end{figure}

The behavior of $|u_x|$ and $|v_x|$ at $t=t_c=0.3050$ is shown in 
Fig.~\ref{gradtcuvellip}. 
We observe at this time that $\| u_x \|_{\infty} \sim 45$ and $\| v_x 
\|_{\infty} \sim 230$ which clearly indicates a point of gradient 
catastrophe for both $u$ and $v$. 

\begin{figure}[htb!]
\begin{center}
\includegraphics[width=0.4\textwidth]{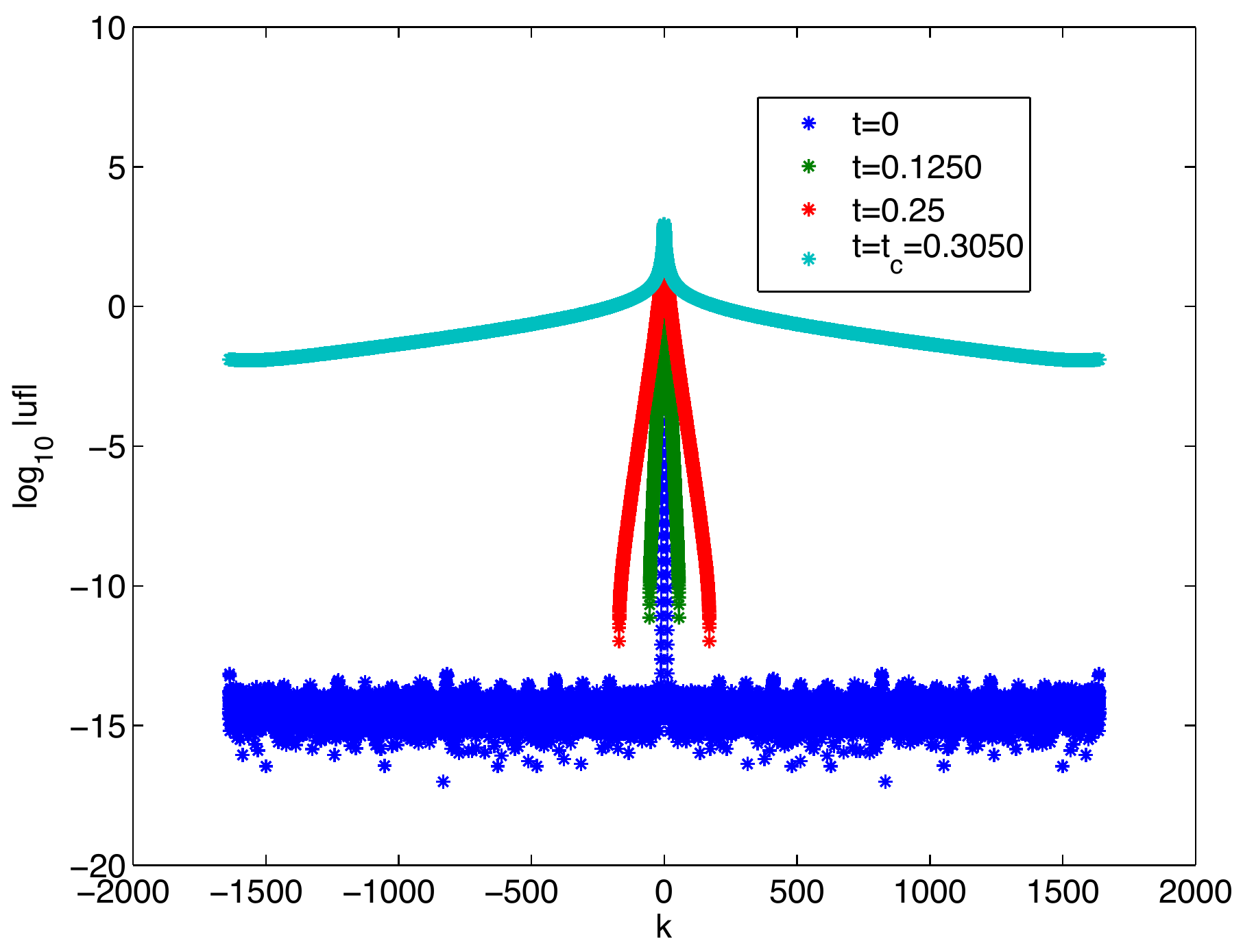}
\includegraphics[width=0.4\textwidth]{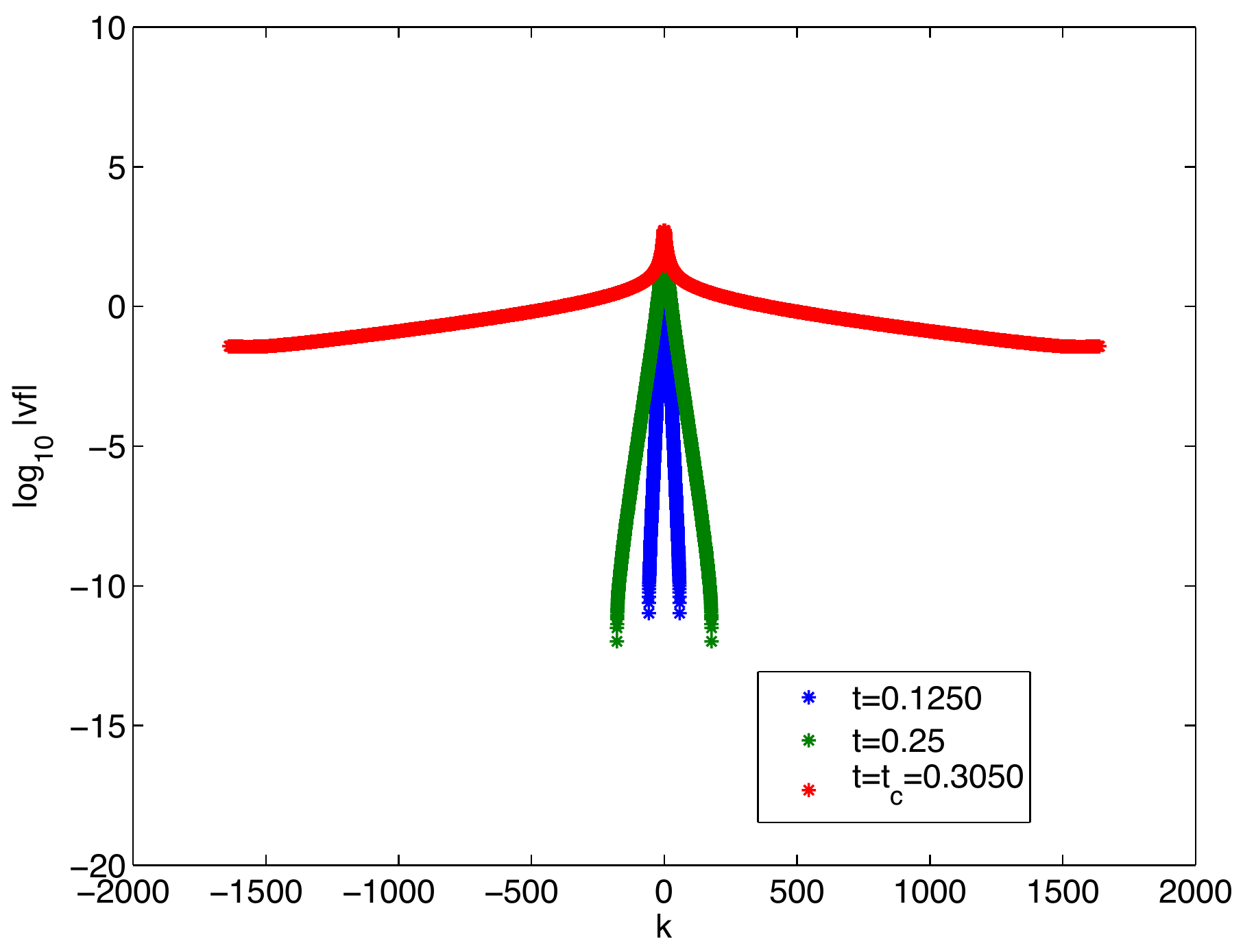}
\caption{Fourier coefficients of $u$ (left), denoted by $uf$, and 
Fourier coefficients of $v$, denoted respectively by $vf$,
at different times, corresponding to the situations in Fig. 
\ref{Solutsellip} and \ref{Solvtsellip}.}
\label{Coefstsuvellip}
\end{center}
\end{figure}

To control the accuracy of the solution, we present in Fig. \ref{Coefstsuvellip}
the Fourier coefficients for the situation in Fig. \ref{Solutsellip} 
and \ref{Solvtsellip}. 
We also verify that the numerically computed energy reaches a precision of $\Delta_E \sim 10^{-13}$ at $t=t_c=0.305$.

\begin{figure}[htb!]
\begin{center}
\includegraphics[width=0.4\textwidth]{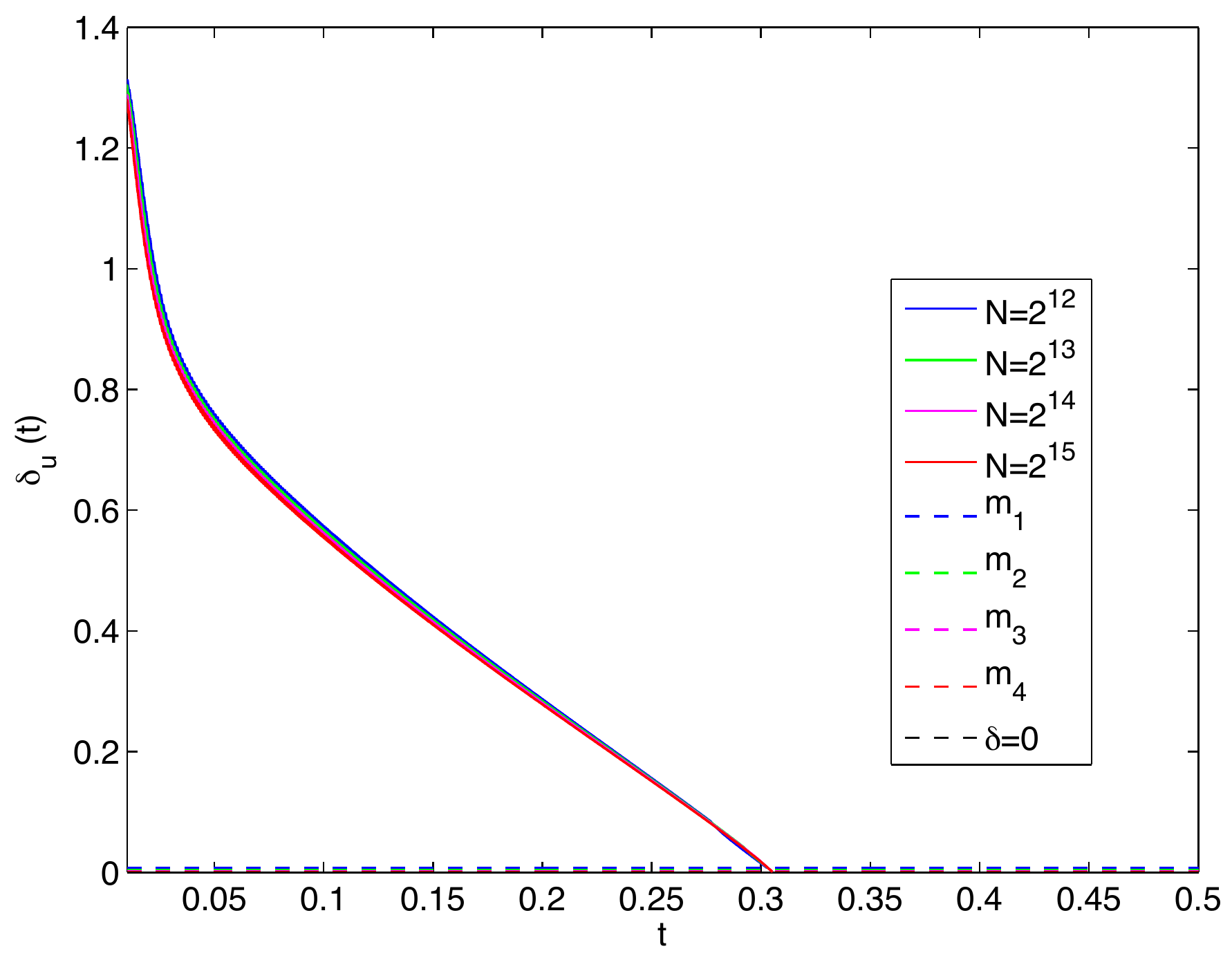}
\includegraphics[width=0.4\textwidth]{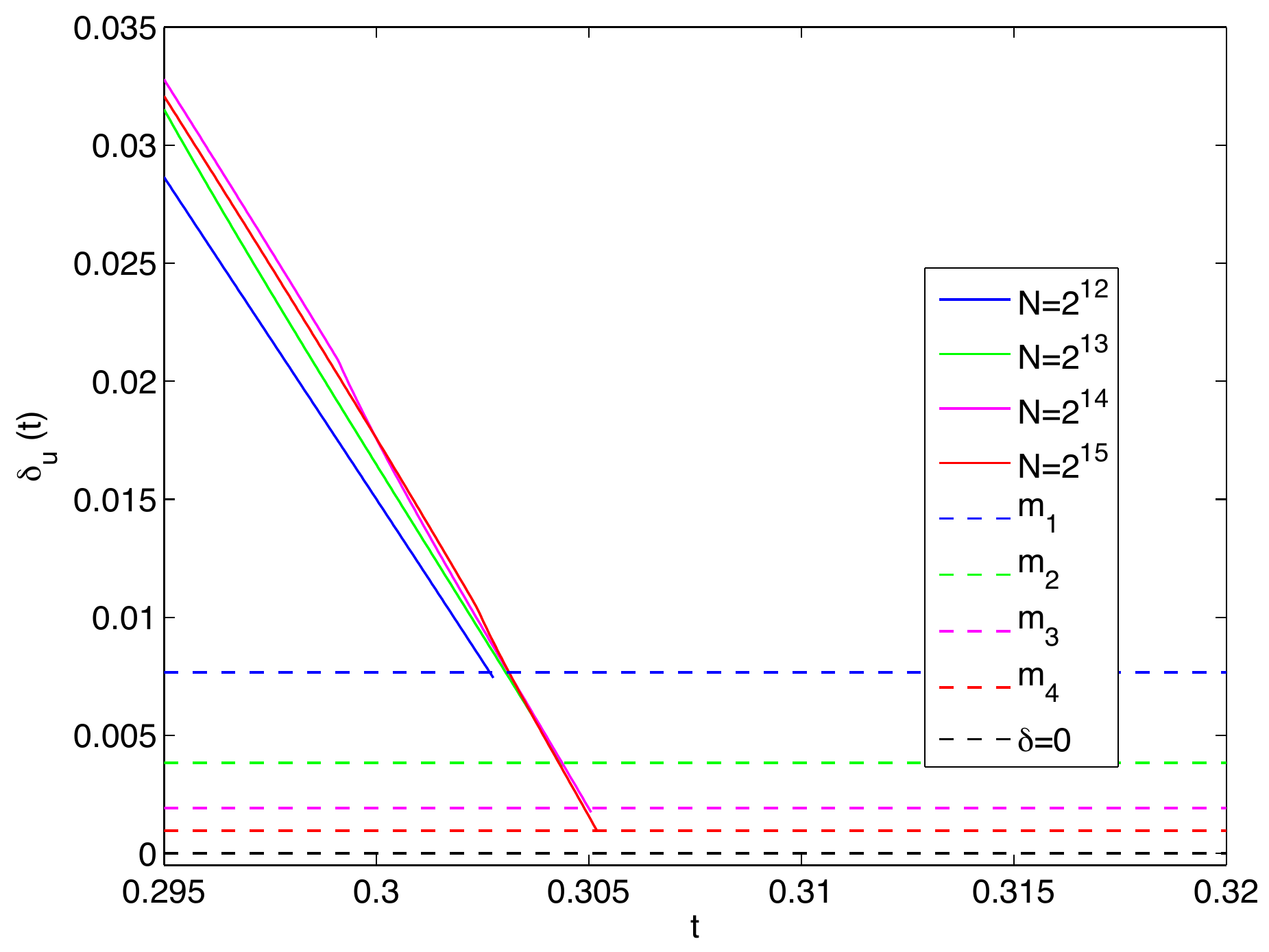}
\caption{Fitting parameter $\delta$ in (\ref{fourasymp2}) for $u$ 
(denoted by $\delta_u$) for the solution of the 1d elliptic 
dispersionless Toda equation (\ref{toda1e0}) with $\rho=-1$ for initial data of the form (\ref{uini1}). 
The fitting is done for $10<k<2*max(k)/3$ for different resolutions 
$N=2^{12},2^{13},2^{14},2^{15}$; on the right a close-up of the 
situation for $t\sim t_{c}$.}
\label{deluNsellip}
\end{center}
\end{figure}

Since the determination of the critical point and the critical 
solution is crucial for the understanding of the behavior of the 
solution to the full Toda equation in the vicinity of the critical 
point, we discuss the used approach in more detail. In Fig. 
\ref{deluNsellip}, we show the time evolution of $\delta_u$ (the 
parameter $\delta$ in (\ref{fourasymp}) for $u$) for different 
resolutions $N$.
We perform the fitting for the same range of $k$ as in the hyperbolic 
case, but stop the computation when $\delta_u$, respectively 
$\delta_v$ are smaller than the smallest distance $m$  
(\ref{mres}) resolved in physical space. 

As already observed in the previous section, both the fitting of the 
Fourier coefficients $uf$ and on $vf$ give (within numerical 
precision) the same critical time, 
$t_c \sim 0.305$ as shown in Table \ref{NtcBuvellip}, which indicates the 
consistency and the reliability of the approach. In Table  \ref{NtcBuvellip}
we also 
present the corresponding values of $B_u(t_c)$ and $B_v(t_c)$ which 
are, as expected in \cite{DSdDS}, not close to the expected 1.5. 
Note  that the  found critical times for $N=2^{14}$ and $N=2^{15}$ 
are almost identical which means that a resolution of $N=2^{14}$ 
points is sufficient for our purposes.  
\begin{table}
\centering
\begin{tabular}{|c|ccc|}
\hline

  $N$ & $t_c$ & $B_u(t_c)$ & $B_v(t_c)$ \\
  \hline
   $2^{12}$ & 0.3027 & 1.2204 & 1.1349 \\
  $2^{13}$ & 0.3044 & 1.1172 & 1.0365\\
   $2^{14}$ &  0.3050 & 1.0575 & 0.9785\\
  $2^{15}$ &  0.3052 & 1.0193 & 0.9424\\
 \hline
  
\end{tabular}
\label{NtcBuvellip}
\caption{Critical times of the solution to the 1d elliptic dispersionless Toda equation for initial data of the form (\ref{uini1})
for several values of $N$. The values of the fitting parameters 
$B_u$, $B_v$ at $t_c$ are also given.}
\end{table}

Thus we find here that the solutions of the elliptic dispersionless 
Toda equation in $1+1$ dimensions show a very similar behavior to 
what was found in \cite{DSdDS} for the semiclassical cubic NLS 
equation. This indicates that the singularity is in both cases of
square root type.

\subsection{One-dimensional elliptic Toda equation in the limit of small dispersion}
In this subsection, 
we study the behavior of solutions of the $1+1$ dimensional Toda 
equation (\ref{todacont3})  for several small values of $\epsilon$ 
and initial data of the form (\ref{uini1}). The solutions will be 
compared near the critical time $t_{c}$ with the critical solution of  
the dispersionless system (\ref{toda1e0}) for the same initial data.

We compute the solution to the  elliptic Toda equation 
(\ref{todacont3}) with $\rho=-1$ for different values of $\epsilon$ for initial data of the form (\ref{uini1}).
The computation is carried out with $N=2^{14}$ points for $x \in 
[-5\pi, 5\pi]$ and time step $\Delta_t=5*10^{-5}$.
We first study the solution up to the critical time of the 
corresponding dispersionless system determined in the previous 
subsection, i.e., $t_c=0.3050$.

\begin{figure}[htb!]
\begin{center}
\includegraphics[width=0.45\textwidth]{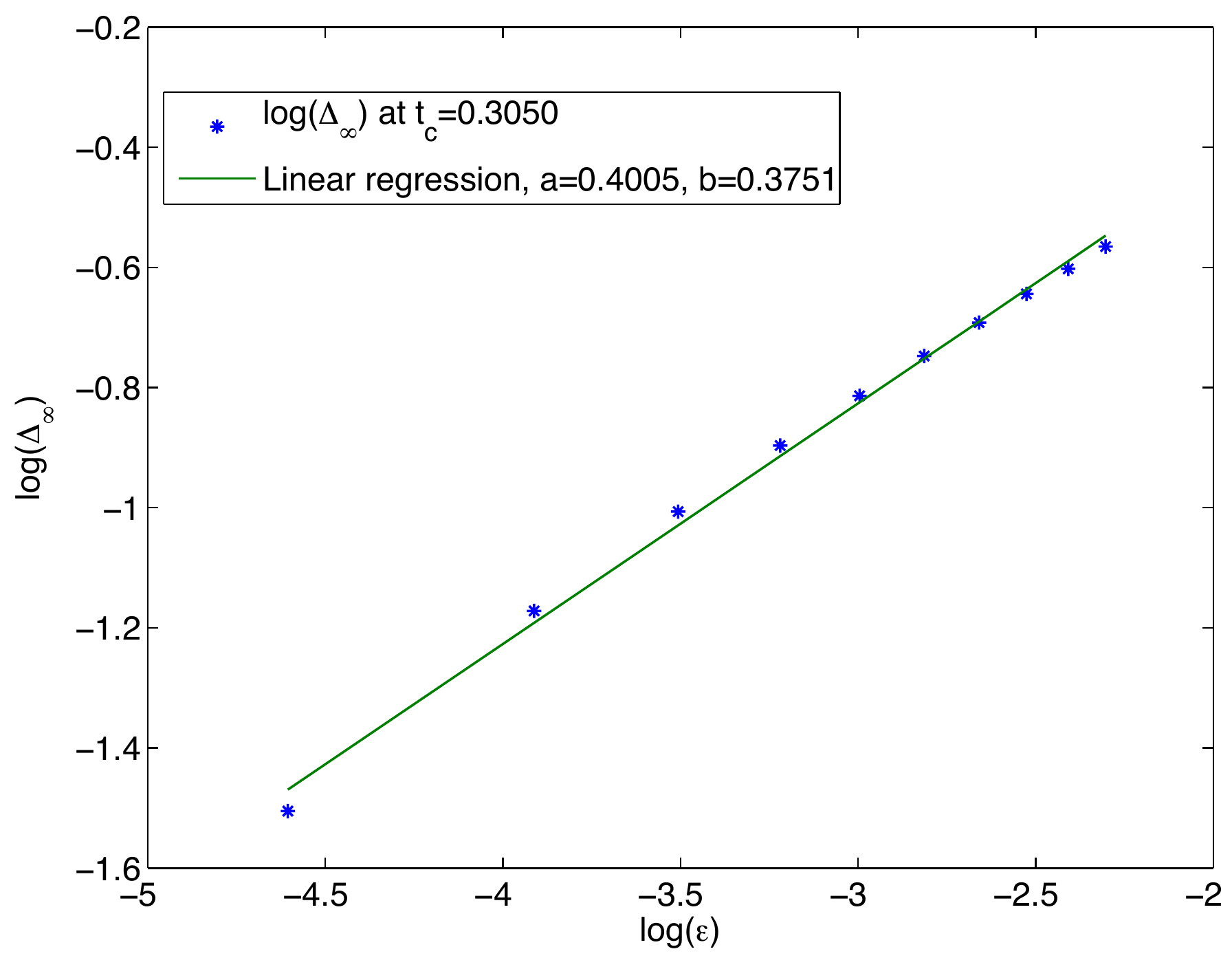}
\caption{$L_{\infty}$ norm  $\Delta_{\infty}$ of the difference 
 between solutions to the 1d elliptic dispersionless Toda and the Toda 
 equation for the 
 initial data (\ref{uini1}) in dependence of $\epsilon$ at $t_c=0.3050$.}
\label{scalingtcellip}
\end{center}
\end{figure}

We are interested in the scaling in $\epsilon$ of the $L_{\infty}$ 
norm $\Delta_{\infty}$ of the difference between the solutions to the 
1d elliptic dispersionless Toda and 
the full Toda equation for the same initial data (\ref{uini1}). This 
norm is shown in Fig. \ref{scalingtcellip} at $t_c=0.3050$ for 
$0.01 \leq \epsilon \leq 0.1$.     

A linear regression analysis ($\log_{10} \Delta_{\infty} = a \log_{10} \epsilon + b$ ) shows that $\Delta_{\infty}$ decreases as 
\begin{align}
\mathcal{O} \left( \epsilon^{0.40} \right)  \sim \mathcal{O} \left( \epsilon^{2/5} \right) \,\, \mbox{at} \,\, t=t_c=0.3050, \,\, \mbox{with}\,\, a=0.4005  \,\,\mbox{and} \,\,  b= 0.3751.
\end{align}
The correlation coefficient is $r = 0.999$. Thus we find the same 
scaling as conjectured for NLS equations, see \cite{DGK,DGK13}.

\begin{figure}[htb!]
\begin{center}
\includegraphics[width=0.45\textwidth]{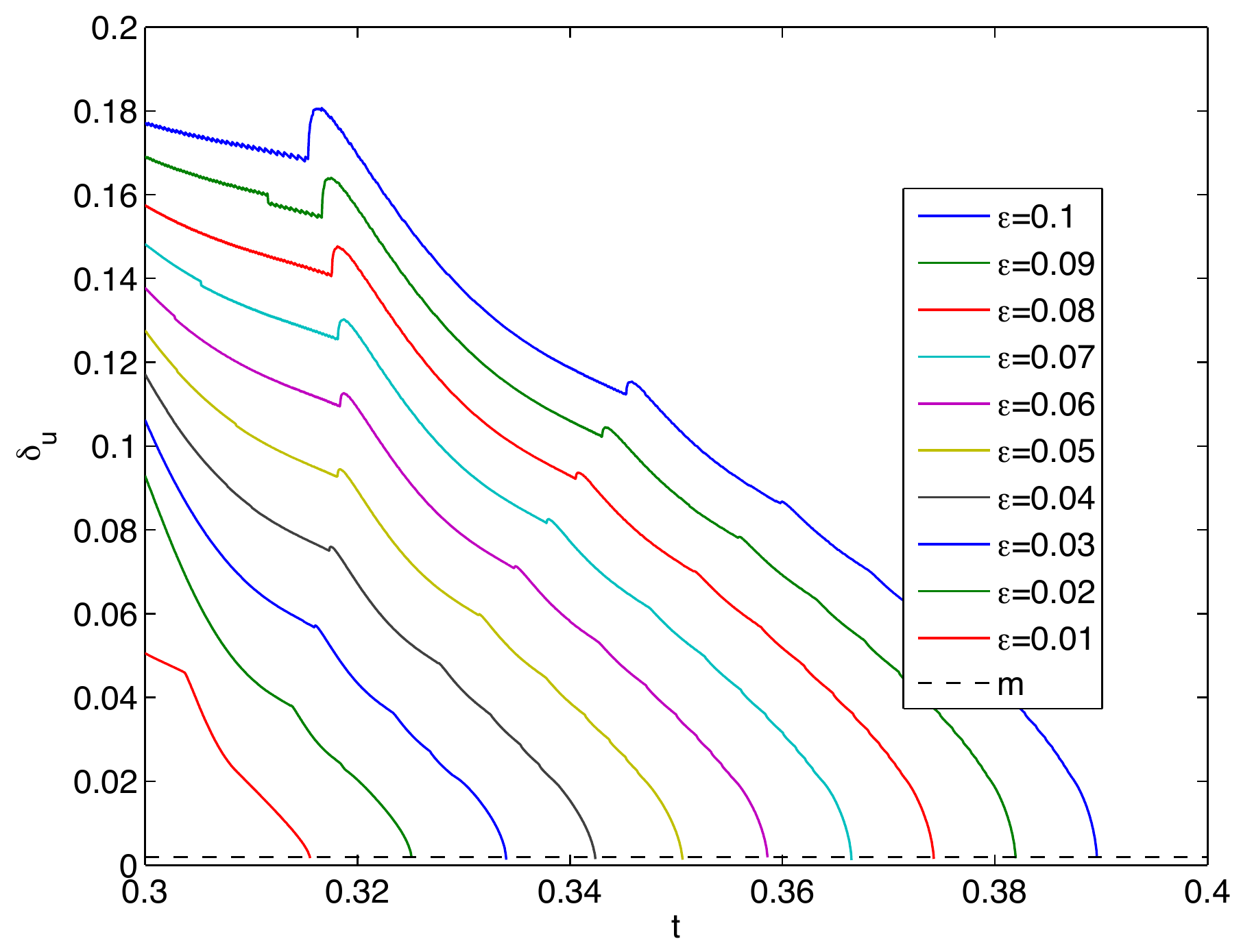}
\includegraphics[width=0.45\textwidth]{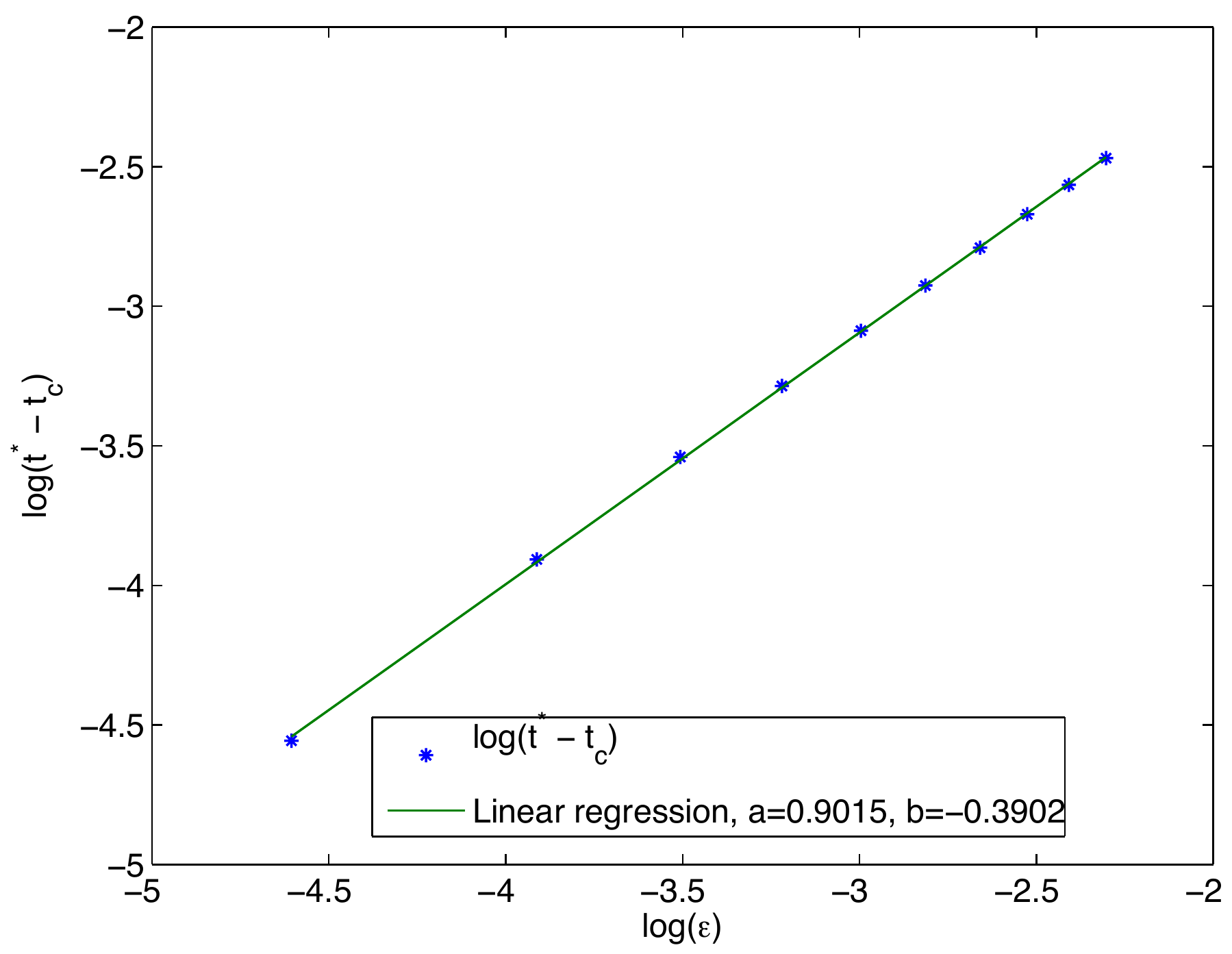}
\caption{Fitting parameter $\delta$ of (\ref{fourasymp}) for the 
solution $u$ (denoted by $\delta_{u}$) of the 1d elliptic Toda equation 
(\ref{todacont3}) with $\rho=-1$ for initial data of the form 
\ref{uini1} on the left, and the $L_{\infty}$ norm   of the difference 
 between blow-up time $t^{*}$ and break-up time $t_c$ in dependence of $\epsilon$ on the right 
 in a loglog plot.}
\label{deluellipeps}
\end{center}
\end{figure}

For times greater than $t_{c}$, we find as in the case  of the $1+1$ 
dispersionless  quintic (or higher power nonlinearity) NLS equation that the solution blows up in finite time. 
We use again the asymptotics of the Fourier coefficients 
(\ref{fourasymp2}) to determine the time of the appearance of a 
singularity on the real axis  as in \cite{DSdDS,KNLSsol}. As the 
blow-up time $t^{*}$ we define the time when the quantity $\delta$ in 
(\ref{fourasymp2}) for $u$ (denoted by $\delta_u$) becomes smaller than the 
smallest resolved distance (\ref{mres}) in physical space. In Fig. 
\ref{deluellipeps} we show the time evolution of $\delta_u$ for several values of $\epsilon$. 

 The thus determined blow-up times $t^{*}$ are given in 
 Table \ref{elblowtimes} for different values of $\epsilon$. The 
 $L_{\infty}$ norm of the difference $t^{*}-t_c$  
 scales as $\epsilon^{0.9}$, see Fig. \ref{deluellipeps} where the 
 results of the linear regression are shown. 
\begin{table}
\centering
\begin{tabular}{|c | cccccccccc|}
\hline
  $\epsilon$ & $0.1$ & $0.09$ & $0.08$ & $0.07$ &  $0.06$ & $0.05$ & $0.04$ & $0.03$ & $0.02$ & $0.01$\\
  \hline
 &&&&&&&&&& \\
  $t^{*}$ &     0.3896 & 0.3819 & 0.3742 & 0.3664 & 0.3586 & 0.3506 & 0.3424 & 0.3340 & 0.3251 & 0.3155 \\
  \hline
 \end{tabular}
\label{elblowtimes}
\caption{Values of the determined blow-up times of the solution of 
the 1d elliptic Toda equation (\ref{todacont3}) with $\rho=-1$ for 
initial data of the form (\ref{uini1}) for several values of $\epsilon$.}
\end{table}

Thus as for focusing  NLS equations,
we find here that  in the limit of small dispersion, i.e., as 
$\epsilon \to 0$, the solutions to the Toda equation 
(\ref{todacont3}) blow up in the $L_{\infty}$ norm at a finite  time 
$t^*$. This time is always greater than the  critical time of the 
break-up of the corresponding solution to the dispersionless system. 
In the limit $\epsilon\to0$, the blow-up time tends to the break-up 
time roughly as $\epsilon^{0.9}$.

%
%
%
%
%
%

\subsection{Dispersionless elliptic two-dimensional Toda equation}
In this subsection we numerically solve the dispersionless elliptic 
Toda equation (\ref{toda2e0}) in $2+1$ dimensions for the initial 
data (\ref{uini2}). It is shown that the same type of singularity is 
found as in the $1+1$ dimensional case. 

\begin{figure}[htb!]
\begin{center}
\includegraphics[width=0.45\textwidth]{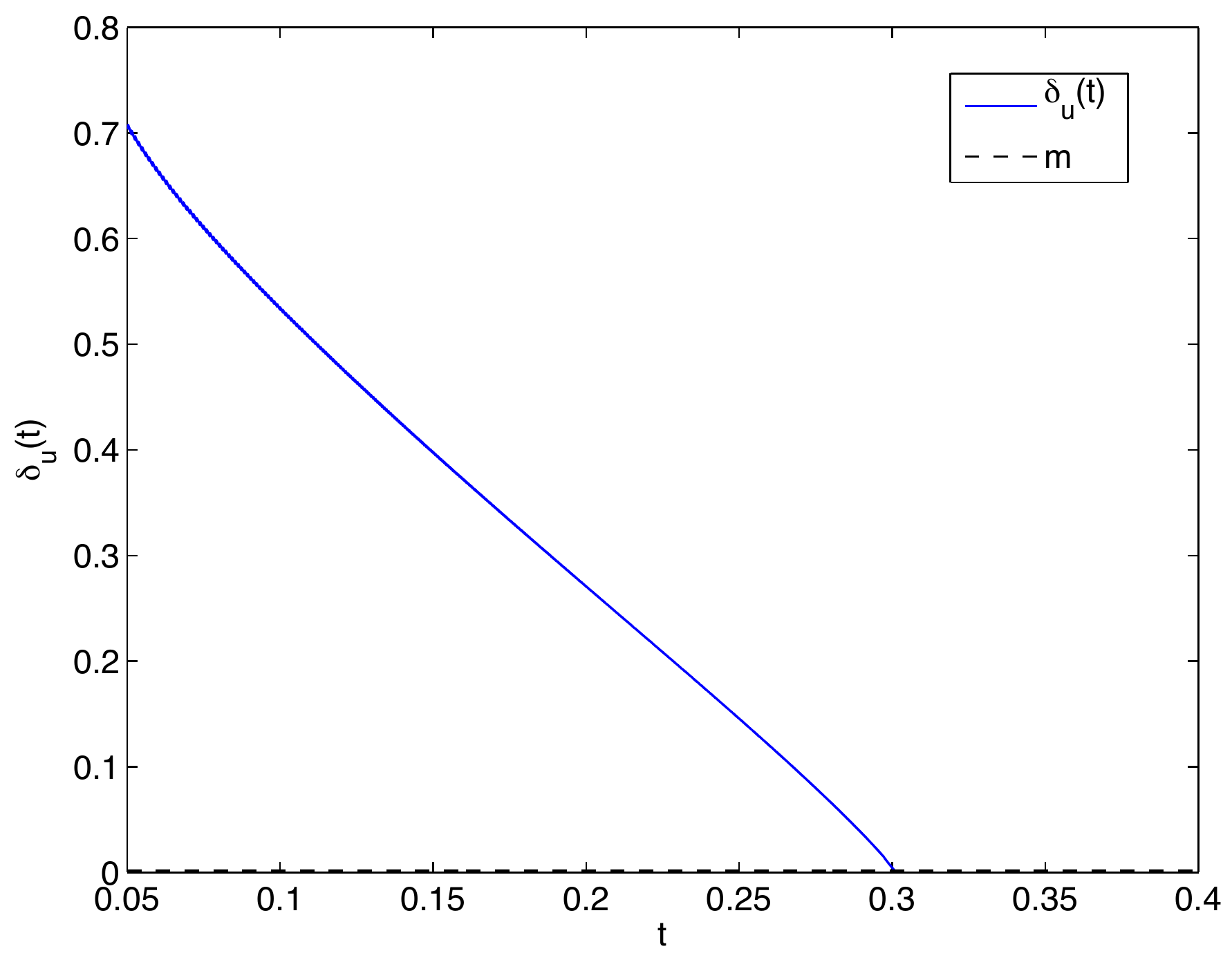}
\caption{Fitting parameter $\delta$ in (\ref{fourasymp}) for $u$ 
(denoted by $\delta_u$) for the solution of the two-dimensional 
elliptic dispersionless Toda equation (\ref{toda2e0}) with $\rho=-1$ 
for initial data of the form (\ref{uini2}). The fitting is done for $10<k<2*max(k)/3$.}
\label{deluellip2d}
\end{center}
\end{figure}

For the dispersionless $2+1$ dimensional Toda equation 
(\ref{toda2e0}) with $\rho=-1$, we use $2^{14} \times 2^{9}$ points 
for $x \times y \in [-5\pi, 5\pi] \times [-5\pi, 5\pi]$ and
the time step is   $\delta_t=5*10^{-5}$. We find that 
the solution of the dispersionless system for initial data of the 
form (\ref{uini2}) develops a singularity at $t_c\sim0.3007$. As in 
the $1+1$ dimensional case, the latter time is obtained from the 
asymptotic behavior of the Fourier coefficients: the Fourier 
coefficients for $u$ are fitted to the asymptotic formula 
(\ref{fourasymp2}) giving $\delta_{u}$. The critical time is defined 
as the time when  $\delta_u$ becomes smaller than the smallest 
resolved distance (\ref{mres}) in physical space, see Fig. \ref{deluellip2d}. 

\begin{figure}[htb!]
\begin{center}
\includegraphics[width=0.45\textwidth]{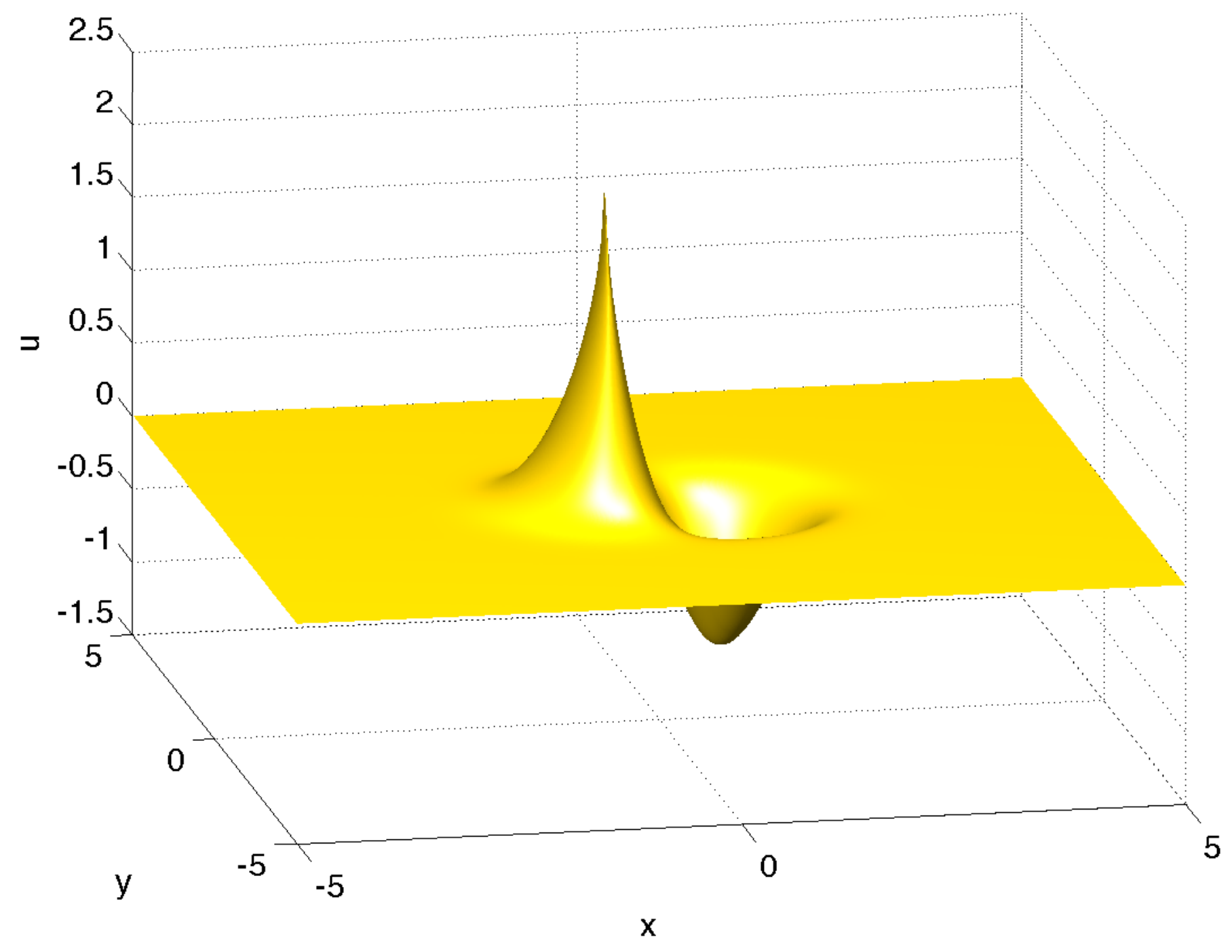}
\includegraphics[width=0.45\textwidth]{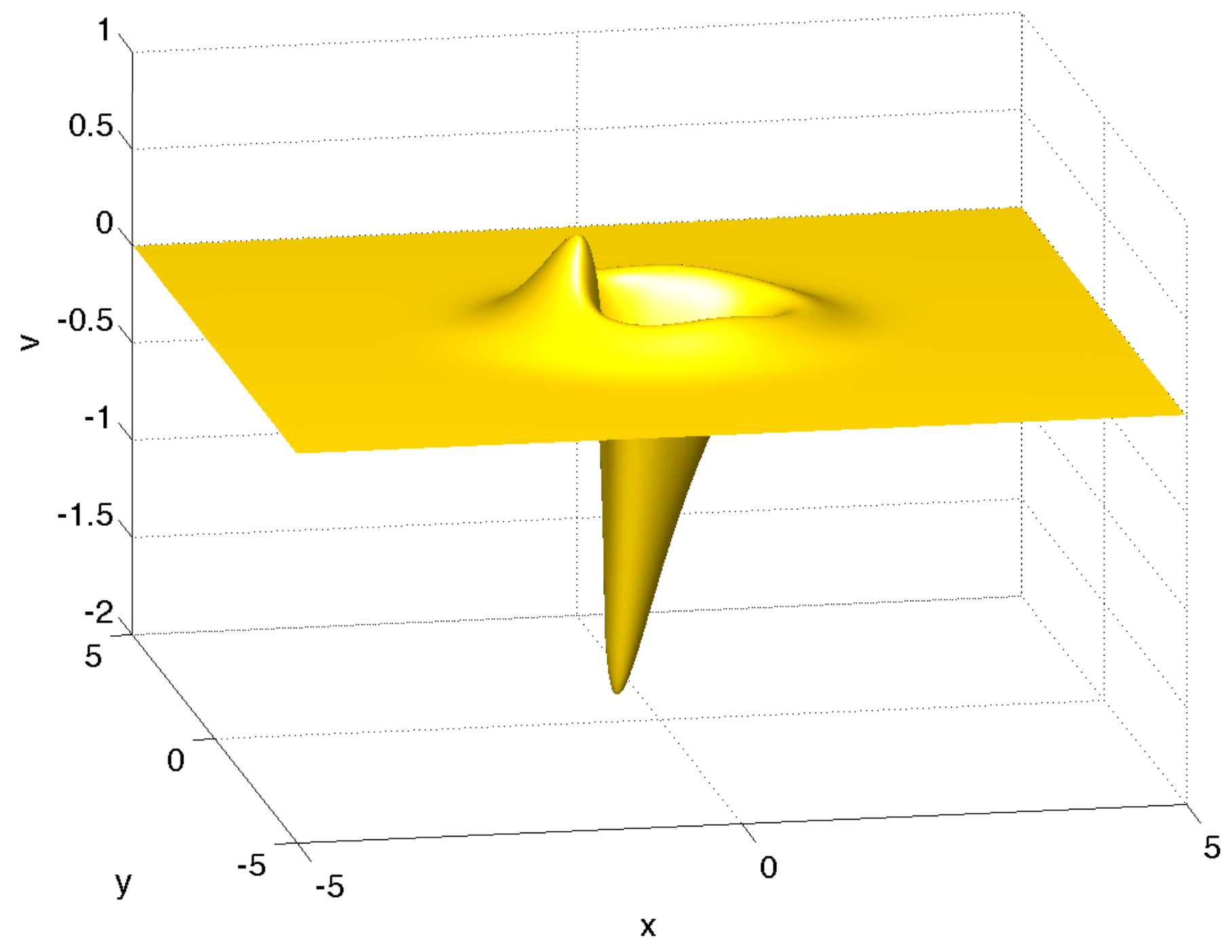}
\caption{Solutions $u$ and $v$ of the elliptic dispersionless Toda equation in 
$2+1$ dimensions (\ref{toda2e0}) with $\rho=-1$ at $t=t_c=0.3007$ for initial data of the form (\ref{uini2}).}
\label{Soluvtsellip2d}
\end{center}
\end{figure}

\begin{figure}[htb!]
\begin{center}
\includegraphics[width=0.4\textwidth]{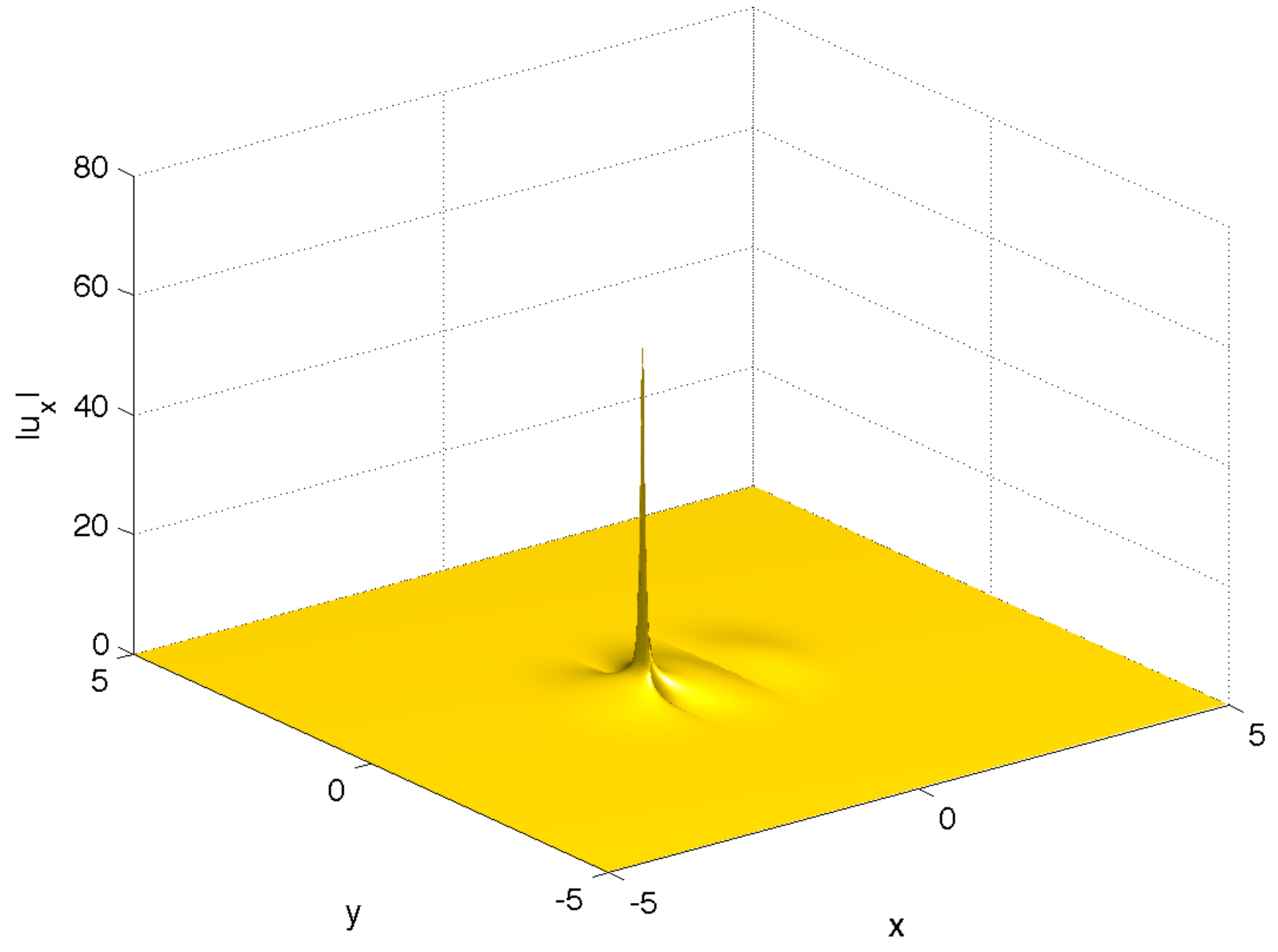}
\includegraphics[width=0.4\textwidth]{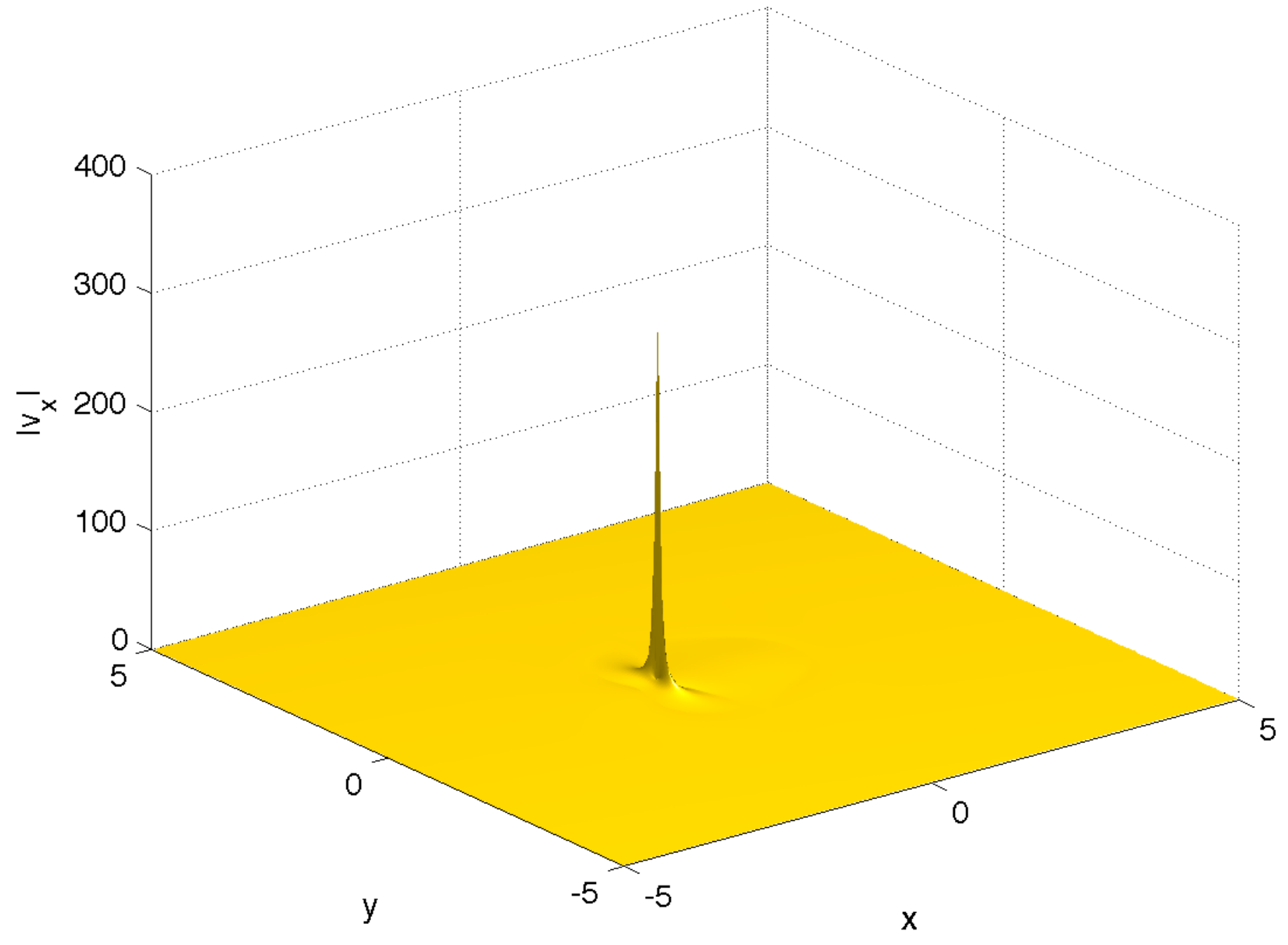}
\caption{$x$ derivative of $u$ on the left, and of $v$ on the right, at $t=t_c=0.3007$; $(u,v)$ being the solution of 
the $2+1$ dimensional elliptic dispersionless Toda equation (\ref{toda2e0}) 
with $\rho=-1$ for initial data of the form (\ref{uini2}).}
\label{gradtcuvellip2d}
\end{center}
\end{figure}

The solutions $u$ and $v$ 
 at $t_c=0.3007$ can be seen in Fig. \ref{Soluvtsellip2d}. Visibly  
 $u$ develops a cusp in $x$-direction as in the one-dimensional case.
%
%
%
%
%
%
%
%
%
%
%
The $x$ derivatives of the solution diverge as can be seen from
the behavior of $|u_x|$ and $|v_x|$ at $t=t_c=0.3007$ in 
Fig.~\ref{gradtcuvellip2d}.  
We observe at this time that $\| u_x \|_{\infty} \sim 60$ and $\| v_x \|_{\infty} \sim 280$. 

\begin{figure}[htb!]
\begin{center}
\includegraphics[width=0.4\textwidth]{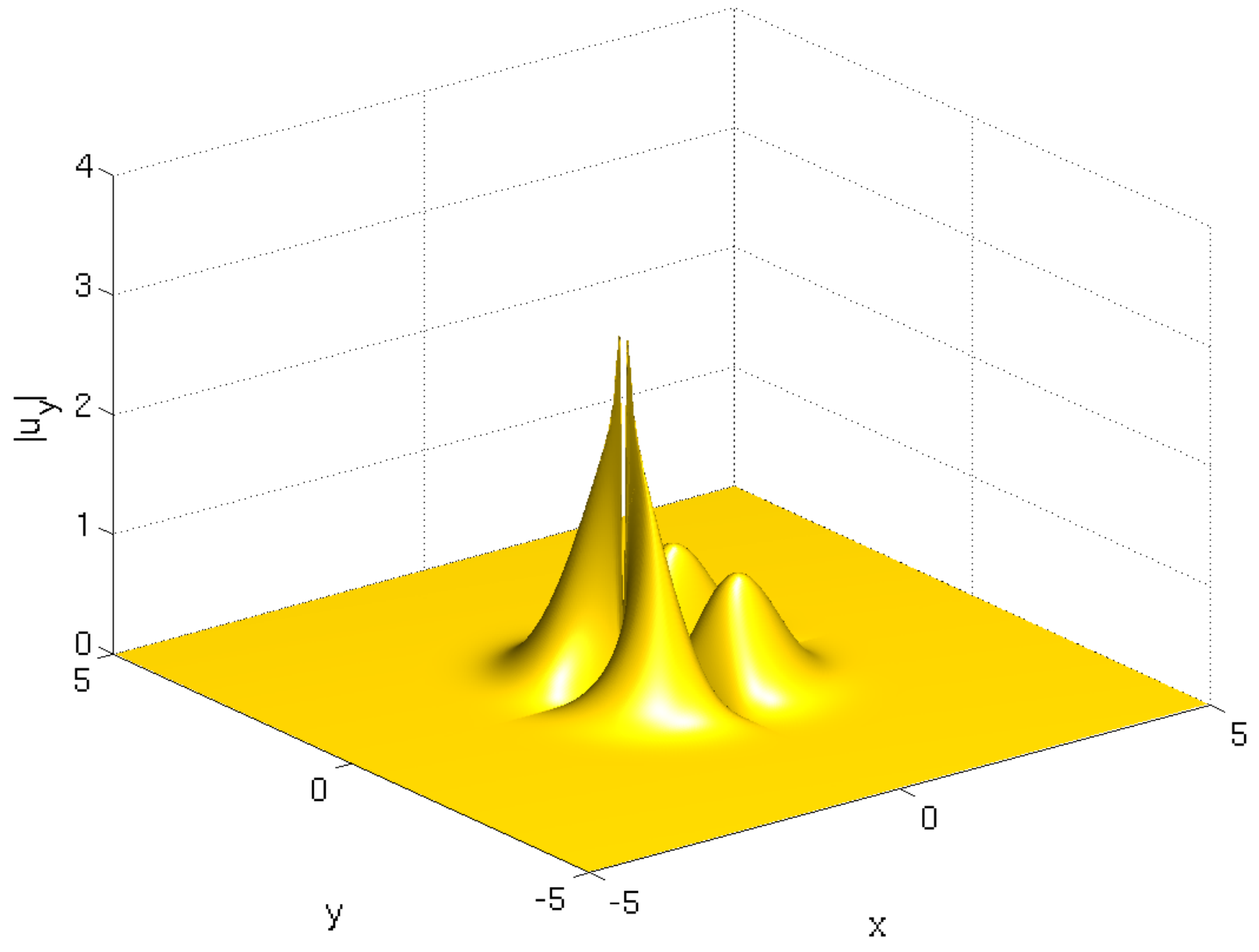}
\includegraphics[width=0.4\textwidth]{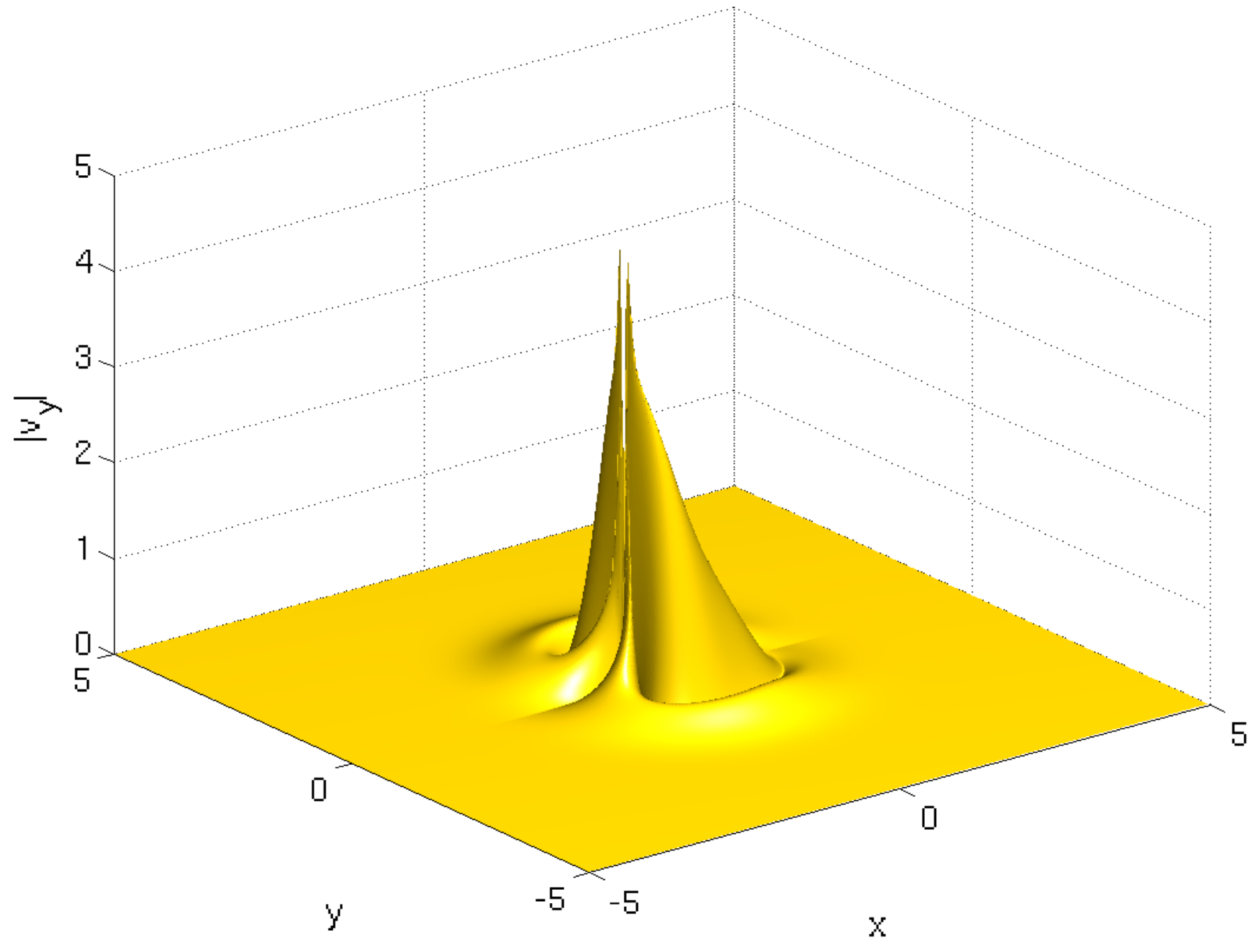}
\caption{$y$ derivative of $u$ on the left, and of $v$ on the right, at $t=t_c=0.3007$; $(u,v)$ being the solution of 
the $2+1$ dimensional elliptic dispersionless Toda equation (\ref{toda2e0}) 
with $\rho=-1$ for initial data of the form (\ref{uini2}).}
\label{gradtcuvyellip2d}
\end{center}
\end{figure}

The singularity is one-dimensional, one finds that the 
$y$-derivatives of $u$ and $v$ remain small at the critical time, see Fig. \ref{gradtcuvyellip2d}.

\begin{figure}[htb!]
\begin{center}
\includegraphics[width=0.4\textwidth]{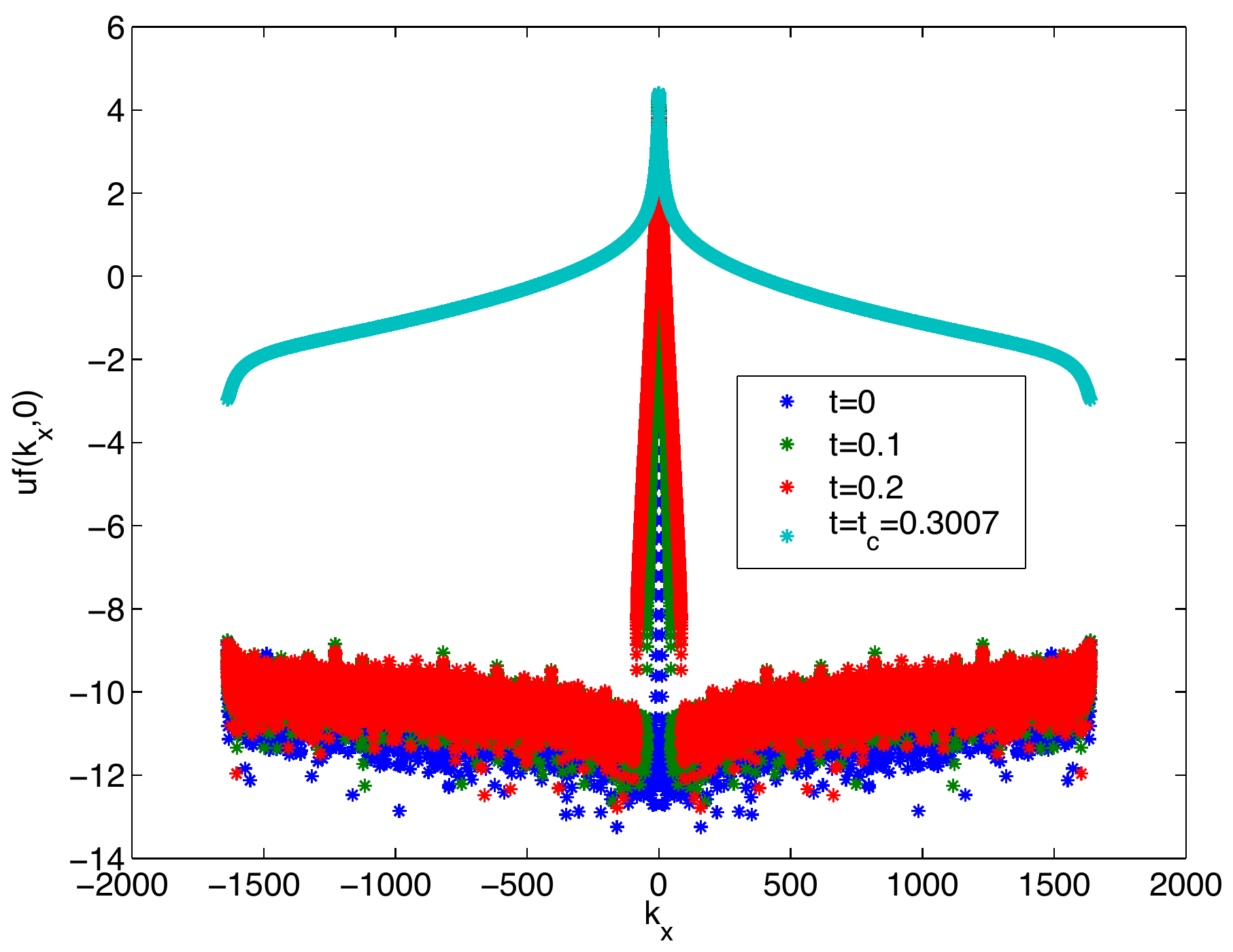}
\includegraphics[width=0.4\textwidth]{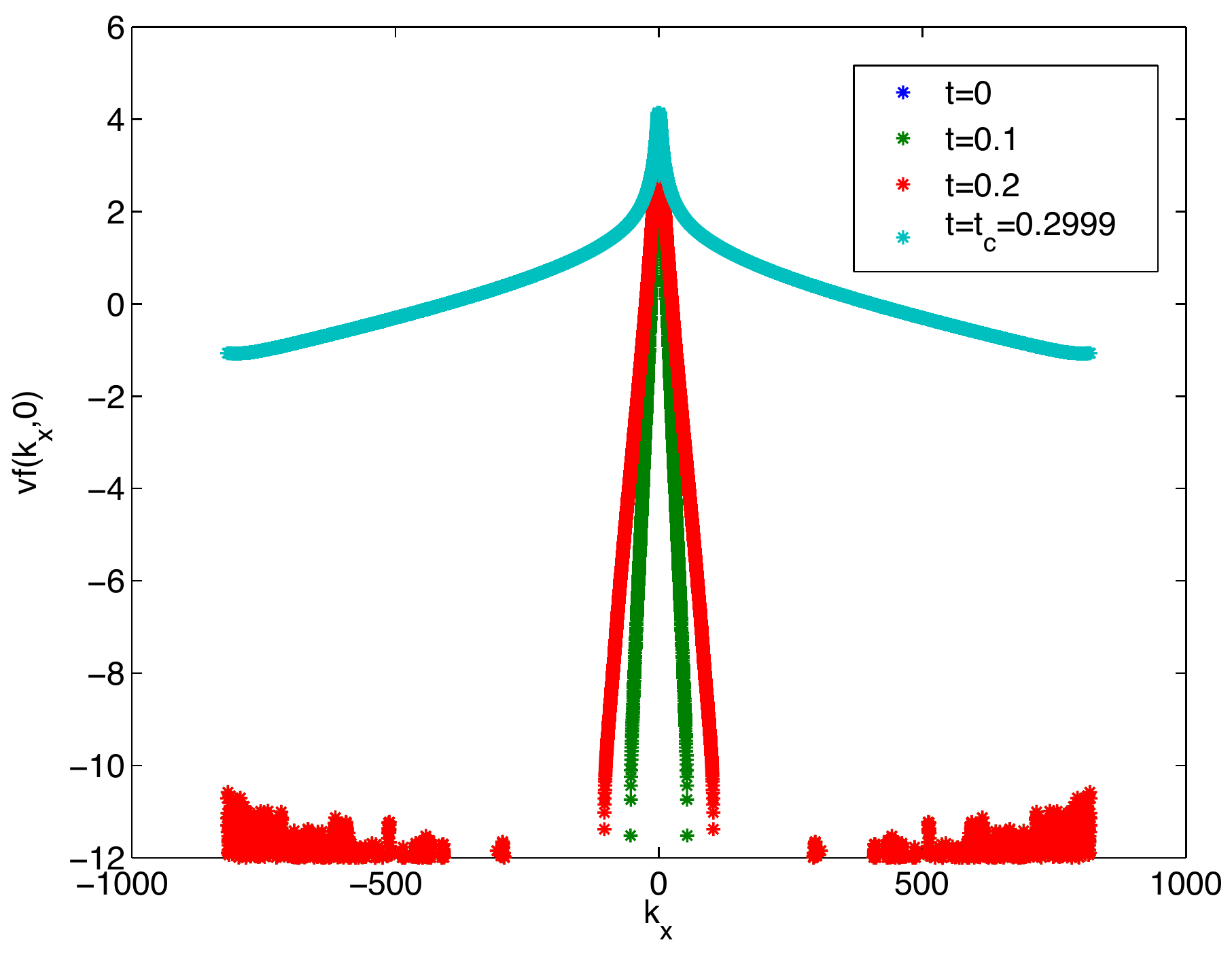}
\caption{Fourier coefficients of $u$ (left), denoted by $uf$, and
 of $v$ (right), denoted by $vf$,
plotted on the $k_x$-axis at different times corresponding to the 
situation in Fig. \ref{Soluvtsellip2d}. }
\label{Coefstsuvellip2d}
\end{center}
\end{figure}

To ensure the accuracy of the numerical solution, we again consider 
the Fourier coefficients of $u$  in Fig. \ref{Coefstsuvellip2d} at 
several times. It can be seen that the solution is numerically well 
resolved up to the formation of the singularity. 

\subsection{Small dispersion limit of the elliptic two-dimensional Toda 
equation}
In this subsection we numerically solve the $2+1$ dimensional 
elliptic Toda equation for the initial data (\ref{uini2}) for small 
nonzero $\epsilon$. We show that the difference between the solution 
to the 2d dispersionless Toda equation and the Toda equation with small 
$\epsilon$ scales as in the $1+1$ dimensional case as 
$\epsilon^{2/5}$ at the critical time $t_{c}$. This is the same behavior as found for the focusing 
DS II equation in \cite{DSdDS}. For times larger 
than $t_{c}$, we find as in the $1+1$ dimensional case an 
$L_{\infty}$ blow-up.

\begin{figure}[htb!]
\begin{center}
\includegraphics[width=0.45\textwidth]{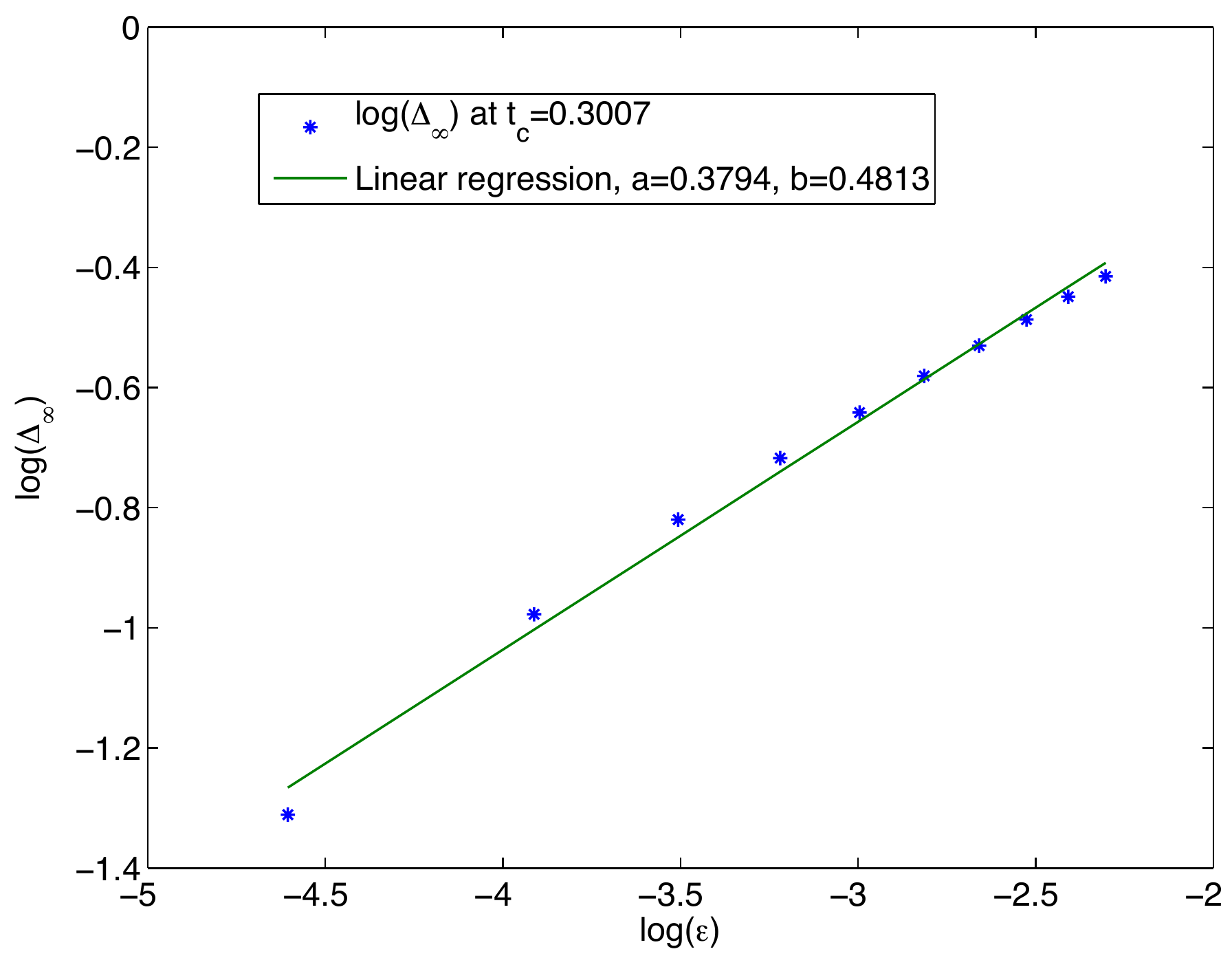}
\caption{$L_{\infty}$ norm  $\Delta_{\infty}$ of the difference 
 between the solutions to $2+1$ dimensional elliptic dispersionless 
 Toda and the Toda equation
 for the initial data (\ref{uini2}) in dependence of $\epsilon$ at 
 $t_c=0.3007$.}
\label{scalingtcellip2d}
\end{center}
\end{figure}

We compute the solution to the elliptic Toda equation 
(\ref{toda2cont}) with $\rho=-1$ for different values of $\epsilon$ 
with  $2^{14} \times 2^{9}$ points 
for $x \times y \in [-5\pi, 5\pi] \times [-5\pi, 5\pi]$ and time step 
$\delta_t=5*10^{-5}$.
We first study the solution until the critical time $t_c=0.3007$ of 
the corresponding dispersionless system. 
We are interested in the scaling in $\epsilon$ of the $L_{\infty}$ 
norm $\Delta_{\infty}$ of the difference between the solution to the 
2d elliptic 
dispersionless Toda equation
(\ref{toda2e0}) and the Toda equation (\ref{toda2cont}) for the same 
initial data (\ref{uini2}). This difference is shown in Fig. 
\ref{scalingtcellip2d} at $t_c=0.3007$ for $0.01 \leq \epsilon \leq 
0.1$.     

A linear regression analysis ($\log_{10} \Delta_{\infty} = a \log_{10} \epsilon + b$ ) shows that $\Delta_{\infty}$ decreases as 
\begin{align}
\mathcal{O} \left( \epsilon^{0.40} \right)  \sim \mathcal{O} \left( \epsilon^{2/5} \right) \,\, \mbox{at} \,\, t=t_c=0.3007, \,\, \mbox{with}\,\, a=0.3794 \,\,\mbox{and} \,\,  b= 0.4813.
\end{align}
The correlation coefficient is $r = 0.999$. 

\begin{figure}[htb!]
\begin{center}
\includegraphics[width=0.45\textwidth]{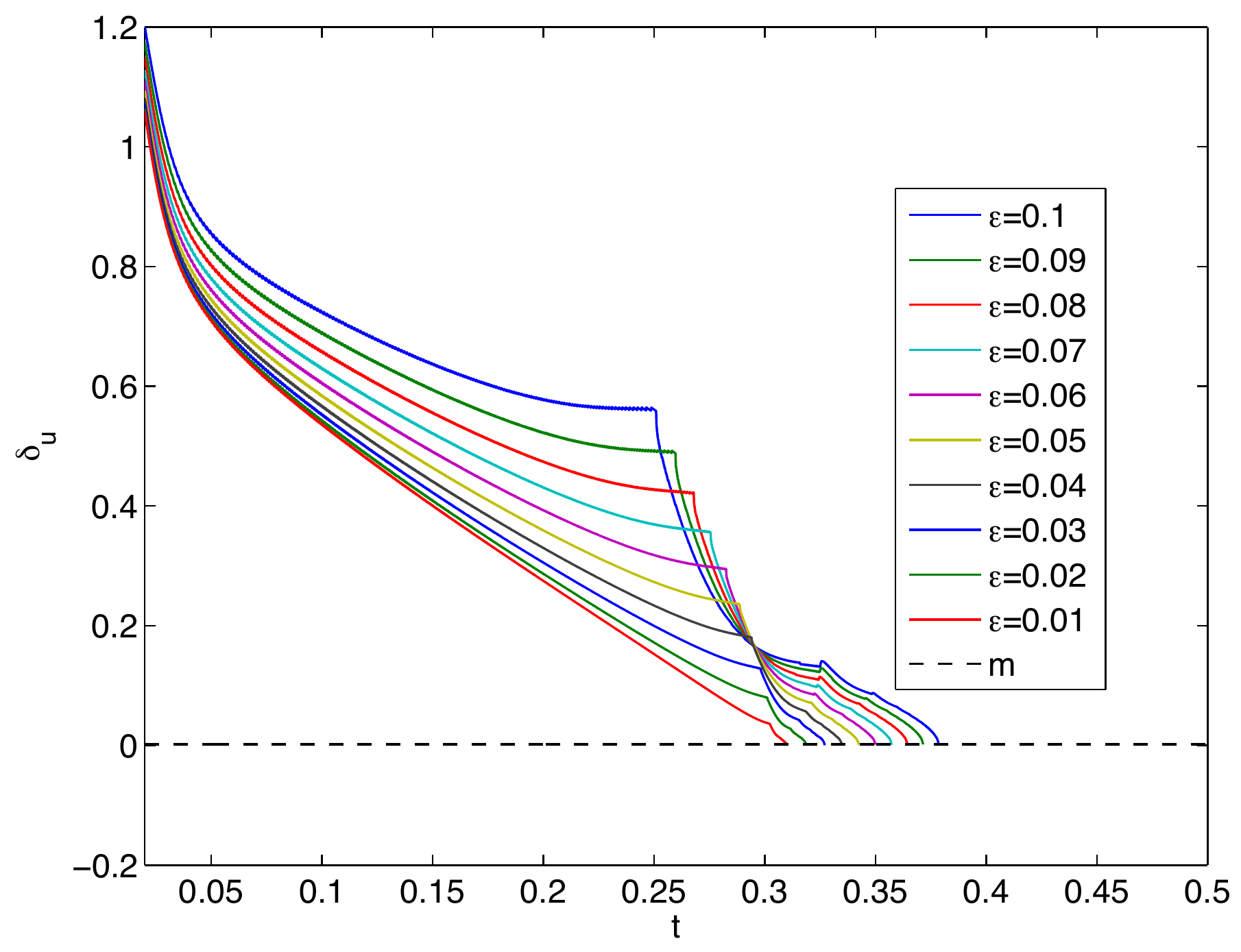}
\includegraphics[width=0.45\textwidth]{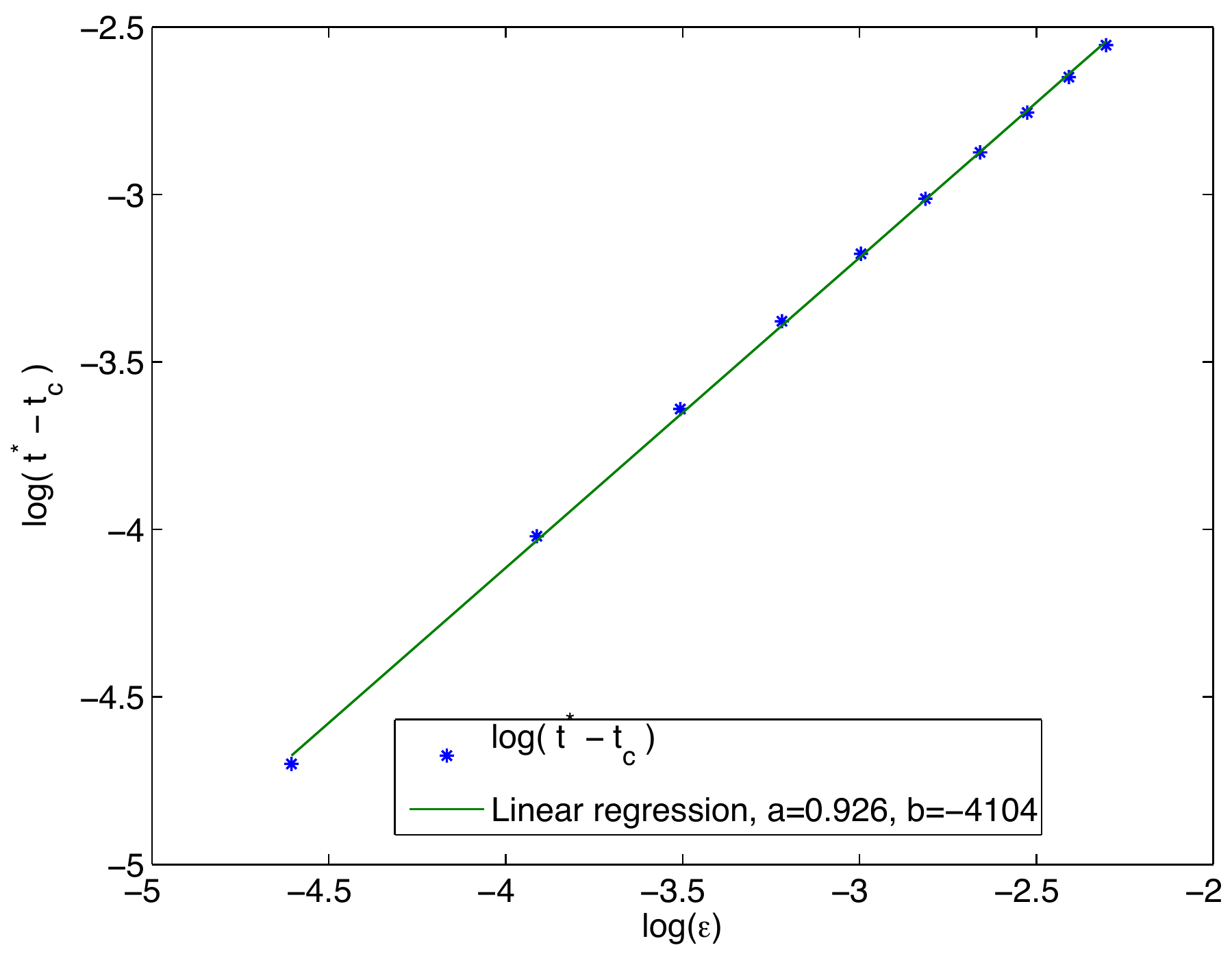}
\caption{Fitting parameter $\delta_u$ (the $\delta$ in 
(\ref{fourasymp}) for $u$) for the solution of the $2+1$ dimensional 
elliptic Toda equation (\ref{toda2cont}) with $\rho=-1$ with initial data of the form \ref{uini2}
on the left, and the $L_{\infty}$ norm   of the difference 
 between blow-up time $t^{*}$ and break-up time $t_c$ in dependence of $\epsilon$ on the right 
 in a loglog plot.}
\label{deluellipeps2d}
\end{center}
\end{figure}

For times $t>t_{c}$ we find as in the $1+1$ dimensional
case that the solution blows up in finite time. 
Again we use the asymptotic analysis of the Fourier coefficients to 
obtain the blow-up time $t^{*}$ as the time  when $\delta_u$ (the 
$\delta$ in (\ref{fourasymp}) for $u$) becomes smaller than the 
smallest resolved distance in physical space (\ref{mres}). In Fig. 
\ref{deluellipeps2d} we show the time evolution of $\delta_u$ for several values of $\epsilon$. 

 The thus determined blow-up times $t^{*}$ are given in 
 Table \ref{elblowtimes2d} for different values of $\epsilon$. The $L_{\infty}$ norm of $(t^{*}-t_c)$ 
 shows a scaling with $\epsilon$ of the form $\epsilon^{0.9}$ as in the 
 $1+1$ dimensional case, see Fig. \ref{deluellipeps2d}. 

\begin{table}
\centering
\begin{tabular}{|c | cccccccccc|}
\hline
  $\epsilon$ & $0.1$ & $0.09$ & $0.08$ & $0.07$ &  $0.06$ & $0.05$ & $0.04$ & $0.03$ & $0.02$ & $0.01$ \\
  \hline
 &&&&&&&&&& \\
  $t^{*}_{\epsilon}$ &      0.3785  &  0.3714  &  0.3643  &  0.3572  &  0.3498 &   0.3424  &  0.3348 & 0.3270  &  0.3186 &   0.3098 \\
  \hline
 \end{tabular}
\label{elblowtimes2d}
\caption{Values of the determined blow-up times of the solutions to 
the 2d elliptic Toda equation (\ref{toda2cont})   with $\rho=-1$ for 
initial data of the form (\ref{uini2}) for several values of $\epsilon$.}
\end{table}

\section{Outlook}
In this paper we have studied numerically the Toda equations in $1+1$ 
and $2+1$ dimensions. The goal was to investigate various critical 
regimes of the equations, especially the continuum limit. It was 
shown that the used numerical tools are able also to address the 
formation of singularities in a reliable way. This allowed to 
identify break-up and blow-up in solutions to these equations and 
even to identify the scaling of the solutions in critical parameters 
in the vicinity of critical points.

We concentrated here on the completely integrable Toda system since 
the numerical results were intended to encourage analytical efforts 
to prove or enhance the found conjectures. Since this is most likely 
for integrable systems for which powerful analytical methods as RHPs 
exist, we only considered integrable cases. The used numerical 
techniques are, however, in no way limited to integrable systems. It 
is the goal of further research to explore not integrable FPU models 
and address similar questions as for Toda in this paper. As in 
\cite{DGK11} for the generalized KdV equation, an interesting issue 
will be to explore to which extent the found features here are 
universal also for non-integrable systems, or how they will change if 
the studied system is not close to integrability. 

\section*{Acknowledgement}
We thank B.~Dubrovin, E.~Ferapontov and A.~Mikhailov for helpful 
discussions and hints.


\def\cprime{$'$}

\end{document}